\documentclass[fleqn,usenatbib]{mnras}

\usepackage{float}

\usepackage[T1]{fontenc}
\usepackage{amssymb}

\usepackage{enumitem}
\newlist{legal}{enumerate}{10}
\setlist[legal]{label*=\arabic*.}

\DeclareRobustCommand{\VAN}[3]{#2}
\let\VANthebibliography\thebibliography
\def\thebibliography{\DeclareRobustCommand{\VAN}[3]{##3}\VANthebibliography}

\usepackage[version=3]{mhchem}
\usepackage{mathtools, cuted}
\usepackage{cuted}       

\usepackage{graphicx}	
\usepackage{amsmath}	

\usepackage{widetext}

\title[Comparative biosignatures]{Comparative biosignatures with systemic retrievals}

\author[T. Constantinou et al.]{
Tereza Constantinou,$^{1}$\thanks{E-mail: tc496@cam.ac.uk}
Oliver Shorttle,$^{1,2}$
Miles Cranmer,$^{1,4,5}$
and Paul B. Rimmer$^{3}$
\\
$^{1}$Institute of Astronomy, University of Cambridge, Madingley Road, Cambridge, CB3 0HA, UK\\
$^{2}$Department of Earth Sciences, University of Cambridge, Downing Street, Cambridge CB2 3EQ, UK\\
$^{3}$Cavendish Astrophysics, University of Cambridge, JJ Thomson Avenue, Cambridge CB3 0HE, UK \\
$^{4}$Department of Applied Mathematics and Theoretical Physics, \\
\hspace{11em}
University of Cambridge, Wilberforce Road, Cambridge CB3 0WA, UK \\
$^{5}$Kavli Institute for Cosmology, University of Cambridge, Madingley Road, Cambridge CB3 0HA, UK
}

\date{Accepted XXX. Received YYY; in original form ZZZ}

\pubyear{2025}

\begin{document}
\label{firstpage}
\pagerange{\pageref{firstpage}--\pageref{lastpage}}
\maketitle
\begin{abstract}
The discovery of inhabited exoplanets hinges on identifying biosignature gases.  JWST can reveal biosignature gases, though current discoveries have yet to evidence life. The central challenge is attribution: how can we confidently identify biogenic sources while ruling out, or deeming unlikely, abiotic explanations?  Attribution is particularly difficult for individual planets, especially given the stochastic abiotic processes that can set atmospheric conditions. To address this, we propose a\ {\emph{comparative}} multi-planet approach centred on systemic retrievals: the analysis of multiple planets within a system to empirically define the `abiotic baseline'. This baseline, constructed from obligate uninhabited planets, serves as a local reference point. Systemic retrievals enable marginalisation over inaccessible, latent, shared abiotic parameters within planet evolution models. This is possible because planets within a system are linked by their birth in the same natal disk, have been irradiated by the same evolving star, and have a linked dynamical history. Observations aligning with the abiotic baseline, where the locally-informed abiotic planet evolution models demonstrate high out-of-sample predictive accuracy, are likely non-biological. Potentially biological anomalies are identified as statistical outliers from the abiotic baseline using Bayesian leave-one-out cross-validation. A comparative biosignature is thus defined: an anomaly where a biotic planetary evolution model provides a superior fit than its abiotic counterpart. Where both abiotic and biotic models yield poor predictive accuracy, the anomaly is flagged as an ``unknown unknown''; a signature of either unconstrained abiotic processes, or life as we don’t yet know it. 
\end{abstract}

\begin{keywords}
astrobiology -- planets and satellites: atmospheres -- planets and satellites: terrestrial planets -- Earth.
\end{keywords}



\section{Introduction}

While Earth remains our only known example of an inhabited planet, it provides the only information on what life's signals might be like.  Observable signals of biological origin, biosignatures \citep{des2008nasa}, may come in several forms \citep[][and references therein]{schwieterman2018exoplanet}.  One biosignature that has long been the focus of the search for life is radio emission.  Radio searches represent a search specifically for technologically capable (`intelligent') life, through its active transmission of electromagnetic radiation into the galaxy \citep{1959NaturCocconi, 2001Tarter, 2024AJChoza}.  However, intelligent life represents only a brief moment in the history of the Earth's biosphere.  Looking towards Earth at a random point over its history, our ability to identify the presence of its biosphere through remote observation is more likely to rest on identifying passive biosignatures: those signals a biosphere imprints on its environment through core metabolic function \citep{2012AsBioSeager, seager2016}.

Passive biosignatures are, however, rarely indisputable evidence of life; abiotic processes may generate the same signals, or modulate the biotic signal. When searched for as gases in a planet's atmosphere, biosignatures need careful interpretation, employing significant geological, atmospheric, and stellar context.  Phosphine is usefully illustrative of this point: in the oxidising atmospheres of rocky planets, phosphine has been suggested to be a powerful biosignature \citep{sousa2020phosphine}. However, in the cold reducing atmosphere of Jupiter, phosphine is expected to be abundant abiotically \citep{prinn1975phosphine, HOWLAND1979301}.  For phosphine, and for biosignatures generally, context is key.

\subsection{Finding biosignatures of biogenic origin}
Our ability to confidently identify a biosignature gas as biotic in origin depends on four crucial factors \citep{seager2025prospects}: 
\begin{enumerate}[align=left]
    \item Establishing the presence of features in the data -- i.e., precise observations, and characterisation of systematic sources of error \citep[e.g.,][]{Schlawin2021};
    \item Unambiguously linking features in the data to the presence of molecules -- i.e., gases need to be sufficiently abundant at atmospheric heights that can be probed, spectral features need to be distinct from other molecules \citep{sousa2019molecular}, and robust data reduction and atmospheric retrieval approaches \citep{constantinou2023early};
    \item Ruling in a biotic explanation for observed molecules -- through knowledge of what biosphere-scale signals life produces through interaction with its environment \citep{Lovelock1974, 2015ESS350001S}; and,
    \item Ruling out an abiotic explanation -- requiring sufficient understanding of the abiotic processes on the planet that could also produce biosignature molecules \citep{Catling2018,Krissansen2022, gillen2023call}.
    \end{enumerate}
Observational limits on biosignature detection are beginning to be overcome, addressing factors (i) and (ii). Advances in instrumentation, chiefly the launch of the James Webb Space Telescope (JWST) \citep[e.g.,][]{beichman2014observations, 2020ATremblay, 2021Gialluca, constantinou2023early} and in the near future the first light of the Extremely Large Telescope \citep[ELT;][]{2023Padovani}, herald a new era in our capacity to characterise exoplanet atmospheres. 

With JWST, biosignature gases are beginning to be reported in the atmospheres of sub-Neptunes, with an early example being methane detected in K2-18 b’s atmosphere \citep{madhusudhan2023carbon}. The same study also reported a possible dimethyl sulfide (DMS) feature, though its significance varied with the data reduction procedure and was questioned by \citet{schmidt2025comprehensive}. A subsequent claim of a DMS detection with JWST/MIRI \citep{madhusudhan2025new} has also been debated \citep{taylor2025there, welbanks2025challenges, luque2025insufficient}. These discussions underscore that there are still important systematics to overcome factors (i) and (ii). Nonetheless, these developments begin to shift the challenge towards disentangling abiotic and biotic atmospheric processes in an effort to better identify the origin of observed gases.

Factors (iii) and (iv) are key, as claiming biogenicity for a given observation will centre on whether abiotic processes can generate the same signature or not. E.g., for the case of biosignature gases, whether abiotic processes can supply them to the atmosphere at sufficient rates to maintain observed abundances, counteracting the species' photolytic and geological sinks. Addressing this requires robust biogeochemical and photochemical models that can simulate abiotic gas production and destruction under varying planetary conditions. Beyond just biosignature \emph{gases}, factors (iii) and (iv) are also critical for seeking objects, patterns or substances that could be created by life, and for distinguishing them from abiotic sources \citep[e.g., assessing candidate biosignatures on Mars;][]{mcmahon2022false}.

Detecting a single biosignature gas is unlikely to provide strong evidence for the presence of life on a planet. For a single datum in a vast possibility space, there will exist many ambiguities in its interpretation and origin \citep{Krissansen2022, 2022arXiv221014293M}. In the one-planet paradigm, a key approach to mitigate this ambiguity is to use more than a single datum: biosignature pairs, i.e,. sets of gases that together evidence disequilibrium chemistry. Observation of such an atmospheric chemical disequilibrium raises the probability of biological activity maintaining (one or more of) these gases to a non-equilibrium state \citep{1965Lovelock, 1967Hitchcock, 1975Lovelock, 1993Sagan}. Otherwise, when left for long enough \citep[which at low temperature could require geological timescales][]{liggins2022growth}, disequilibrium gases will react towards equilibrium.  For example, the simultaneous detection of abundant methane (\ce{CH4}) and carbon dioxide (\ce{CO2}) in a habitable-zone rocky exoplanet may be a biosignature, especially in the absence of CO, as, in this context, abiotic sources would struggle to sustain large methane fluxes \citep{Krissansen2018}. While the strategy described above seeks evidence for the biogenicity of signals within a single atmosphere, the multi-planet paradigm we propose here offers a more informed context, by the definition of a system-wide baseline to identify relative anomalies.

Whether biosignature gases are observed singly, or as ensembles in planetary atmospheres, there always remains the question over the efficacy of abiotic processes to have created the same signal, or to be modulating the signal observed. To attribute a signal to life would be an extraordinary claim, and as noted by Carl Sagan ``extraordinary claims require extraordinary evidence'' \citep{truzzi1978extraordinary, sagan2011broca}. Existing biosignature definitions cover a spectrum of required evidence levels. At the most conservative end, a biosignature is a phenomenon for which biological processes are a known possible explanation, and whose potential abiotic origins have been reasonably explored and ruled out \citep{gillen2023call}. A less strict definition is ``a biosignature is any substance, group of substances, or phenomenon that provides evidence of life'' \citep{Catling2018}. 

Adopting a Bayesian approach, \citet{Catling2018} elaborate a scheme to calculate the posterior probability of life’s existence, given the observational evidence. In this scheme, a biosignature would be the observational evidence that yields a high likelihood of life. More universal still is the concept from \citet{cleland2019moving} of searching for ``potentially biological anomalies (as opposed to life per se) using tentative (vs. defining) criteria''. We here combine the \citet{cleland2019moving} approach of looking for anomalies with the \citet{Catling2018} approach of calculating posterior probabilities of life explaining the anomalous data.

\subsection{Comparative biosignatures}
We can identify anomalies from two different paradigms of abiotic baseline definitions \citep{mcmahon2024astrobiology}.

First, we could have a baseline defined by the best-fit physical-chemical model of planetary processes. Here, the anomaly is defined by the failure of a model reproduction; either because of the presence of life (biology) on the planet, or could be due to the absence of some abiotic process from the model.  Either way, initial analysis of the data under this paradigm points to a `potentially biological anomaly'.

Alternatively, our expectations about a planet's characteristics could be set empirically, by a large number of observations of similar planets and formalising statistical relationships between their observable properties. In this case, we would have distributions of, e.g., atmospheric mixing ratios for the major species in such planets' atmospheres according to their mass and instellation, and a particular planet may appear anomalous by lying far from this empirical distribution. Under the assumption that life is rare, our empirical distribution is the abiotic baseline, and the planet once again appears to host an anomaly of potentially biological origin. Such examples are detailed in Appendix \ref{appendix:biosig}.

The diagnostic power of such an empirical calibration is best illustrated by the tentative detection of DMS on K2-18b \citep{madhusudhan2025new}. In a single-planet paradigm, such a detection is compelling simply because current abiotic models lack efficient production pathways for the molecule. A comparative approach, however, imposes a stricter empirical test: if DMS were observed on a physically similar but definitively uninhabitable neighbour, the signal would be subsumed into the system’s common abiotic baseline. The claim for biogenicity is thus weakened not by the discovery of new laboratory chemistry, but purely by the molecule’s ubiquity in situ.

The two paradigms above for searching for life each have strengths and weaknesses. Two key weaknesses of the physics-chemistry informed approach are (1) that we require sufficiently process-complete models to form a useful baseline, and (2) physical processes can be stochastic and historical, with no observational fingerprint today (e.g., giant impacts).  

The empirical approach to defining the abiotic baseline also has two key weaknesses.  (1), is its reliance on the assumption that life is rare.  At least with abiotic models we can know the processes of life are not captured by them, but if life is widespread, its presence on planets in the sample will contaminate the abiotic baseline so defined. The consequence is that it will be harder to find life with such an abiotic baseline. (2), and more fundamentally, the dimensionality of the parameter space relevant for capturing planetary abiotic processes is large. A large number of planets would therefore need to be discovered and characterised to have an empirical distribution with useful predictive power.  A feat that for rocky worlds may lie beyond even the next generation of planet characterising telescopes \citep{quanz2022large}.  

To mitigate these challenges in defining the abiotic baseline, we draw on the strengths of both model-based and empirical approaches and introduce the concept of `comparative biosignatures': potentially biological anomalies identified by their deviation from a system-wide abiotic baseline.

This baseline is informed by planetary evolution models, but calibrated by observations of planetary neighbours, effectively using the rest of the system as an empirical anchor for the local context \citep{Catling2018}. Such planetary evolution models would encapsulate the chemistry, physics, and geology that have shaped the planet. Identifying and ruling out non-biological signals is a key step in life detection \citep{green2021call, gillen2023call}; under this framework, planets aligning with the locally defined baseline \citep{mcmahon2024astrobiology} are likely shaped by abiotic processes. Significant outliers, however, provide a statistical definition for the `potentially biological anomalies’ of \citet{cleland2019moving}, which can then be evaluated against \emph{biotic} planetary evolution models to assess their biogenicity. If the outliers are in agreement with \emph{biotic} planetary evolution models and be accounted for by life's influence on an environment, it constitutes a comparative biosignature;  if it defies both abiotic and biotic models, it represents an anomaly of unknown origin.

\begin{figure}
	\includegraphics[width=1.1\columnwidth]{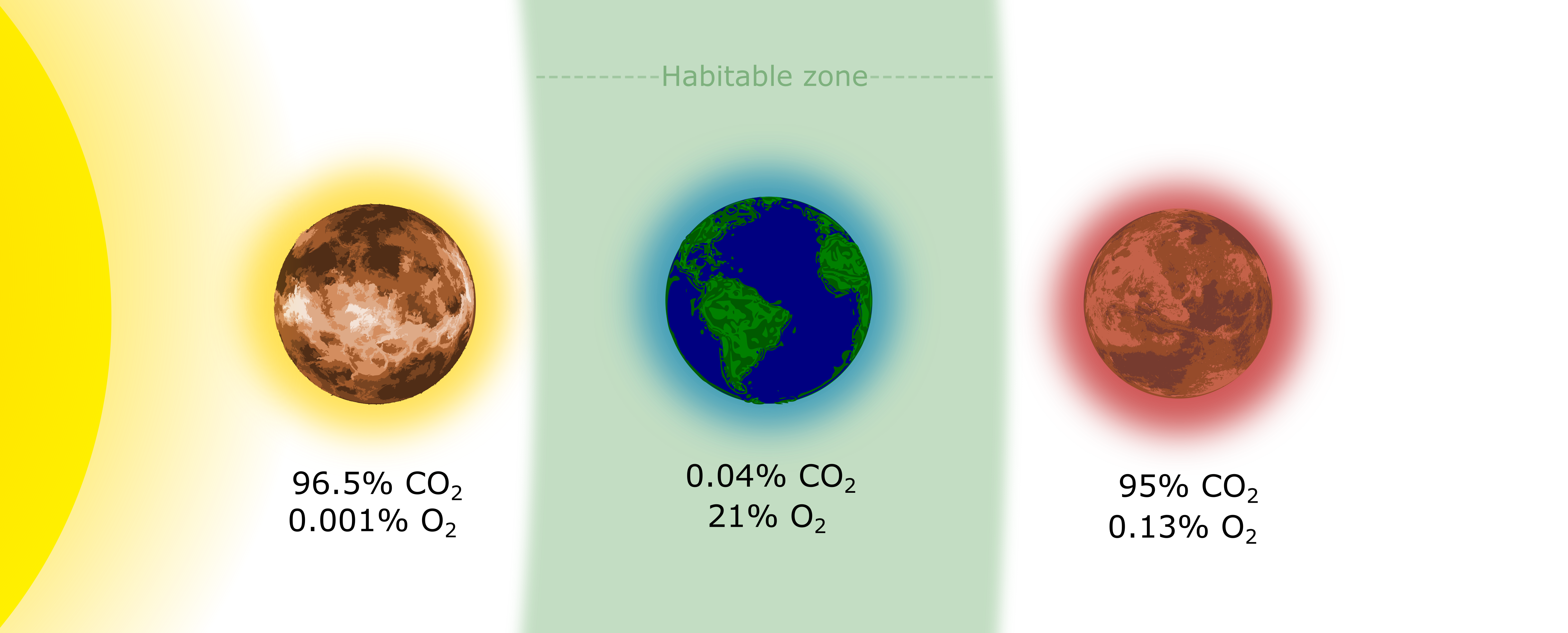}
    \caption{Earth's atmospheric composition as an example of comparative biosignatures and habsignatures. Earth, situated within the habitable zone (green annulus sector), emerges as an outlier. Compared to the abiotic baseline established by Venus and Mars, Earth exhibits significantly lower \ce{CO2} levels \citep{thienen2007ingeology} and considerably higher \ce{O2} levels (indicated below each planet). All of these gases have plausible abiotic origins, but the abiotic baseline informs our expectation and identifies Earth as anomalous.
}
\label{fig:solar_syst}
\end{figure}

Our Solar System itself demonstrates the insights drawn from this comparison. Earth's abundant atmospheric \ce{O2} and depleted \ce{CO2} \citep{Triaud2024} emerge as distinct outliers when compared to the abiotic baseline established by Venus and Mars (Fig. \ref{fig:solar_syst}). Purely abiotic models can successfully reproduce the atmospheric states of our neighbours, yet cannot predict Earth's composition. This anomaly can only be reconciled when the models are augmented with biology. This is canonically illustrated by the anomalously high \ce{O2} driven by oxygenic photosynthesis, overwhelming the input of reductive volcanic gases \citep{holland2002volcanic}, and the depleted \ce{CO2}, which necessitates a biotic drawdown mechanism \citep{Triaud2024}.

\subsection{Paper structure}

This article presents the comparative biosignature framework. Section \ref{sec:bayes} establishes the core methodology: we operationalise Bayesian approaches,  motivate the need for an empirical `abiotic baseline' (Section \ref{sec:baseline}), detail the statistical procedure for its construction using a Hierarchical Bayesian Model (Section \ref{sec:build}), and explain how this baseline is then used to identify and assess comparative biosignatures (Section \ref{sec:bio_from_baseline}). We then explore the framework's broader utility (Section \ref{sec:ab}), expand it to `habsignatures' (Section \ref{sec:habio}), and conclude with a summary and outlook (Section \ref{sec:summary}).

\section{Bayesian analysis of comparative biosignatures}\label{sec:bayes}
Proposed by \citet{Catling2018} and \citet{walker2018exoplanet}, the Bayesian approach to biosignature identification enables the assignment of confidence levels to putative detections of life.  The confidence levels are set by the Bayesian posterior probability of life on an exoplanet $p(\text{life} \: | \: \text{data, context})$ as determined given (1) the observed spectral or photometric `data' of an exoplanet that may contain biosignatures, and (2) the stellar and planetary `context' of the exoplanet that is relevant to the possibility of life producing the set of data. 

For a binary hypothesis of life as either being present or absent on an individual exoplanet and $p(\text{data, context})>0$, Bayes' theorem is applied \cite[as in][]{Catling2018} to express the posterior
\begin{equation}\label{eq:work_through}
\begin{aligned}
&p(\text{life} \:  | \: \text{data, context}) = \\
&p(\text{life} \:  | \: \text{D, C}) = \\
&\frac{ p(\text{D, C} \:  | \: \text{life}) p(\text{life}) }{p(\text{D, C} \:  | \: \text{life}) p(\text{life}) + p(\text{D, C} \:  | \: \text{no life}) p(\text{no life})},
\end{aligned}
\end{equation}
where `data' is represented by `D' and context by `C' for brevity. Using the rules of conditional probability, Equation (\ref{eq:work_through}) can be expressed as 
\begin{equation}\label{eq:bayes_02}
\begin{aligned}
&p(\text{life} \:  | \: \text{D, C}) =  \\
&\frac{p(\text{D} \: | \: \text{C, life}) p(\text{life} \: | \: \text{C})}{p(\text{D} | \text{C, life})p(\text{life} | \text{C}) +p(\text{D}| \text{C, no life})p(\text{no life} | \text{C})}
\end{aligned}
\end{equation}
In this form, $p(\text{life} \:  | \: \text{D, C})$ represents the posterior probability of life being present on a planet, weighted by the prior probability of inhabitation, and the likelihood that the data occur in the given environmental context of an inhabited planet.   

The relevant context in $p(\text{life} \: | \: \text{C})$ are the planetary and systemic parameters measured (e.g., planet mass, semi-major axis, climate, stellar flare rate) that influence abiogenesis and habitability. Conversely, $p(\text{no life} \: | \: \text{C})$ gives the probability of no life being present on a planet given the context. These priors incorporate knowledge about the probability of the origins of life as a function of the observed astrophysical and geochemical environment and the subsequent habitability of planets to support and sustain life.  

The habitable zone (HZ) \citep{1993Kasting} is often considered a proxy for habitability based on our presumption of life's need for liquid water. Within the priors $p(\text{life} \: | \: \text{C})$ and $p(\text{no life} \: | \: \text{C})$, the HZ is incorporated as a probability density function \citep{Zsom2015}: planets within the HZ possess a higher probability of harbouring (Earth-like) life, with higher $p(\text{life} \: | \: \text{C})$ (and so lower $p(\text{no life} \: | \: \text{C})$). Treating the HZ as a probabilistic prior, rather than a binary filter for the search for life, enables the potential detection of non-Earth-like life; a signal that deviates sufficiently from the abiotic baseline can compensate for the low prior probability of inhabitation outside the HZ to yield a high posterior probability of life.

Two likelihoods are also considered in Equation \ref{eq:bayes_02}. The first, $p(\text{D} \:  | \:  \text{C, life})$, gives the probability that the observed data occur in the astrophysical context of the planet, given that the planet is inhabited. The second likelihood, $p(\text{D} \:  | \:  \text{C, no life})$, represents the probability of the data occurring given the astrophysical context of the planet and that the exoplanet has no life. This second likelihood incorporates the idea of an abiotic false positive detection of life. 

The likelihood $p(\text{D} \mid \text{C, no life})$ is evaluated using abiotic planetary evolution models ($\mathcal{M}$), with latent parameters calibrated against the system-wide abiotic baseline to capture the local empirical context. Similarly, $p(\text{D} \mid \text{C, life})$ is evaluated using models extended to include the presence of life ($\mathcal{M}_L$). 

We discuss the methods in the context of exoplanet atmospheric spectra as a specific use case to aid with clarity, but the mathematics and application is in principal more general. For spectra, the forward planetary evolution models can be used to generate synthetic exoplanet spectra or photometric data ($\mathcal{S}_L$ for models with life, $\mathcal{S}$ for abiotic models). Comparing this synthetic data to actual observations allows us to estimate the likelihood that each model (where life is another process in the model, \ref{sec:bio_from_baseline}) explains the planet’s atmosphere  \citep{Catling2018} for the given context C, 
\begin{equation}
\begin{aligned}
 &p(\text{D} \mid \text{C}, \text{life}) = \prod_i^N p(\text{D}_i \mid \mathcal{S}_{L,i}, \text{C}), \\
 &p(\text{D} \mid \text{C}, \text{no life}) = \prod_i^N p(\text{D}_i \mid \mathcal{S}_i, \text{C}),
    \label{eq:life_likelihood}
\end{aligned}
\end{equation}
where $D_i$ is the $i^\text{th}$ spectroscopic or photometric data point, such that D = \{D$_i$\}$_{i=1}^N$ and each D$_i \in  \mathcal{D}$, where $\mathcal{D}$ is the space of possible data. The development of such biotic and abiotic planetary evolution models is thus vital for quantitative assessment of biosignature probabilities.

\subsection{Planetary evolution models}\label{sec:models}
To apply the comparative framework, we require models capable of predicting a planet’s atmospheric state as the outcome of its geological and astrophysical history. For any single planet, key historical parameters, such as its initial volatile inventory, cumulative stellar irradiation, and impact record, are fundamentally unobservable. This lack of constraint creates a vast parameter space where abiotic processes cannot be uniquely determined. Embedding planetary evolution models within a \emph{systemic retrieval} methodology allows these unobservables to be statistically constrained.

\begin{figure}
	\includegraphics[width=\columnwidth]{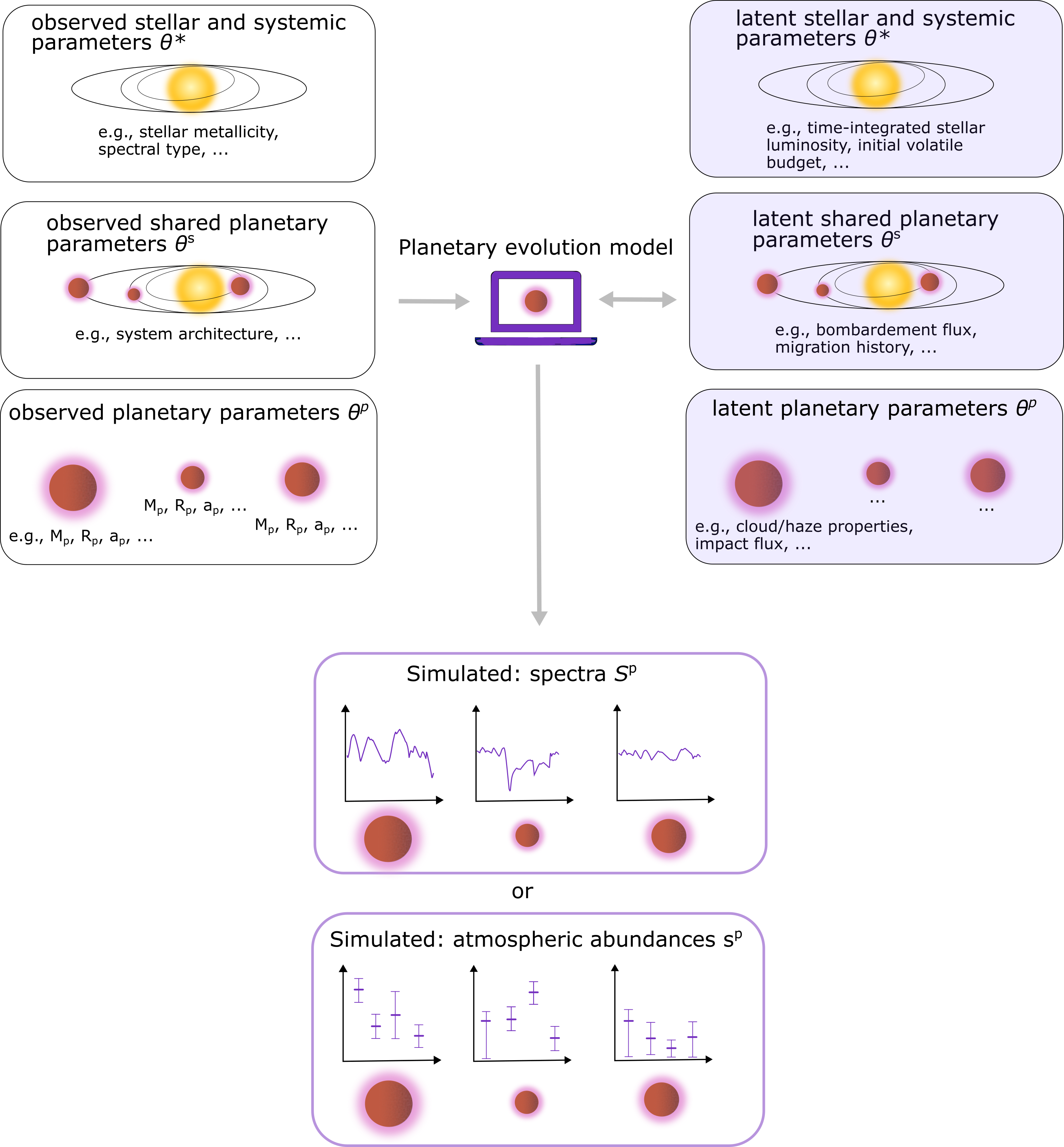}
    \caption{Schematic of the Geochemical Planet Evolution Model for Systemic Retrievals. This workflow illustrates how observations of a planetary system are used to constrain a forward model of planetary evolution. The model takes directly observed stellar and systemic parameters $\theta^*$, shared planetary parameters $\theta^s$ and planetary parameters $\theta^p$, such as mass (M$_p$), radius (R$_p$), and semi-major axis (a$_p$), and observed stellar and systemic parameters ($\theta^*$) as inputs. A key feature of this systemic approach is its ability to simultaneously model and marginalise over latent parameters --- crucial properties that are hidden from direct observation. These include latent planetary properties (e.g., cloud properties, impact flux) and, most importantly, latent system-shared parameters $\theta^*$ and  $\theta^s$ that are inaccessible when analysing planets in isolation. The geochemical model links all these parameters, both observed and latent, to produce simulated observables for each planet in the system. The final output can be either simulated spectra (S$^p$) or the inferred atmospheric abundances ($x^p$) with their associated uncertainties, which are then compared to actual observations.}

\label{fig:geo_model}
\end{figure}

At the core of this approach lies a Hierarchical Bayesian Model (HBM) that analyses all planets in a system simultaneously, using their shared context to constrain latent parameters (Fig.~\ref{fig:geo_model}). Processes unconstrained for a single planet often imprint coherent patterns across an entire system. Planetary composition, for example, reflects stellar composition modulated by an unknown efficiency of element retention \citep{wang2019enhanced, 2023Spaargaren}. Atmospheric escape depends on both planetary gravity \citep{Kubyshkina} and the star’s early UV output, itself unobserved and highly variable \citep{2012Owen, 2013ApJOwen, France_2016}. Bombardment history, which sculpts atmospheres through volatile delivery and loss, is likewise a system-level outcome of early disk mass and architecture \citep{schlichting2015atmospheric, wyatt2008evolution, wyatt2016design}. Modelling these shared effects yields an empirically anchored picture of a system’s abiotic context.

Realising this framework requires us to solve the inverse problem of planetary evolution: inferring a planet's history from its present state. Central to this is the deployment of a comprehensive forward model, incorporating existing methods of advanced atmospheric chemistry solvers \citep[e.g.,][]{tsai2017vulcan, rimmer2021hydroxide}, coupled atmosphere–interior evolution models \citep[e.g.,][]{krissansen2022predictions}, volcanic degassing models \citep[e.g.,][]{liggins2022growth}, and long-term biogeochemical cycling frameworks such as COPSE \citep{lenton2018copse}.

With the same core model, the method of inference differs based on the observational interface. We envision two complementary implementations:

\textbf{Abundance-Space Inference}. Here, the evolutionary model is fitted directly to atmospheric profiles of chemical abundances (rather than raw exoplanet spectra). This implementation serves as model validation, leveraging the data-rich environment of the Solar System where we possess model-independent constraints of chemical profiles from in-situ sampling and remote sensing. By conditioning our evolutionary models on these robust constraints, which often capture information like vertical or temporal variability unavailable for exoplanets, we can rigorously stress-test the underlying models without the confounding variables of radiative transfer or interpreting ambiguous spectra. Any inability of the model to reproduce these observed abundances within other Solar System planets isolates a specific gap in our theoretical framework, highlighting precisely where our parametrisation of physical or chemical processes is incomplete. For examples of abundance-space inference on exoplanet systems, including \ce{O2}, \ce{CH4}, and \ce{PH3}, see Appendix \ref{appendix:biosig}.

\textbf{Raw Data-Based Inference}, representing the long-term objective for exoplanetary science. In this end-to-end inversion, the evolutionary model is coupled to a radiative transfer code and fitted directly to the observed spectrum. An inference algorithm adjusts the model's inputs until the synthetic spectrum best matches the raw data, allowing for the direct inference of systemic processes while eliminating biases introduced by intermediate retrieval steps. 

It is tempting to postpone the application of such frameworks until our models are deemed `complete'. However, the framework is designed to work with evolving, imperfect models; its diagnostic power lies in identifying where those models fail. When observations cannot be reconciled with our best-fit abiotic models, the discrepancy isolates a clear target for refinement, pinpointing exactly where our physical or chemical understanding is incomplete, or where the inclusion of biology may be necessary.

The real bottle-neck lies in computational costs: performing a retrieval on the planetary evolution model. This is presently not possible, even for models that would serve as constituents. Unlike standard atmospheric retrievals that solve for a static state, evolutionary retrievals must solve for the dynamic history that produced it. The challenge lies not only in constructing sufficiently comprehensive coupled models but in developing the computational capacity to perform inference on very computationally expensive models.

The Solar System is the proving ground for this approach. Although we present an integrated methodology, applying Abundance-Space Inference locally is the immediate priority. Our system is replete with the puzzles the framework is designed to resolve, from Venusian ozone \citep{calder2025abiotic} to Uranian methane \citep{moses2020atmospheric}. These anomalies allow us to test abiotic baselines with high-precision data --- a prerequisite for extending the framework to exoplanets.

\subsection{When is a baseline most valuable?}\label{sec:baseline}
\begin{figure*}
	\includegraphics[width=\textwidth]{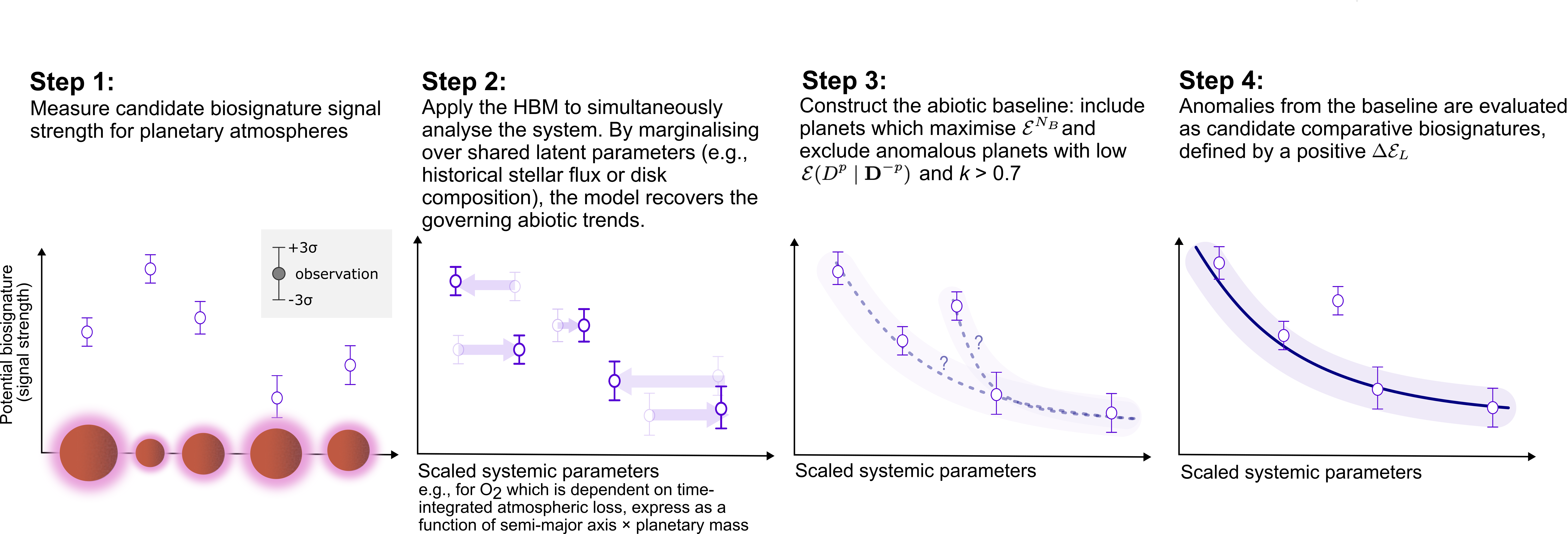}
    \caption{Procedural construction of the abiotic baseline. Step 1: Atmospheric observations are collated for the planetary system. While the inference framework operates on high-dimensional data (e.g., spectra or abundance profiles), these are visualised here as projected scalar signal strengths. Step 2: The Hierarchical Bayesian Model marginalises over shared latent parameters to constrain system-wide abiotic trends. Step 3: The abiotic baseline is iteratively defined by excluding high-influence outliers (Pareto $k > 0.7$). Step 4: These anomalies are evaluated against biotic models; those yielding $\Delta \mathcal{E}_L > 4$ are classified as comparative biosignatures.}
    \label{fig:build}
\end{figure*}

When considering the utility of the abiotic baseline, three distinct classes of candidate biosignature data emerge based on whether their abiotic formation pathways can be marginalised over:
\begin{itemize}[align=left]
    \item \textbf{Class I, Deterministic data:} These are candidate biosignatures whose abiotic origin can be confidently predicted from first principles, or with established models of planetary geochemistry and formation. In such cases, all relevant parameters influencing the abiotic disposition of these gases are either directly observable or well constrained, allowing for robust predictions independent of other planets in the system. However, quantifying the uncertainty of a biosignature requires an assessment of the extent to which we have explored the relevant possibility space \citep{vickers2023confidence}. Establishing an abiotic baseline serves as a critical test: if biosignature trends across a system align with predicted deterministic relationships, it suggests the abiotic processes are well-mapped. A clear baseline emerging when observables are scaled to systemic parameters (as in Steps 2 and 3 of Fig. \ref{fig:build}) indicates that abiotic contributions are well understood.

    \item \textbf{Class II, Data dependent on latent parameters or historic events:} These candidate biosignatures depend on abiotic processes governed by parameters that cannot be directly observed for individual planets --- either because they are inaccessible (e.g., exogenic dust flux), historical (e.g., formation conditions), or not yet known \citep[e.g., UV stellar spectrum, as in the case for Trappist 1, ][]{wilson2021mega}. A key example is cometary bombardment: though the flux varies predictably with planet mass, system architecture, and distance from the snow line \citep[e.g.,][]{Anslow}, the total amount of past and present cometary material remains unobservable. These latent parameters can only be inferred indirectly through models linking them to observable planetary properties. In establishing the abiotic baseline, we are calibrating for these latent parameters by leveraging data from planets where there is no ambiguity of their observable being of abiotic origin (Section \ref{sec:build}). Given the limited observational constraints that can be placed on exoplanets and their systems, such biosignatures are likely common. 
    
    \item \textbf{Class III, Stochastic data:} These candidate biosignatures either lack predictable relationships with planetary or systemic parameters, or result from inherently unpredictable events. Lacking a predictive model, they have no clear abiotic reference point. However, when stochastic processes affect multiple planets in a system, comparative planetology can help quantify their impact. For example, stellar flares increase atmospheric escape rates across all planets in a system \citep{2021MNRASAtri}.  If such events are modulated by known parameters --- such as atmospheric mass loss scaling with planetary mass, radius, and distance from the star \citep{2021MNRASAtri} --- the affected biosignature data can be treated similarly to Class II, depending on a latent variable, here the time-integrated stellar luminosity. In such cases, observables follow a predictable system-wide trend. Where stochastic events show no clear modulation with system parameters (e.g., giant impacts), comparative planetology has limited utility in finding life. Instead, it can provide a framework to assesses confidence in anomalies: high intra-system variance suggests abiotic randomness rather than biology, cautioning against over-interpretation. Alternatively, a lack of expected trends may indicate flawed assumptions or poor parameter selection (e.g., the x-axis in Fig. \ref{fig:build}), flagging the need for a reassessment of the model and its underlying variables.

\end{itemize}

\begin{figure}
	\includegraphics{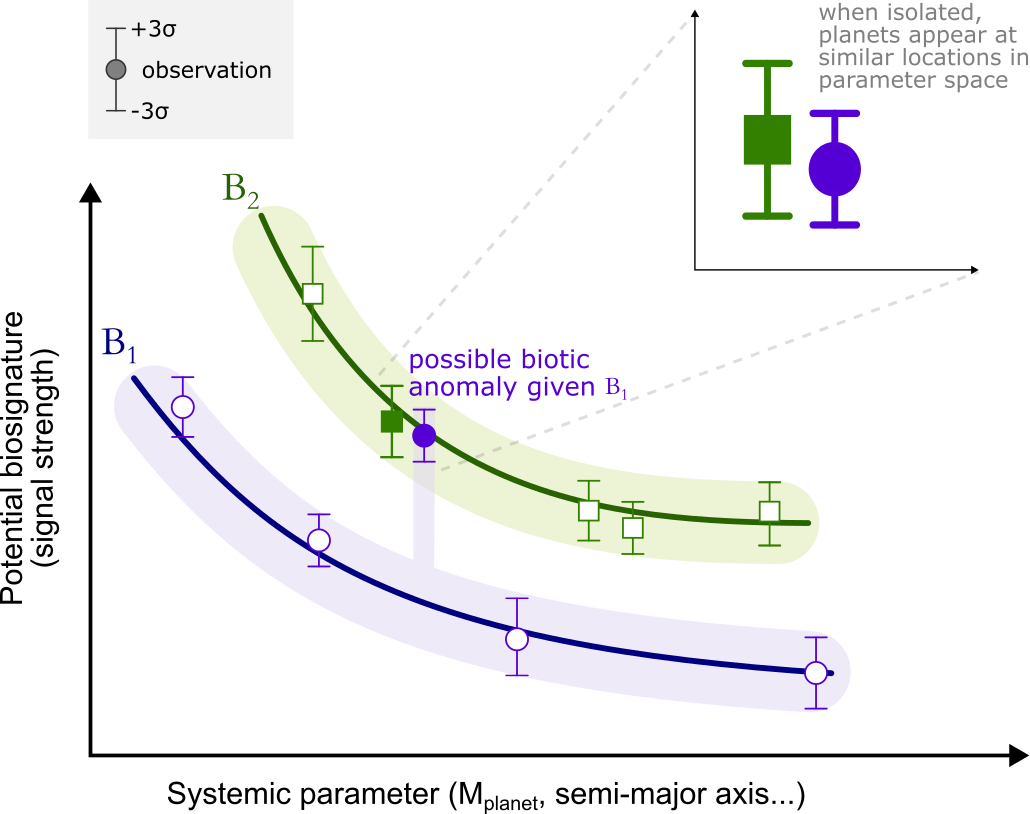}
    \caption{Illustration of comparative biosignatures using abiotic baselines. Planetary systems 1 and 2 are represented by their respective abiotic baselines, B$_1$ (purple line) and B$_2$ (green line). Candidate biosignature (or signal strength) observed for planets within each system are shown: purple circles for System 1 and green squares for System 2.  The two filled symbols are planets within the habitable zones of their respective systems.  The x-axis represents a systemic parameter (e.g., planetary mass, orbital distance); the y-axis represents observational data, which for illustration purposes are simplified to a single data point per planet. The top right panel highlights how even if two planets occupy a similar region in parameter space (i.e., show signs of biosignatures), their contextualisation within their respective planetary system baselines differs. For the case of the HZ planet from B$_2$ (filled green square), the planet's biosignature signal strength falls within agreement of the system's baseline. Conversely, for the HZ planet from B$_1$ (purple filled circle) it deviates significantly from its system's abiotic baseline, so presents a possible biotic anomaly. Such anomalies are comparative biosignatures, potentially indicative of the planets in the system having been modified by life.}
    \label{fig:comparative}
\end{figure}

Knowledge of the abiotic baseline accounts for the unknown stochastic variations between planetary systems, and for system-wide astrophysical or planetary processes that cannot be remotely constrained.  As exemplified in Fig. \ref{fig:comparative}, two planets might initially appear alike (e.g., high \ce{O2} levels, both situated in the HZ, comparable bulk densities) . However, it becomes clear that only one of the planets deviates from the  abiotic baseline of its host system -- the other has been affected by processes common to all planets in its system. It is through this comparative lens that we can discern anomalies that may hint at processes beyond the purely abiotic.

\subsection{Establishing the abiotic baseline}\label{sec:build}
The abiotic baseline is constructed by performing a \textbf{systemic retrieval} of the underlying physical parameters, $\theta$, that govern the system's evolution. The HBM organises parameters into three levels. Stellar and systemic parameters ($\theta^\ast$) describe properties influencing all planets, e.g., stellar metallicity or spectral type. Shared planetary parameters ($\theta^s$) capture system-level formation conditions, e.g., disk composition, migration history. Planet-specific parameters ($\theta^p$) encode individual features, such as planet mass, orbital distance and cloud/haze properties. For a system of $N_p$ planets,
\begin{equation}
    \theta = (\theta^\ast, \theta^s, \{\theta^p\}_{p=1}^{N_p}),
\end{equation}
where each $\theta$ lies within the total parameter space
\begin{equation}
    \theta \in \Theta \equiv \Theta^\ast \times \Theta^s \times (\Theta^p)^{N_p}.
\end{equation}

The planetary evolution model $\mathcal{M}$ serves as the forward model for the systemic retrieval. A key component of $\theta$ is the vector of simulated atmospheric abundances, $\mathbf{x} = \{x^p\}_{p=1}^{N_p}$. Within the fully-integrated framework, these abundances are latent variables inferred by the planetary evolution model $\mathcal{M}$. The model uses the inferred $\theta$ to generate a synthetic observable, e.g., a spectrum for each planet, as demonstrated in Figure \ref{fig:geo_model}. This synthetic observable is then compared to the actual observational data, $\mathbf{D} = \{D^p\}_{p=1}^{N_p}$. 

The posterior distribution for the system-wide and planet-specific latent parameters $\theta$, given the observed data $\mathbf{D}$, can be determined using Bayes' theorem by
\begin{equation}\label{eq:parameter}
\begin{aligned}
p({\theta} \:  | \: \mathbf{D} \text{, } \mathcal{M}) =  \frac{p(\mathbf{D} \: | \: {\theta},  \mathcal{M}  ) p({\theta}  \: | \: \mathcal{M})}{p({ \mathbf{D} | \: \mathcal{M})}} ,
\end{aligned}
\end{equation}
Performing this retrieval, inferring $p(\theta | \mathbf{D}, \mathcal{M})$, recovers the latent parameters that describe the system’s abiotic behaviour, effectively calibrating the model against observed atmospheric data.

Further observational constraints are drawn from $C_{\text{obs}}$, the set of measurable parameters (e.g., stellar spectrum) not captured by the planet-specific observations of $\mathbf{D}$. 

The likelihood $p(\mathbf{D} | \theta, \mathcal{M})$ in Equation \ref{eq:parameter} quantifies the probability of all the observed data given the model parameters. The abiotic model $\mathcal{M}$ with parameters $\theta$ generates a synthetic spectrum $S^p$ for each planet. Assuming the noise in each observational data point $D_i$ (e.g., each spectral bin) is independent, the likelihood is the product of the probabilities for each data point:
\begin{equation} p(\mathbf{D} \: | \: {\theta} , \mathcal{M} ) = \prod_p^{N_p} \prod_i p(D_i^p \: | \: {\theta} , \mathcal{M} ) , \label{eq:likelihood_parameter} \end{equation}
where the likelihood of each single data point $p(D_i^p \: | \: {\theta} , \mathcal{M} )$ is evaluated by comparing the observed $D_i^p$ to the synthetic $S_i^p$. This is weighted by priors $p(\theta | \mathcal{M})$. The priors can be informed from theoretical expectations by knowledge of the process, e.g., the time-integrated luminosity of the star informed by knowledge of stellar evolution and stellar type. The framework does not prescribe specific likelihood functions, but rather provides a general hierarchical structure adaptable to future analysis. 

The evidence $p(\mathbf{D} | \mathcal{M})$ in Equation \ref{eq:parameter} serves as normalisation,
\begin{equation}\label{eq:evidence_parameter}
    p(\mathbf{D} | \mathcal{M}) = \int p(\mathbf{D} | \theta, \mathcal{M}) \, p(\theta | \mathcal{M}) \, d\theta.
\end{equation}
Bayesian sampling methods such as nested or MCMC algorithms \citep{skilling2004nested, sharma2017markov} can evaluate this integral.

To distinguish between planets within the baseline and departures from it, we introduce Bayesian leave-one-out cross-validation (LOO-CV). For each planet $p$, the model $\mathcal{M}$ is fitted to all other planets ($\mathbf{D}^{-p}$), and its predictive ability tested on the omitted planet. The expected log pointwise predictive density (elpd$_\mathrm{LOO}$) measures how well the model predicts $D^p$:
\begin{equation}
\begin{aligned}
\mathcal{E}(D^{p} \mid \mathbf{D}^{-p}) &= \log p({D}^p \mid \mathbf{D}^{-p}, \theta, \mathcal{M})\\
&= \int p({D}^p | {\theta}, \mathcal{M}) p({\theta} | \mathbf{D}^{-p}, \mathcal{M}) d{\theta}.
\label{eq:elpd_loo}
\end{aligned}
\end{equation}

The likelihood $p({D}^p| {\theta}, \mathcal{M})$ is calculated as in Equation \ref{eq:likelihood_parameter}; 
\begin{equation}
   p({D}^p \: | \: {\theta} , \mathcal{M}  ) = \prod_i   p({D}^p_i \: | \: {\theta} , \mathcal{M}  ).
    \label{eq:likelihood_parameter_xp}
\end{equation}
comparing the model-generated synthetic spectrum for planet $p$ to its observed data points $D_i^p$, using the forward model $\mathcal{M}$ with parameters $\theta$. The LOO posterior distribution of the input parameters $p({\theta} | \mathbf{D}^{-p}, \mathcal{M})$ is inferred by fitting the model as in Equation \ref{eq:parameter}, 
\begin{equation}\label{eq:parameter_baselined}
\begin{aligned}
p({\theta} \:  | \: \mathbf{D}^{-p} \text{, } \mathcal{M}) =  \frac{p(\mathbf{D}^{-p} \: | \: {\theta},  \mathcal{M}  ) p({\theta}  \: | \: \mathcal{M})}{p({ \mathbf{D}^{-p} | \: \mathcal{M})}} ,
\end{aligned}
\end{equation}
critically only using data $\mathbf{D}^{-p}$: all observed data, excluding that of planet $p$. 

An intuitive way to interpret elpd$_{\text{LOO}}$ is as a measure of how ``expected” a planet’s atmospheric data is, given the trends learned from all other planets. If the model trained on $\mathbf{D}^{-p}$ predicts $D^p$ well, the $\mathcal{E}_L(D^{p} \mid \mathbf{D}^{-p})$ will be high, indicating that planet $p$ is statistically typical under the abiotic model. Conversely, if the model performs poorly in predicting $D^p$, the $\mathcal{E}_L(D^{p} \mid \mathbf{D}^{-p})$ will be low, suggesting that the planet’s atmosphere deviates from the patterns captured by the abiotic framework.

High $\mathcal{E}(D^p | \mathbf{D}^{-p})$ values indicate that $D^p$ conforms to system-wide abiotic trends, so $p$ is a planet in agreement with the abiotic baseline, whereas low values identify statistical outliers. This is especially powerful when applied to `peas in a pod' systems \citep{weiss2022architectures}; their characteristic regularity implies that latent constraints derived from the abiotic baseline will be highly predictive of the target planet, minimising parameter degeneracy.

The model performance for all $N_p$ planets is the average score
\begin{equation}
     \mathcal{E}^{N_p} = \frac{1}{N_p}\sum_{p}^{N_p}     \mathcal{E}(D^{p} \mid \mathbf{D}^{-p}).
    \label{eq:elpd_sum}
\end{equation}
The abiotic baseline is not assumed; it is constructed through an iterative process guided by the Pareto $k$ diagnostic, a measure of each planet's statistical influence on the model fit. The procedure begins by fitting the abiotic model $\mathcal{M}$ to all $N_p$ planets. Then, the $\mathcal{E}(D^p | \mathbf{D}^{-p})$ and the Pareto $k$ value are calculated for each planet. According to established statistical guidance \citep[Appendix~\ref{appendix:limitations};][]{vehtari2024pareto}, any planet with a high Pareto $k$ value ($k > 0.7$) is considered an influential outlier that is poorly predicted by the model trained on its peers. The planet with the highest $k$ value above this threshold is flagged as an anomaly and removed from the dataset. The model is then re-fit to the remaining planets, and this process is repeated until no influential outliers remain.

The final set of $N_B$ planets constitutes the robust abiotic baseline --- a self-consistent group whose properties are well-described by the abiotic model. The overall performance of this baseline is given by $\mathcal{E}^{N_B} = \frac{1}{N_B} \sum_{p=1}^{N_B} \mathcal{E}(D^{p} \mid \mathbf{D}^{-p})$. 

The computation of an exact $\mathcal{E}^{N_B}$ score requires fitting the model to infinite data. Consequently, the underlying assumption is that data $\mathbf{D}_B = \{D_i\}_{i=1}^{\text{data from N}_B}$ applied in calculating $\mathcal{E}^{N_B}$ are drawn from a complete population of all abundances from all abiotic planets, allowing the calculation of the standard error (SE) of the $\mathbf{D}_B$ data points drawn from $N_B$. The SE of the estimated $\mathcal{E}^{N_B}$ score is thus given by
\begin{equation}
    \text{SE}( \mathcal{E}^{N_B}) = \sqrt{N_B \: \text{Var}_{p=1}^{N_B} (\mathcal{E}(D^{p} \mid \mathbf{D}^{-p}))},
    \label{eq:elpd_loo_se}
\end{equation}
where Var is the Variance Operator. 

Any planets not included within the abiotic baseline, the set of $N_B$ planets, can subsequently be evaluated as potentially biotic anomalies \citep{cleland2019moving}.

\begin{figure}
	\includegraphics[width=\columnwidth]{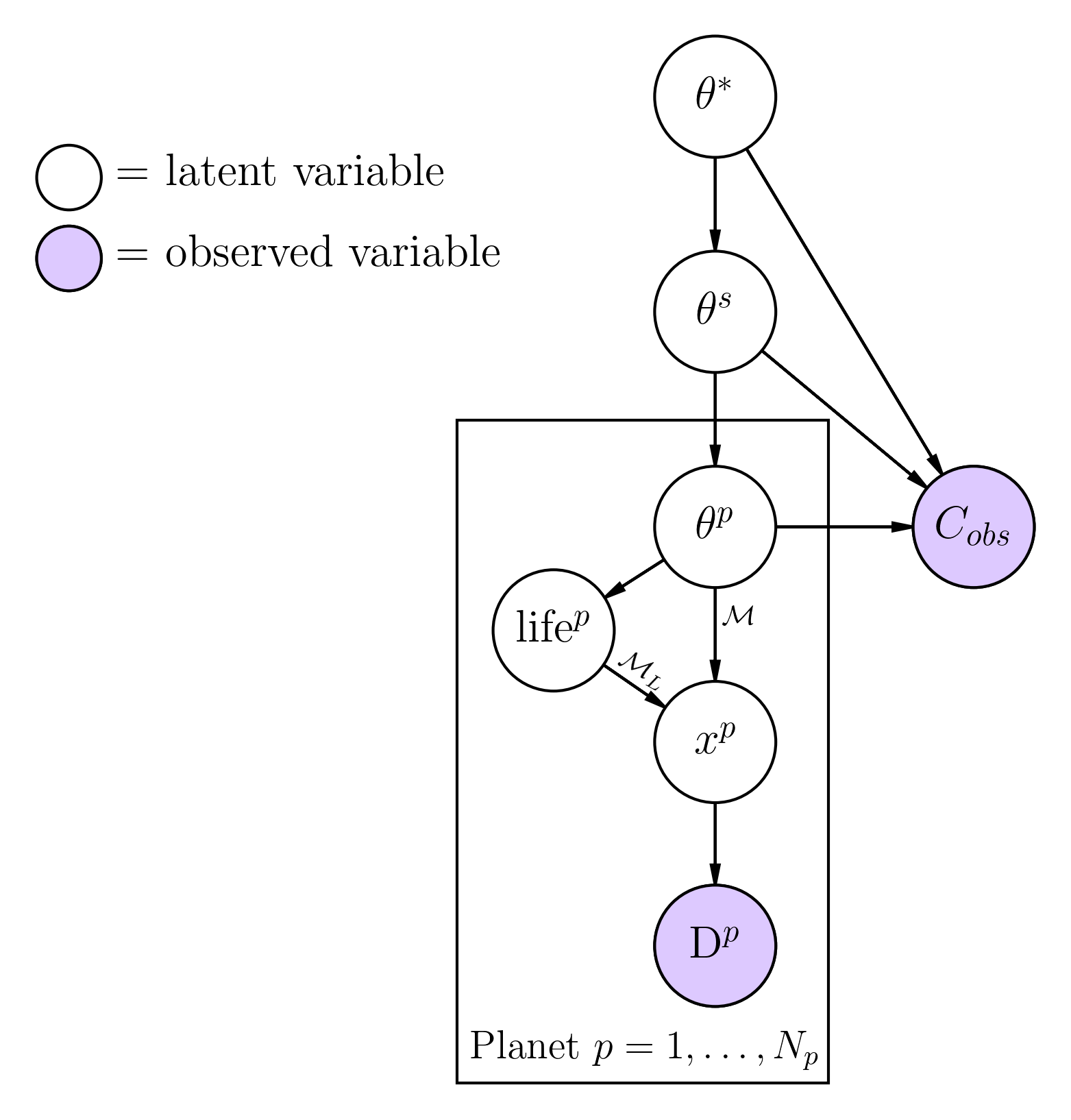}
    \caption{The Bayesian hierarchical model linking latent parameters to observables. Stellar and systemic parameters ($\theta^*$) influence shared planetary parameters ($\theta^s$), which in turn determine planet-specific parameters ($\theta^p$) for each planet $p$ of $N_p$. These parameters collectively shape the observed stellar and planetary context, $C_{\text{obs}}$ (as referenced in Section \ref{sec:bayes}). The presence or absence of life on a planet (life$^p$), along with all $\theta$ parameters, influence the abundances of atmospheric species ($x^p$), which in turn determine the planet’s observed spectral or photometric data (D$^p$). The models $\mathcal{M}$ and $\mathcal{M}_L$ represent abiotic and biotic planet evolution forward models that simulate the observations.
}
    \label{fig:model}
\end{figure}

\subsection{Comparative biosignatures from an abiotic baseline}\label{sec:bio_from_baseline}
\begin{figure*}
	\includegraphics[width=0.9\textwidth]{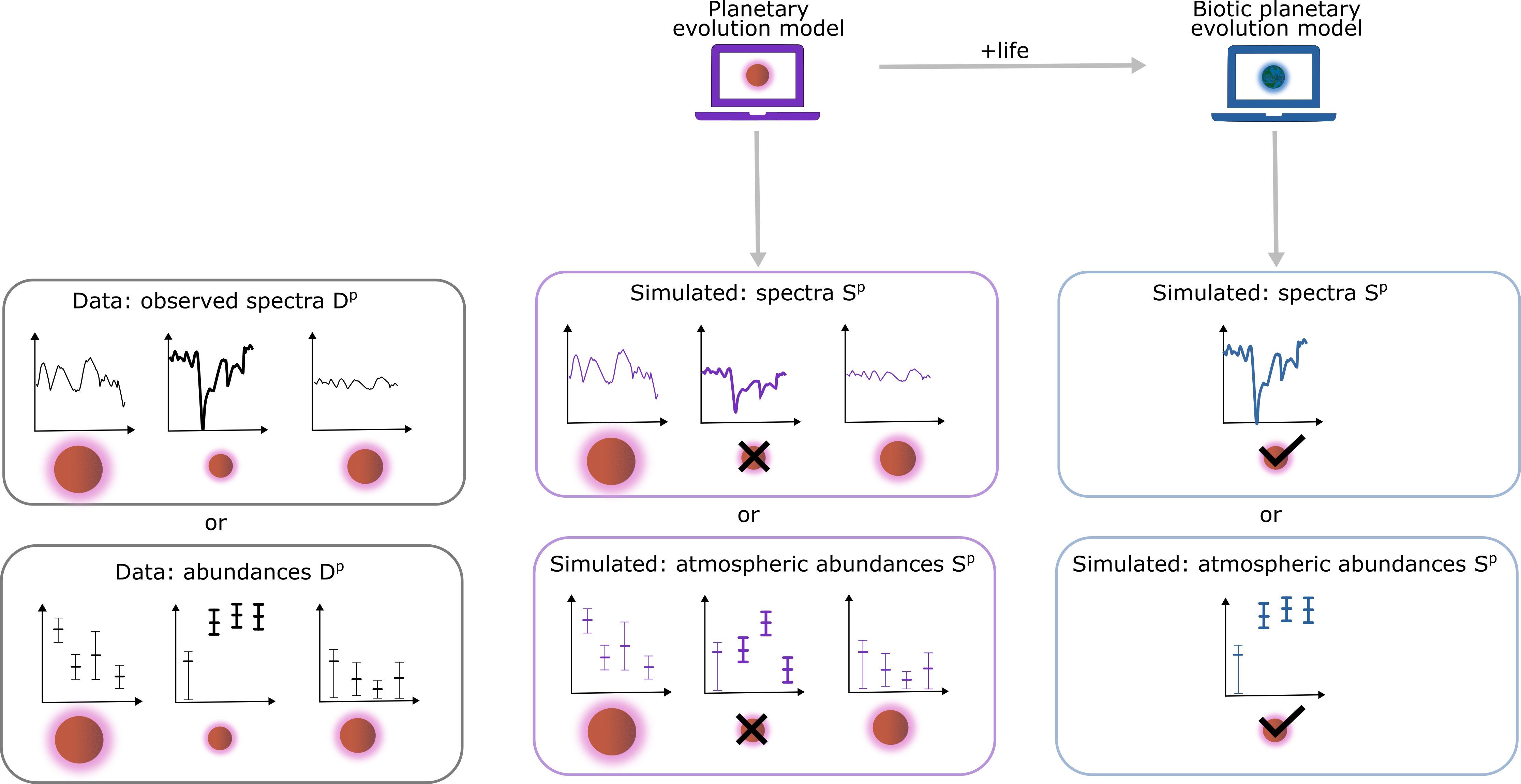}
\caption{Schematic illustrating the process of assessing a potential biosignature. First, an abiotic planetary evolution model is compared against the data (observations in black box, planetary evolution model outputs in purple box), and fitted to the non-anomalous planets in the system, excluding the potential outlier (in bold). This process establishes the abiotic baseline and allows for marginalisation over the latent systemic parameters $\theta$. Next, this calibrated abiotic model is augmented with biological processes, creating a biotic planetary evolution model. This model, now informed by the constrained $\theta$, is used to predict the atmospheric properties of the anomalous planet (outputs in purple box). The workflow is shown for both spectral data (top row) and the abundance-space inference approach with retrieved atmospheric abundances (bottom row).
}
\label{fig:bio_geo_model}
\end{figure*}

To determine whether a baseline anomaly is due to biology or unknown abiotic mechanisms, it is tested against both abiotic and biotic planetary evolution models $\mathcal{M}$ and $\mathcal{M}_L$, the latter of which incorporate life-driven processes (Figure \ref{fig:bio_geo_model}).

Biotic planetary models simulate life by introducing metabolisms that couple biological activity with geochemical and physical processes. Metabolic functions can be implemented as biogenic gas fluxes, constrained by temperature, nutrient availability, stoichiometry, redox balance, and energy sources \citep[e.g.,][]{claire2006biogeochemical, gebauer2017evolution}.  The key biological parameters --- biomass and metabolic diversity --- govern the extent and nature of atmospheric modification. By treating these as free parameters added to $\mathcal{M}$, after marginalising over $\theta$ (the abiotic latent parameters in $\mathcal{M}$; Section \ref{sec:build}), a broad spectrum of possible biotic atmospheric chemistries can be explored.

For an anomalous planet $p$, the predictive performance of biotic planetary evolution models $\mathcal{M}_L$ is compared against abiotic models  $\mathcal{M}$ to assess whether biological processes provide a better fit. The expected log posterior predictive density for observational data $D^p$  of planet $p$, is calculated by
\begin{equation}
\begin{aligned}
  \mathcal{E}_L(D^{p} \mid \mathbf{D}^{-p}, \vec{\lambda}) &= \log p({D}^p \mid \mathbf{D}^{-p}, \theta, \vec{\lambda}, \mathcal{M}_{L})\\
&= \int p({D}^p | {\theta},\vec{\lambda}, \mathcal{M}_L) p({\theta} | \mathbf{D}^{-p}, \vec{\lambda}, \mathcal{M}) d{\theta}.
    \label{eq:elpd_loo_life}
    \end{aligned}
\end{equation}
for a given set of life parameters ${\vec{\lambda}}$: metabolic diversity, of varying combinations; and biomass, from absent, to the maximal allowed by the thermodynamic and stoichiometric limits of the environment. The permissible biomass is dictated by the available redox energy and nutrient fluxes --- constraints which are themselves derived from the empirically informed abiotic context $\theta$. 

The likelihood $p({D}^p | {\theta},\vec{\lambda}, \mathcal{M}_L)$ is calculated as in Equation \ref{eq:likelihood_parameter}; 
\begin{equation}
   p({D}^p \: | \: {\theta} ,\vec{\lambda}, \mathcal{M}_L  ) =\prod_i^{N_p}   p({D}^p_i \: | \: {\theta},\vec{\lambda} , \mathcal{M}_L  ),
    \label{eq:likelihood_parameter_xa}
\end{equation}
where the likelihood of each data point $p(D^p_i | \theta, \vec{\lambda}, \mathcal{M}_L)$ is evaluated by comparing the observed data $D_i^p$ (e.g., flux in a spectral bin) to the corresponding point in the synthetic spectrum, $S_i^p$, generated by the forward biotic model $\mathcal{M}_L$ with input parameters $\theta$ and $\vec{\lambda}$.

The distribution of input parameters, $ p({\theta} | \mathbf{D}^{-p}, \mathcal{M})$ in Equation \ref{eq:elpd_loo_life}, is inferred from the obligate abiotic dataset $\mathbf{D}^{-p}$ using Equation \ref{eq:parameter}. Crucially, this inference is performed using the abiotic forward model $\mathcal{M}$, as the abiotic model is a subset of the biotic model. The abiotic parameters ${\theta}$ must first be determined within the abiotic baseline framework before being applied to the biotic model.   

To determine whether biotic or abiotic processes best explain an anomalous observation, the difference in elpd$_{\text{LOO}}$ scores between the abiotic model and the best-fitting biotic model is computed. The broad parameter space of \( {\vec{\lambda}} \) spans a wide range of $ \mathcal{E}_L(D^{p} \mid \mathbf{D}^{-p}, \vec{\lambda})$ values. To ensure the best model comparison, the biotic model's elpd$_{\text{LOO}}$ is taken as the highest $\mathcal{E}_L(D^{p} \mid \mathbf{D}^{-p}, \vec{\lambda})$ value, where $  \mathcal{E}_L(D^{p} \mid \mathbf{D}^{-p}) = \max(  \mathcal{E}_L(D^{p} \mid \mathbf{D}^{-p}, \vec{\lambda}))$, assuming no constraints on \( {\vec{\lambda}} \). The elpd$_{\text{LOO}}$ difference between biotic and abiotic models is given by:
\begin{equation}
    \Delta \mathcal{E}_L = \mathcal{E}_L(D^{p} \mid \mathbf{D}^{-p}) - \mathcal{E}(D^{p} \mid \mathbf{D}^{-p}).
    \label{eq:elpd_difference}
\end{equation}
A value of $\Delta \mathcal{E}_L > 4$ indicates significant support for the biotic model ($\mathcal{M}_L$) over the abiotic alternative \citep{VehtariFAQ, sivula2025uncertainty}, thereby defining the anomaly as a comparative biosignature. However, this threshold must be interpreted with caution due to the statistical limitations detailed in Appendix \ref{appendix:limitations}.

For comparative biosignatures, $ \Delta \mathcal{E}_L $ can be combined with other metrics to strengthen the inference (Appendix \ref{appendix:limitations}), like the Bayesian posterior probability of life $p(\text{life} \:  | \: \text{D, C})$ (Equation \ref{eq:bayes_02}).  The two metrics offer complementary insights: $ \Delta \mathcal{E}_L $ evaluates whether biotic or strictly abiotic processes better simulate the anomalous data, and $p(\text{life} \:  | \: \text{D, C})$ quantifies the Bayesian posterior probability that life exists given the data observed and the given context of the exoplanet. Together, these tools refine biosignature detection by identifying planets where biotic explanations are statistically favoured.

\begin{figure*}
	\includegraphics[width=0.9\textwidth]{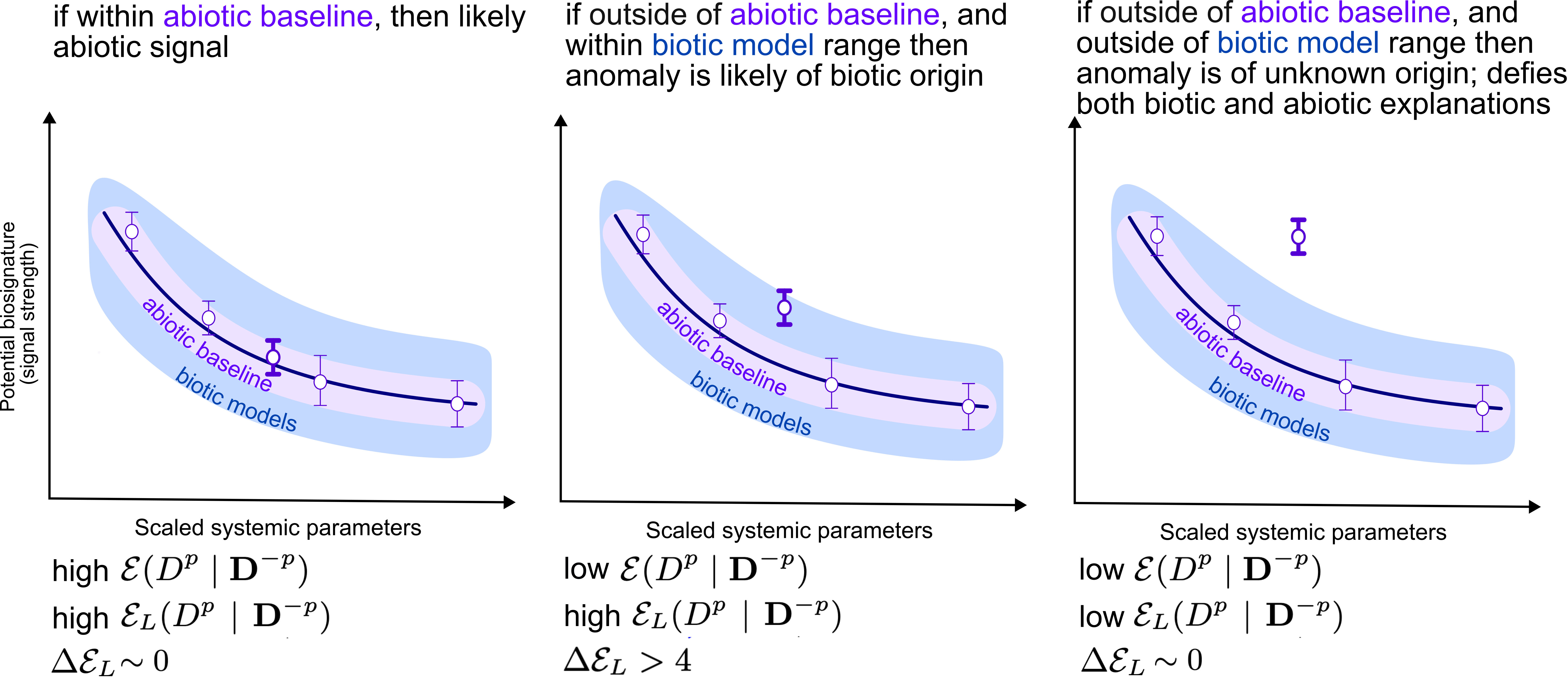}
    \caption{ Three possible scenarios in the evaluation of potential biosignature data on planet $p$ (bold data point) using an established abiotic baseline. While the inference framework operates on the full dimensionality of observational data (e.g., transmission spectra), this schematic projects the data onto a scalar `signal strength' for visualisation. Left Subplot: Observed planetary properties align with the abiotic baseline (purple line and area), resulting in a similarly high $\mathcal{E}(D^{p} \mid \mathbf{D}^{-p})$  (and $ \Delta \mathcal{E}_L \sim 0)$ for all planets . The observables are most likely shaped by known abiotic mechanisms, further evidenced by a minimal $p(\text{life} \:  | \: \text{D, C})$ and a higher $p(\text{not life} \: | \: \text{data, context})$ for all planets. Middle Subplot: One planet's observable (bold circle) deviates from the abiotic baseline but fits within Earth-like life models (blue area), leading to a $\Delta \mathcal{E}_L > 4$ for the anomalous planet $p$.  Here, life becomes the most plausible explanation for the anomaly, and further analysis is required to rule out alternative possibilities. Right Subplot: An anomaly arises that cannot be explained by current abiotic models or Earth-like biosignatures, with similarly low $\mathcal{E}(D^{p} \mid \mathbf{D}^{-p})$ and $\mathcal{E}_L(D^{p} \mid \mathbf{D}^{-p})$, such that $   \Delta \mathcal{E}_L \sim 0$. In this case, neither elpd$_{\text{LOO}}$ nor $p(\text{life/no life} \:  | \: \text{D, C})$ offer reliable confidence assessments, suggesting an unknown abiotic process or the potential for life as we do not yet know it. In this case, it is not clear which of the biotic or abiotic models should be favoured.}
\label{fig:three_options}
\end{figure*}

Three distinct cases can emerge as interpretations for the candidate biosignature, illustrated in Fig. \ref{fig:three_options}. 
\begin{legal}
     \item \textbf{All planets follow the abiotic baseline} (left subplot).In this case, the observations for all planets are well-explained by the abiotic model, yielding high $\mathcal{E}(D^{p} \mid \mathbf{D}^{-p})$ and low $k$ values across the system. The biotic model offers no significant improvement in predictive accuracy, meaning $\Delta \mathcal{E}_L \approx 0$. When the data provides no preference, the outcome of a Bayesian analysis $p(\text{life/ no life} \: | \: \text{C})$ is driven by the priors. A conservative prior that life is rare will therefore yield a high posterior probability for `no life'. This outcome confirms that the system is behaving as expected for a lifeless one.
     
    \item \textbf{An outlier deviates from the abiotic baseline, but aligns with biotic models} (middle subplot). Here, an anomalous planet exhibits a $ \Delta \mathcal{E}_L > 4$, indicating that $\mathcal{M}_L$ provides a better fit to the data compared to $\mathcal{M}$ This constitutes a comparative biosignature, suggesting that the planet’s atmospheric properties are best explained by biotic processes.  This scenario provides the strongest case for life detection, but further analysis, including a significantly higher $p(\text{life} \:  | \: \text{D, C})$ than $p(\text{no life} \:  | \: \text{D, C}$) is required for the inference from ``biotic processes as the best explanation" to a confident claim of life \citep[e.g.,][]{gillen2023call}.

    \item \textbf{An anomaly unexplained by both abiotic and biotic models} (right panel).  An anomaly is identified that defies explanation by \textit{either} our abiotic or our standard biotic planetary evolution models. Both models yield low $\mathcal{E}$ scores and high Pareto $k$ values, indicating that both fail to predict the data. The biotic model offers no improvement, so $\Delta \mathcal{E}_L \approx 0$. This is a diagnostic success, as it reveals the limits of our current understanding. The cause could be a novel, unrecognised abiotic process, an incorrectly marginalised parameter, an unconstrained process unique to this planet within the system, or it could be life as we do not yet know it. Such an outcome is an invitation for new theory and targeted observations.
\end{legal}

We emphasise that without the establishment of the abiotic baseline as a reference point, the detection of such an outlier would otherwise be challenging. In the case given in Fig. \ref{fig:three_options}, we see that the potentially biotic anomaly does not lie outside the envelope of signals seen across the planetary system. It is only in the context of its location within the system, whether spatial or otherwise, that the planet's atmosphere stands out as anomalous.

\section{Utility of the abiotic baseline}
\label{sec:ab}

\subsection{Constructing Super-systems \label{sec:super}}
The ultimate goal of the comparative biosignature approach is to extend beyond the confines of individual planetary systems by incorporating data from as many planets as possible into a single dataset: a `super-system'. Expanding the number of planets used to construct the abiotic baseline increases its statistical power, and provides a broader context for interpreting biosignature candidates. (Computational costs would also increase, Appendix \ref{appendix:compute}.)

Although this is future-orientated, anticipating a wealth of planetary observations, it also has immediate relevance. Systems may have limited observations, often constrained by the allocation of telescope time across multiple shorter projects rather than singular, in-depth observations. For such cases with limited planet characterisations, combining data from multiple systems can compensate for observational gaps. Pooling data from several systems enhances our ability to detect meaningful patterns or outliers indicative of biological activity. This is provided that the relevant latent parameters are well understood and carefully marginalised over.

When a particular physiochemical planetary process behaves consistently across systemic parameters, constructing a super-system can be straightforward. A key example is the cosmic shoreline \citep{zahnle2017cosmic, xue2024jwst}, which divides Solar System bodies into those with and without atmospheres when arranged by insolation and escape velocity \citep{zahnle2017cosmic}. This relationship may apply generally, and pinpointing the location of the cosmic shoreline beyond the Solar System is a central goal of current exoplanet science. In contrast, when a process varies between systems, observations can be rescaled to account for these differences before being combined into a single super-system dataset for building the abiotic baseline (Sec. \ref{sec:build}, Fig. \ref{fig:build}).

However, super-systems must be constructed with care. Differences between systems are often multi-parameter dependent, and overlooking critical variables can obscure important trends. In some cases, examining a single system may yield clearer insights than aggregating multiple systems, as added data could introduce confounding factors, such as system-specific stochasticity, which scramble otherwise clear signals. 

The inclusion of more data points in constructing the abiotic baseline can be tested against basic Bayesian principles: that additional data should either improve or maintain the derived likelihoods, but never reduce them. Therefore, if super-systems are responsibly constructed, they will always enhance or preserve the reliability of the abiotic baseline.

\begin{figure*}
	\includegraphics[width=\textwidth]{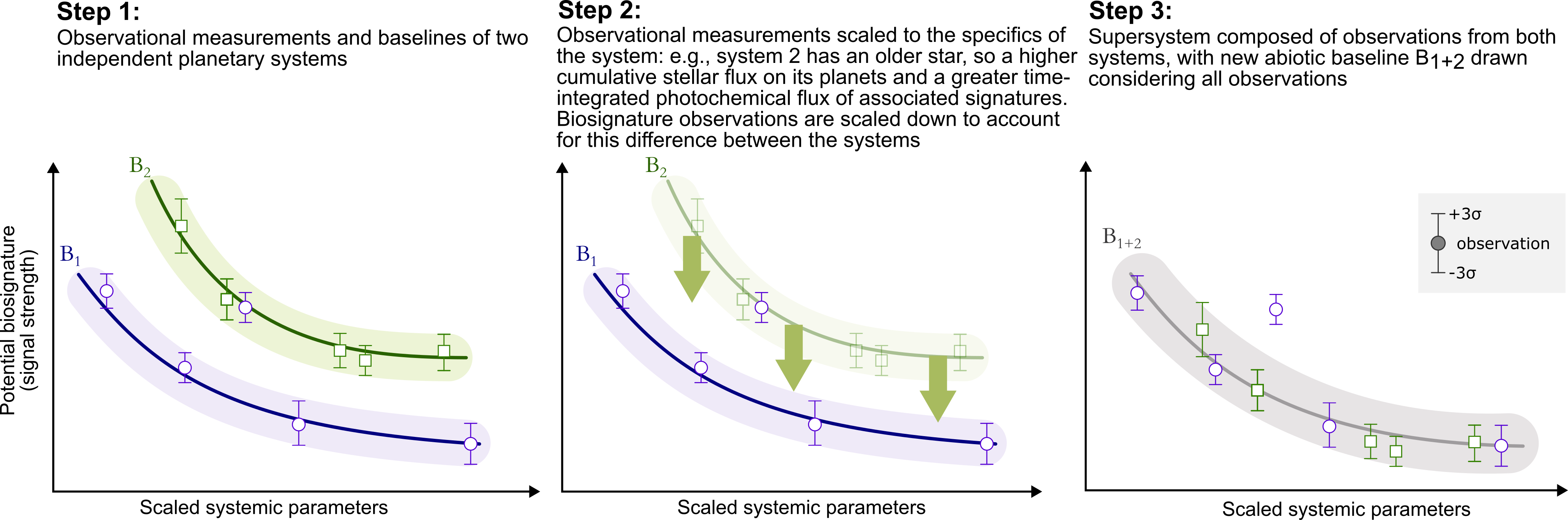}
    \caption{Methodology for synthesising the abiotic baseline for a ``super-system''. Step 1: Independent baselines ($B_1, B_2$) are established for distinct planetary systems, conditioned on their local stellar and dynamical contexts. Step 2: Observables are rescaled against governing systemic parameters to normalise for inter-system variability (e.g., correcting for differing stellar luminosities or disk compositions). Step 3: The harmonised datasets are merged to construct a unified super-system baseline ($B_{1+2}$). This aggregation expands the statistical support of the abiotic model, providing a broader and more robust reference frame for evaluating biosignature candidates.}
\label{fig:supersystems}
\end{figure*}

\section{Habsignatures}
\label{sec:habio}

The presence of surface liquid water has traditionally been the cornerstone of the identification of potentially habitable exoplanets. However, planetary habitability is a complex interplay of intrinsic (e.g., plate tectonics, magnetic fields, atmospheric composition) and extrinsic (e.g., orbital dynamics, stellar activity) factors.

In the absence of direct water detection, atmospheric composition offers valuable clues. Habsignatures --- introduced by \citet{Triaud2024} as observables indicative of a hydrosphere --- can provide such empirical diagnostics. Here we detail how elpd$_{\text{LOO}}$ can be applied to the detection of comparative habsignatures. 

Oceans, and their imprints on the atmosphere from surface-atmosphere interactions, are themselves abiotic processes, so would be in agreement with the abiotic baseline as previously defined. When searching to identify oceans, and discern them from other processes and their imprints on observables, the baseline needs redefining: instead of being composed of planets with no evidence of life, it is instead composed of planets with no evidence of oceans. This ocean-less baseline is built similarly to the abiotic baseline (Section \ref{sec:build}), where the $\mathcal{E}^{N_B}$ is iteratively evaluated to include $N_B$ planets that maximise the metric. 

With care to disentangle oceans from models, given challenges due to their interconnectedness with processes like tectonics, volcanism, and weathering,  abiotic planetary evolution models can be categorised as either with oceans ($\mathcal{M}$) or with no oceans ($\mathcal{M}_{nO}$). Planets whose data yield anomalously low $\mathcal{E}_{nO}(D^{p} \mid \mathbf{D}^{-p})$ values suggest poor model fit, suggesting that the data cannot be explained by the other abiotic processes included in the model. 

(Biotic) planetary evolution models can include oceans by simulating their role as reservoirs and regulators of chemical exchanges, accounting for processes like solute transport, mineral precipitation, and gas exchange with the atmosphere. These models often integrate ocean chemistry, circulation, and interactions with the seafloor and continental inputs \citep[e.g.,][]{caldeira2003anthropogenic, Fennel}.  

Similar to evaluating whether models with or without life better explain observations, the same approach can be applied to models with or without oceans, to now also calculate
\begin{equation}
\begin{aligned}
  \mathcal{E}_O(D^{p} \mid \mathbf{D}^{-p}, \vec{w}) &= \log p({D}^p \mid \mathbf{D}^{-p}, \theta, \vec{w}, \mathcal{M}_{O})\\
&= \int p({D}^p | {\theta},\vec{w}, \mathcal{M}_O) p({\theta} | \mathbf{D}^{-p}, \vec{w}, \mathcal{M}_{nO}) d{\theta}
    \label{eq:elpd_loo_ocean}
    \end{aligned}
\end{equation}
for a given set of ocean parameters ${\vec{w}}$, such as the ocean's total mass, pH, and age. The best-fit ocean model 
is then used for comparison against the ocean-less model, defining $  \mathcal{E}_O(D^{p} \mid \mathbf{D}^{-p}) = \max(  \mathcal{E}_O(D^{p} \mid \mathbf{D}^{-p}, \vec{w}))$. This comparison can assess the explanatory power of each model by evaluating the difference in expected log predictive density across the ocean and ocean-less model
\begin{equation}
    \Delta \mathcal{E}_O = \mathcal{E}_O(D^{p} \mid \mathbf{D}^{-p}) - \mathcal{E}_{nO}(D^{p} \mid \mathbf{D}^{-p}).
    \label{eq:elpd_difference_ocean}
\end{equation}
Here, $\Delta \mathcal{E}_O > 4$ when the data is best-explained by the model including the ocean, than the model without.   

The posterior probability of a planet's observable data being shaped by an ocean ($O$, or $nO$ for no ocean for brevity) can also be determined by
\begin{equation}\label{eq:bayes_ocean}
\begin{aligned}
&p(O \:  | \: \text{D,C}) = \\
&\frac{p(\text{D} \: | \: O \text{, C}) p(O \: | \: \text{C})}{p(\text{D | $O$, C})p(O\text{ | C}) +p(\text{D | $nO$, C})p(nO\text{ | C}) }.
\end{aligned}
\end{equation}
The priors considered here $p(O \: | \: \text{C})$ and $p(nO\: | \: \text{C})$ are the probability of an ocean being present or absent on the target planet, given stellar and systemic information that establish the context, e.g., proximity to the liquid-water HZ. For each planet, $p(O \: | \: \text{C})$ will always be greater than $p(\text{life} \: | \: \text{C})$, given that the presence of an ocean on a planet is itself a prior that is correlated with a greater probability of life as we know it. 

The two likelihoods $p(\text{D} \:  | \:  \text{C,} O)$, and $p(\text{D}\:  | \:  \text{C,} nO)$, in Equation \ref{eq:bayes_ocean} pertain to the probability that the observed data occur in the astrophysical context of the planet, given that the planet has an ocean or does not have an ocean, respectively. The likelihoods are computed similarly to Equations \ref{eq:life_likelihood}, with the ocean-less model $\mathcal{M}_{nO}$ in place of the life-less model  $\mathcal{M}$, and the planetary evolution model  $\mathcal{M}$ in place of the life more  $\mathcal{M}_L$.  $\mathcal{M}$ is used to determine $ p(\text{D}  \mid \text{C}, O)$, and $\mathcal{M}_{nO}$ is used for $ p(\text{D}  \mid \text{C}, nO)$. (For cases where D$_i$ is not Gaussian, the likelihood can be modified accordingly.) 

Given the strong link between habitability and the potential for life, we also differentiate between habsignatures (indicating suitable conditions for life) and habiosignatures (indicating both habitable conditions and biological activity) \citep{Triaud2024}. In this framework, a comparative habsignature is characterised by a  $ \Delta \mathcal{E}_O > 4$ in Equation \ref{eq:elpd_difference_ocean}.  A comparative ha\emph{bio}signature is instead characterised by a
\begin{equation}
    \Delta   \mathcal{E}_{O,L}= \mathcal{E}_{O, L}(D^{p} \mid \mathbf{D}^{-p}) - \mathcal{E}_{nO}(D^{p} \mid \mathbf{D}^{-p}).
    \label{eq:elpd_difference_ocean_life}
\end{equation}
of value greater than 4, where biotic planetary evolution models M$_{nO, L}$ have life incorporated (a varied range of biomass), but no oceans. 

This framework provides a systematic way to assess the roles of oceans and life in shaping a planet's observable characteristics, enabling a clearer understanding of their relative contributions to habitability and biosignature detection.

Recently, a `comparative' habsignature was suggested in the form of atmospheric carbon depletion \citep{Triaud2024}. Through a comparative approach, a terrestrial planet with an atmospheric carbon abundance measurably lower than others within the same system (i.e., an outlier in C atmospheric abundance) could evidence the presence of a hydrosphere (and/or of a biosphere). Specific case studies of such abundance-space habsignatures, including detailed scenarios for \ce{CO2} and \ce{NH3} depletion, are explored in Appendix \ref{appendix:habsig}.

\section{Summary}\label{sec:summary}

The search for life beyond Earth depends on identifying biosignature gases in exoplanetary atmospheres, yet distinguishing biogenic signals from abiotic processes remains a formidable challenge. Here, we present a comparative, multi-planet framework that bridges the philosophically-driven search for ``potentially biological anomalies'' proposed by \citet{cleland2019moving}, with Bayesian approaches to biosignature assessment, such as \citet{Catling2018}. 

The abiotic baseline is the set of planets within a system whose observations follow expected patterns set by abiotic processes. The calibration of an abiotic baseline (by local calibration of the planetary evolution models describing them) provides a system-specific abiotic reference point from which potentially abiotic anomalies can be discerned. The abiotic baseline is defined by the planets whose observed signals can be confidently attributed to non-biological processes. Abiotic planetary evolution models are then fit to the abiotic baseline, using Bayesian inference to determine the posterior distributions of otherwise observationally-inaccessible model parameters such as historical stellar luminosity or initial volatile inventories, enabling complete calibration of abiotic planetary evolution models. 

Potentially biological anomalies are observables that deviate from the empirically-defined abiotic baseline. These anomalies are assessed using elpd$_{\text{LOO}}$ scores, which quantify the out-of-sample predictive accuracy of models. An anomaly is most convincingly biological when the elpd$_{\text{LOO}}$ score of a biotic model exceeds that of its abiotic counterpart in explaining the observed atmospheric composition. Comparative biosignatures are thus defined as anomalies for which biotic models provide a better fit than abiotic models, as evidenced by higher elpd$_{\text{LOO}}$ scores. However, if both abiotic and biotic models perform poorly,  each yielding low elpd$_{\text{LOO}}$ scores relative to other planets in the same system, the anomaly is flagged as an unknown unknown: a signal that cannot be confidently attributed to known abiotic chemistry or existing biological theory. Such anomalies may point to previously unconsidered abiotic mechanisms or to life as we do not yet understand it, and highlight areas where both modelling frameworks and conceptual boundaries require expansion.

As with any metric, elpd$_{\text{LOO}}$ has limitations, and can be complementarily used in parallel with Bayesian posterior probabilities of life to provide a more complete assessment. These posterior probabilities reflect a weighted assessment of the prior probability of life originating on and inhabiting a planet, given the probability that the data occur in the given stellar and planetary context.

To operationalise this framework, we advocate for the creation and validation of planetary evolution models against Solar System planets. By employing `Abundance-Space Inference', we can condition our evolutionary models on the high-fidelity, model-independent constraints available for solar system planets. This approach allows us to stress-test the underlying physics of planetary evolution against hard data, and isolate theoretical deficiencies. Such deficiencies will guide targeted re-observations to probe the unexplained chemistry and systematically refine the abiotic baseline. Establishing and validating abiotic baselines locally is thus a prerequisite for applying these tools with confidence to the lower-resolution regime of exoplanets.

The ultimate objective is to apply this framework to exoplanetary systems via `Raw Data-Based Inference', performing end-to-end inversions directly from observed spectra. The primary implementation challenge lies in the computational cost of model inversion. Unlike standard atmospheric retrievals that solve for a static state, evolutionary retrievals must solve for the dynamic history that produced it. Developing the computational capacity to perform inference on such complex, coupled models, or the optimisation required to render them tractable over vast parameter spaces, represents the critical technical frontier for the next generation of biosignature science.

This comparative systems-first approach calls for a paradigm shift in the search for life, particularly in strategies for observational proposals like those for JWST, and in designing mission objectives. Instead of focussing solely on individual exoplanet targets, missions such as LIFE \citep{quanz2022large} and HWO \citep{Harada_2024} should prioritise system-level exploration --- examining planets even beyond the habitable zone --- to better understand the abiotic environments of the planets being characterised. This is especially the case for systems like TRAPPIST-1, where uniform planetary architectures simplify calibration and the larger number of planets provides the statistical power needed to robustly compare models \citep{sivula2020uncertainty}.

Multi-planet observations offer a critical advantage: an empirical constraint of abiotic processes across the system. As observational capabilities advance, applying this framework across diverse planetary systems will allow us to recognise a genuine biological anomaly when we finally encounter one.

\section*{Data availability}
 No new data were used by or were generated for this work.
 
\section*{Acknowledgements}
T.C. thanks the Science and Technology Facilities Council (STFC) for the PhD studentship (grant reference ST/X508299/1).  T.C. thanks Christopher Moore and Christian Kirkham for helpful discussions on statistics, Richard Anslow for helpful discussions on meteoritic bombardment,  Daniel Angerhausen for helpful discussions on biosignatures and space missions, and Edouard Barrier for helpful discussions on climate models. O.S. acknowledges support from STFC grant UKRI1184.
M.C. thanks Bob Carpenter for helpful clarifying discussions on some subtleties of the paper.




\bibliographystyle{mnras}
\bibliography{comparative_biosignatures}

@article{holland2002volcanic,
  title={Volcanic gases, black smokers, and the Great Oxidation Event},
  author={Holland, Heinrich D},
  journal={Geochimica et Cosmochimica acta},
  volume={66},
  number={21},
  pages={3811--3826},
  year={2002},
  publisher={Elsevier}
}

@article{wyatt2016design,
  title={How to design a planetary system for different scattering outcomes: giant impact sweet spot, maximising exocomets, scattered disks},
  author={Wyatt, MC and Bonsor, A and Jackson, AP and Marino, S and Shannon, A},
  journal={Monthly Notices of the Royal Astronomical Society},
  pages={stw2633},
  year={2016},
  publisher={Oxford University Press}
}

@article{wyatt2008evolution,
  title={Evolution of debris disks},
  author={Wyatt, Mark C},
  journal={Annu. Rev. Astron. Astrophys.},
  volume={46},
  number={1},
  pages={339--383},
  year={2008},
  publisher={Annual Reviews}
}

@ARTICLE{Catling2018,
       author = {{Catling}, David C. and {Krissansen-Totton}, Joshua and {Kiang}, Nancy Y. and {Crisp}, David and {Robinson}, Tyler D. and {DasSarma}, Shiladitya and {Rushby}, Andrew J. and {Del Genio}, Anthony and {Bains}, William and {Domagal-Goldman}, Shawn},
        title = "{Exoplanet Biosignatures: A Framework for Their Assessment}",
      journal = {Astrobiology},
         year = 2018,
        month = jun,
       volume = {18},
       number = {6},
        pages = {709-738},
          doi = {10.1089/ast.2017.1737},
archivePrefix = {arXiv},
       eprint = {1705.06381},
 primaryClass = {astro-ph.EP},
       adsurl = {https://ui.adsabs.harvard.edu/abs/2018AsBio..18..709C}
}

@article{domagal2014abiotic,
  title={Abiotic ozone and oxygen in atmospheres similar to prebiotic Earth},
  author={Domagal-Goldman, Shawn D and Segura, Ant{\'\i}gona and Claire, Mark W and Robinson, Tyler D and Meadows, Victoria S},
  journal={The Astrophysical Journal},
  volume={792},
  number={2},
  pages={90},
  year={2014},
  publisher={IOP Publishing}
}

@article{kasting2005methane,
  title={Methane and climate during the Precambrian era},
  author={Kasting, James F},
  journal={Precambrian Research},
  volume={137},
  number={3-4},
  pages={119--129},
  year={2005},
  publisher={Elsevier}
}

@article{tian2014high,
  title={High stellar FUV/NUV ratio and oxygen contents in the atmospheres of potentially habitable planets},
  author={Tian, Feng and France, Kevin and Linsky, Jeffrey L and Mauas, Pablo JD and Vieytes, Mariela C},
  journal={Earth and Planetary Science Letters},
  volume={385},
  pages={22--27},
  year={2014},
  publisher={Elsevier}
}

@article{gao2015stability,
  title={Stability of CO2 atmospheres on desiccated M dwarf exoplanets},
  author={Gao, Peter and Hu, Renyu and Robinson, Tyler D and Li, Cheng and Yung, Yuk L},
  journal={The Astrophysical Journal},
  volume={806},
  number={2},
  pages={249},
  year={2015},
  publisher={IOP Publishing}
}

@article{offre2013archaea,
  title={Archaea in biogeochemical cycles},
  author={Offre, Pierre and Spang, Anja and Schleper, Christa},
  journal={Annual review of microbiology},
  volume={67},
  pages={437--457},
  year={2013},
  publisher={Annual Reviews}
}

@article{harman2015abiotic,
  title={Abiotic O2 levels on planets around F, G, K, and M stars: possible false positives for life?},
  author={Harman, CE and Schwieterman, EW and Schottelkotte, James C and Kasting, JF},
  journal={The Astrophysical Journal},
  volume={812},
  number={2},
  pages={137},
  year={2015},
  publisher={IOP Publishing}
}

@ARTICLE{Wordsworth2014,
       author = {{Wordsworth}, Robin and {Pierrehumbert}, Raymond},
        title = "{Abiotic Oxygen-dominated Atmospheres on Terrestrial Habitable Zone Planets}",
      journal = {\apjl},
         year = 2014,
        month = apr,
       volume = {785},
       number = {2},
          eid = {L20},
        pages = {L20},
          doi = {10.1088/2041-8205/785/2/L20},
archivePrefix = {arXiv},
       eprint = {1403.2713},
 primaryClass = {astro-ph.EP},
       adsurl = {https://ui.adsabs.harvard.edu/abs/2014ApJ...785L..20W}
}

@article{Segura2003,
author = {Segura, Ant\'{\i}gona and Krelove, Kara and Kasting, James F. and Sommerlatt, Darrell and Meadows, Victoria and Crisp, David and Cohen, Martin and Mlawer, Eli},
title = {Ozone Concentrations and Ultraviolet Fluxes on Earth-Like Planets Around Other Stars},
journal = {Astrobiology},
volume = {3},
number = {4},
pages = {689-708},
year = {2003},
doi = {10.1089/153110703322736024},
    note ={PMID: 14987475},
URL = { https://doi.org/10.1089/153110703322736024},
eprint = { https://doi.org/10.1089/153110703322736024}
}

@ARTICLE{Meadows2018,
       author = {{Meadows}, Victoria S. and {Reinhard}, Christopher T. and {Arney}, Giada N. and {Parenteau}, Mary N. and {Schwieterman}, Edward W. and {Domagal-Goldman}, Shawn D. and {Lincowski}, Andrew P. and {Stapelfeldt}, Karl R. and {Rauer}, Heike and {DasSarma}, Shiladitya and {Hegde}, Siddharth and {Narita}, Norio and {Deitrick}, Russell and {Lustig-Yaeger}, Jacob and {Lyons}, Timothy W. and {Siegler}, Nicholas and {Grenfell}, J. Lee},
        title = "{Exoplanet Biosignatures: Understanding Oxygen as a Biosignature in the Context of Its Environment}",
      journal = {Astrobiology},
         year = 2018,
        month = jun,
       volume = {18},
       number = {6},
        pages = {630-662},
          doi = {10.1089/ast.2017.1727},
archivePrefix = {arXiv},
       eprint = {1705.07560},
 primaryClass = {astro-ph.EP},
       adsurl = {https://ui.adsabs.harvard.edu/abs/2018AsBio..18..630M}
}

@article{planavsky2014evidence,
  title={Evidence for oxygenic photosynthesis half a billion years before the Great Oxidation Event},
  author={Planavsky, Noah J and Asael, Dan and Hofmann, Axel and Reinhard, Christopher T and Lalonde, Stefan V and Knudsen, Andrew and Wang, Xiangli and Ossa Ossa, Frantz and Pecoits, Ernesto and Smith, Albertus JB and others},
  journal={Nature Geoscience},
  volume={7},
  number={4},
  pages={283--286},
  year={2014},
  publisher={Nature Publishing Group UK London}
}

@article{lyons2014rise,
  title={The rise of oxygen in Earth’s early ocean and atmosphere},
  author={Lyons, Timothy W and Reinhard, Christopher T and Planavsky, Noah J},
  journal={Nature},
  volume={506},
  number={7488},
  pages={307--315},
  year={2014},
  publisher={Nature Publishing Group UK London}
}

@article{planavsky2014low,
  title={Low Mid-Proterozoic atmospheric oxygen levels and the delayed rise of animals},
  author={Planavsky, Noah J and Reinhard, Christopher T and Wang, Xiangli and Thomson, Danielle and McGoldrick, Peter and Rainbird, Robert H and Johnson, Thomas and Fischer, Woodward W and Lyons, Timothy W},
  journal={science},
  volume={346},
  number={6209},
  pages={635--638},
  year={2014},
  publisher={American Association for the Advancement of Science}
}

@article{reinhard2017false,
  title={False negatives for remote life detection on ocean-bearing planets: lessons from the early Earth},
  author={Reinhard, Christopher T and Olson, Stephanie L and Schwieterman, Edward W and Lyons, Timothy W},
  journal={Astrobiology},
  volume={17},
  number={4},
  pages={287--297},
  year={2017},
  publisher={Mary Ann Liebert, Inc. 140 Huguenot Street, 3rd Floor New Rochelle, NY 10801 USA}
}

@article{Itcovitz_2022,
doi = {10.3847/PSJ/ac67a9},
url = {https://dx.doi.org/10.3847/PSJ/ac67a9},
year = {2022},
month = {may},
publisher = {The American Astronomical Society},
volume = {3},
number = {5},
pages = {115},
author = {Jonathan P. Itcovitz and Auriol S. P. Rae and Robert I. Citron and Sarah T. Stewart and Catriona A. Sinclair and Paul B. Rimmer and Oliver Shorttle},
title = {Reduced Atmospheres of Post-impact Worlds: The Early Earth},
journal = {The Planetary Science Journal}
}

@article{burgess1996mechanism,
  title={Mechanism of molybdenum nitrogenase},
  author={Burgess, Barbara K and Lowe, David J},
  journal={Chemical reviews},
  volume={96},
  number={7},
  pages={2983--3012},
  year={1996},
  publisher={ACS Publications}
}

@incollection{thienen2007ingeology,
 author = "van Thienen, P. et al.",
 booktitle = "Geology and Habitability of Terrestrial Planets",
 editor = {Fishbaugh, {K}. {E}. et al.},
 volume = {24},
 chapter = {5},
 year = {2007},
publisher = {Springer New York},
doi = {10.1007/978-0-387-74288-5}
}

@incollection{rascio2008biological,
  title={Biological nitrogen fixation},
  author={Rascio, Nicoletta and LA ROCCA, Nicoletta and others},
  booktitle={Encyclopedia of ecology},
  volume={1},
  pages={412--419},
  year={2008},
  publisher={Elsevier}
}

@ARTICLE{2010Horner,
       author = {{Horner}, J. and {Jones}, B.~W. and {Chambers}, J.},
        title = "{Jupiter - friend or foe? III: the Oort cloud comets}",
      journal = {International Journal of Astrobiology},
         year = 2010,
        month = jan,
       volume = {9},
       number = {1},
        pages = {1-10},
          doi = {10.1017/S1473550409990346},
archivePrefix = {arXiv},
       eprint = {0911.4381},
 primaryClass = {astro-ph.EP},
       adsurl = {https://ui.adsabs.harvard.edu/abs/2010IJAsB...9....1H}
}

@ARTICLE{2009Horner,
       author = {{Horner}, J. and {Jones}, B.~W.},
        title = "{Jupiter - friend or foe? II: the Centaurs}",
      journal = {International Journal of Astrobiology},
         year = 2009,
        month = apr,
       volume = {8},
       number = {2},
        pages = {75-80},
          doi = {10.1017/S1473550408004357},
archivePrefix = {arXiv},
       eprint = {0903.3305},
 primaryClass = {astro-ph.EP}
}

@ARTICLE{2012IHorner,
       author = {{Horner}, J. and {Jones}, B.~W.},
        title = "{Jupiter - friend or foe? IV: the influence of orbital eccentricity and inclination}",
      journal = {International Journal of Astrobiology},
         year = 2012,
        month = jul,
       volume = {11},
       number = {3},
        pages = {147-156},
          doi = {10.1017/S1473550412000043},
archivePrefix = {arXiv},
       eprint = {1111.3144},
 primaryClass = {astro-ph.EP},
       adsurl = {https://ui.adsabs.harvard.edu/abs/2012IJAsB..11..147H}
}

@article{Triaud2024,
author = {Triaud, Amaury H M J and de Wit, Julien and Klein, Frieder and Turbet, Martin and Rackham, Benjamin V and Niraula, Prajwal and Glidden, Ana and Jagoutz, Oliver E and Pe{\v{c}}, Matej and Petkowski, Janusz J and Seager, Sara and Selsis, Franck},
doi = {10.1038/s41550-023-02157-9},
issn = {2397-3366},
journal = {Nature Astronomy},
number = {1},
pages = {17--29},
title = {{Atmospheric carbon depletion as a tracer of water oceans and biomass on temperate terrestrial exoplanets}},
url = {https://doi.org/10.1038/s41550-023-02157-9},
volume = {8},
year = {2024}
}

@article{wordsworth2018redox,
  title={Redox evolution via gravitational differentiation on low-mass planets: implications for abiotic oxygen, water loss, and habitability},
  author={Wordsworth, RD and Schaefer, LK and Fischer, RA},
  journal={The Astronomical Journal},
  volume={155},
  number={5},
  pages={195},
  year={2018},
  publisher={IOP Publishing}
}

@article{turbet2018modeling,
  title={Modeling climate diversity, tidal dynamics and the fate of volatiles on TRAPPIST-1 planets},
  author={Turbet, Martin and Bolmont, Emeline and Leconte, Jeremy and Forget, Fran{\c{c}}ois and Selsis, Franck and Tobie, Gabriel and Caldas, Anthony and Naar, Joseph and Gillon, Micha{\"e}l},
  journal={Astronomy \& Astrophysics},
  volume={612},
  pages={A86},
  year={2018},
  publisher={EDP Sciences}
}

@article{sleep2005dioxygen,
  title={Dioxygen over geological time},
  author={Sleep, Norman H},
  journal={Metal Ions in Biological Systems, Volume 43-Biogeochemical Cycles of Elements},
  pages={49--73},
  year={2005},
  publisher={CRC Press}
}

@article{blankenship2010early,
  title={Early evolution of photosynthesis},
  author={Blankenship, Robert E},
  journal={Plant physiology},
  volume={154},
  number={2},
  pages={434--438},
  year={2010},
  publisher={American Society of Plant Biologists}
}

@INPROCEEDINGS{Zsom2015,
       author = {{Zsom}, Andras},
        title = "{A population-based Habitable Zone perspective}",
    booktitle = {IAU General Assembly},
         year = 2015,
       volume = {29},
        month = aug,
          eid = {2253795},
        pages = {2253795},
       adsurl = {https://ui.adsabs.harvard.edu/abs/2015IAUGA..2253795Z}
}

@ARTICLE{1982Hunten,
       author = {{Hunten}, D.~M.},
        title = "{Thermal and nonthermal escape mechanisms for terrestrial bodies}",
      journal = {\planss},
         year = 1982,
        month = aug,
       volume = {30},
       number = {8},
        pages = {773-783},
          doi = {10.1016/0032-0633(82)90110-6},
       adsurl = {https://ui.adsabs.harvard.edu/abs/1982P&SS...30..773H}
}

@ARTICLE{2008Johnson,
       author = {{Johnson}, R.~E. and {Combi}, M.~R. and {Fox}, J.~L. and {Ip}, W. -H. and {Leblanc}, F. and {McGrath}, M.~A. and {Shematovich}, V.~I. and {Strobel}, D.~F. and {Waite}, J.~H.},
        title = "{Exospheres and Atmospheric Escape}",
      journal = {\ssr},
         year = 2008,
        month = aug,
       volume = {139},
       number = {1-4},
        pages = {355-397},
          doi = {10.1007/s11214-008-9415-3},
       adsurl = {https://ui.adsabs.harvard.edu/abs/2008SSRv..139..355J}
}

@ARTICLE{LandB2015,
       author = {{Luger}, R. and {Barnes}, R.},
        title = "{Extreme Water Loss and Abiotic O2Buildup on Planets Throughout the Habitable Zones of M Dwarfs}",
      journal = {Astrobiology},
         year = 2015,
        month = feb,
       volume = {15},
       number = {2},
        pages = {119-143},
          doi = {10.1089/ast.2014.1231},
archivePrefix = {arXiv},
       eprint = {1411.7412},
 primaryClass = {astro-ph.EP},
       adsurl = {https://ui.adsabs.harvard.edu/abs/2015AsBio..15..119L}
}

@article{Anslow,
author = {Anslow, R. J.  and Bonsor, A.  and Rimmer, P. B. },
title = {Can comets deliver prebiotic molecules to rocky exoplanets?},
journal = {Proceedings of the Royal Society A: Mathematical, Physical and Engineering Sciences},
volume = {479},
number = {2279},
pages = {20230434},
year = {2023},
doi = {10.1098/rspa.2023.0434},
URL = {https://royalsocietypublishing.org/doi/abs/10.1098/rspa.2023.0434},
eprint = {https://royalsocietypublishing.org/doi/pdf/10.1098/rspa.2023.0434}
}

@article{CLEMENT2019778,
title = {The early instability scenario: Terrestrial planet formation during the giant planet instability, and the effect of collisional fragmentation},
journal = {Icarus},
volume = {321},
pages = {778-790},
year = {2019},
issn = {0019-1035},
doi = {https://doi.org/10.1016/j.icarus.2018.12.033},
url = {https://www.sciencedirect.com/science/article/pii/S0019103518306262},
author = {Matthew S. Clement and Nathan A. Kaib and Sean N. Raymond and John E. Chambers and Kevin J. Walsh}
}

@article{wang2019enhanced,
  title={Enhanced constraints on the interior composition and structure of terrestrial exoplanets},
  author={Wang, Haiyang S and Liu, Fan and Ireland, Trevor R and Brasser, Ramon and Yong, David and Lineweaver, Charles H},
  journal={Monthly Notices of the Royal Astronomical Society},
  volume={482},
  number={2},
  pages={2222--2233},
  year={2019},
  publisher={Oxford University Press}
}

@article{Krissansen2018,
author = {Joshua Krissansen-Totton  and Stephanie Olson  and David C. Catling },
title = {Disequilibrium biosignatures over Earth history and implications for detecting exoplanet life},
journal = {Science Advances},
volume = {4},
number = {1},
pages = {eaao5747},
year = {2018},
doi = {10.1126/sciadv.aao5747},
URL = {https://www.science.org/doi/abs/10.1126/sciadv.aao5747},
eprint = {https://www.science.org/doi/pdf/10.1126/sciadv.aao5747}}

@article{Gronoff,
author = {Gronoff, G. and Arras, P. and Baraka, S. and Bell, J. M. and Cessateur, G. and Cohen, O. and Curry, S. M. and Drake, J. J. and Elrod, M. and Erwin, J. and Garcia-Sage, K. and Garraffo, C. and Glocer, A. and Heavens, N. G. and Lovato, K. and Maggiolo, R. and Parkinson, C. D. and Simon Wedlund, C. and Weimer, D. R. and Moore, W. B.},
title = {Atmospheric Escape Processes and Planetary Atmospheric Evolution},
journal = {Journal of Geophysical Research: Space Physics},
volume = {125},
number = {8},
pages = {e2019JA027639},
doi = {https://doi.org/10.1029/2019JA027639},
url = {https://agupubs.onlinelibrary.wiley.com/doi/abs/10.1029/2019JA027639},
eprint = {https://agupubs.onlinelibrary.wiley.com/doi/pdf/10.1029/2019JA027639},
note = {e2019JA027639 10.1029/2019JA027639},
year = {2020}
}

@article{shematovich2018escape,
  title={Escape of planetary atmospheres: physical processes and numerical models},
  author={Shematovich, Valerii Ivanovich and Marov, M Ya},
  journal={Physics-Uspekhi},
  volume={61},
  number={3},
  pages={217},
  year={2018},
  publisher={IOP Publishing}
}

@article{lundin2007planetary,
  title={Planetary magnetic fields and solar forcing: Implications for atmospheric evolution},
  author={Lundin, Rickard and Lammer, Helmut and Ribas, Ignasi},
  journal={Space Science Reviews},
  volume={129},
  pages={245--278},
  year={2007},
  publisher={Springer}
}

@ARTICLE{Kral2018,
       author = {{Kral}, Quentin and {Wyatt}, Mark C. and {Triaud}, Amaury H.~M.~J. and {Marino}, Sebastian and {Th{\'e}bault}, Philippe and {Shorttle}, Oliver},
        title = "{Cometary impactors on the TRAPPIST-1 planets can destroy all planetary atmospheres and rebuild secondary atmospheres on planets f, g, and h}",
      journal = {\mnras},
         year = 2018,
        month = sep,
       volume = {479},
       number = {2},
        pages = {2649-2672},
          doi = {10.1093/mnras/sty1677},
archivePrefix = {arXiv},
       eprint = {1802.05034},
 primaryClass = {astro-ph.EP},
       adsurl = {https://ui.adsabs.harvard.edu/abs/2018MNRAS.479.2649K}
}

@article{catling2001biogenic,
  title={Biogenic methane, hydrogen escape, and the irreversible oxidation of early Earth},
  author={Catling, David C and Zahnle, Kevin J and McKay, Christopher P},
  journal={Science},
  volume={293},
  number={5531},
  pages={839--843},
  year={2001},
  publisher={American Association for the Advancement of Science}
}

@ARTICLE{2021Gialluca,
       author = {{Gialluca}, Megan T. and {Robinson}, Tyler D. and {Rugheimer}, Sarah and {Wunderlich}, Fabian},
        title = "{Characterizing Atmospheres of Transiting Earth-like Exoplanets Orbiting M Dwarfs with James Webb Space Telescope}",
      journal = {\pasp},
         year = 2021,
        month = may,
       volume = {133},
       number = {1023},
          eid = {054401},
        pages = {054401},
          doi = {10.1088/1538-3873/abf367},
archivePrefix = {arXiv},
       eprint = {2101.04139},
 primaryClass = {astro-ph.EP},
       adsurl = {https://ui.adsabs.harvard.edu/abs/2021PASP..133e4401G}
}

@article{HOWLAND1979301,
title = {The role of phosphorus in the upper atmosphere of Jupiter},
journal = {Icarus},
volume = {37},
number = {1},
pages = {301-306},
year = {1979},
issn = {0019-1035},
doi = {https://doi.org/10.1016/0019-1035(79)90135-0},
url = {https://www.sciencedirect.com/science/article/pii/0019103579901350},
author = {Glenn R. Howland and Paul Harteck and Robert R. Reeves}
}

@article{madhusudhan2023carbon,
  title={Carbon-bearing molecules in a possible hycean atmosphere},
  author={Madhusudhan, Nikku and Sarkar, Subhajit and Constantinou, Savvas and Holmberg, M{\aa}ns and Piette, Anjali AA and Moses, Julianne I},
  journal={The Astrophysical Journal Letters},
  volume={956},
  number={1},
  pages={L13},
  year={2023},
  publisher={IOP Publishing}
}

@article{catling2020archean,
  title={The archean atmosphere},
  author={Catling, David C and Zahnle, Kevin J},
  journal={Science advances},
  volume={6},
  number={9},
  pages={eaax1420},
  year={2020},
  publisher={American Association for the Advancement of Science}
}

@article{sharma2017markov,
  title={Markov chain Monte Carlo methods for Bayesian data analysis in astronomy},
  author={Sharma, Sanjib},
  journal={Annual Review of Astronomy and Astrophysics},
  volume={55},
  number={1},
  pages={213--259},
  year={2017},
  publisher={Annual Reviews}
}

@article{KASTING2005119,
title = {Methane and climate during the Precambrian era},
journal = {Precambrian Research},
volume = {137},
number = {3},
pages = {119-129},
year = {2005},
note = {Stable Isotopes, Life and Early Earth History},
issn = {0301-9268},
doi = {https://doi.org/10.1016/j.precamres.2005.03.002},
url = {https://www.sciencedirect.com/science/article/pii/S030192680500032X},
author = {James F. Kasting}
}

@article{jackson2020increasing,
  title={Increasing anthropogenic methane emissions arise equally from agricultural and fossil fuel sources},
  author={Jackson, Robert B and Saunois, Marielle and Bousquet, Philippe and Canadell, Josep G and Poulter, Benjamin and Stavert, Ann R and Bergamaschi, Peter and Niwa, Y and Segers, Arjo and Tsuruta, Aki},
  journal={Environmental Research Letters},
  volume={15},
  number={7},
  pages={071002},
  year={2020},
  publisher={IOP Publishing}
}

@article{crutzen1975solar,
  title={Solar proton events: Stratospheric sources of nitric oxide},
  author={Crutzen, Paul J and Isaksen, Ivar SA and Reid, George C},
  journal={Science},
  volume={189},
  number={4201},
  pages={457--459},
  year={1975},
  publisher={American Association for the Advancement of Science}
}

@article{gordon2022hitran2020,
  title={The HITRAN2020 molecular spectroscopic database},
  author={Gordon, Iouli E and Rothman, Laurence S and Hargreaves, RJ and Hashemi, R and Karlovets, Ekaterina Vladimirovna and Skinner, FM and Conway, Eamon K and Hill, Christian and Kochanov, Roman V and Tan, Y and others},
  journal={Journal of quantitative spectroscopy and radiative transfer},
  volume={277},
  pages={107949},
  year={2022},
  publisher={Elsevier}
}

@article{mcmahon2024astrobiology,
  title={Astrobiology: life detection and the abiotic baseline},
  author={McMahon, Sean and Cockell, Charles},
  journal={Astronomy \& Geophysics},
  volume={65},
  number={1},
  pages={1--23},
  year={2024},
  publisher={Oxford University Press Oxford, UK}
}

@ARTICLE{Kubyshkina,
       author = {{Kubyshkina}, D. and {Fossati}, L.},
        title = "{The mass-radius relation of intermediate-mass planets outlined by hydrodynamic escape and thermal evolution}",
      journal = {\aap},
         year = 2022,
        month = dec,
       volume = {668},
          eid = {A178},
        pages = {A178},
          doi = {10.1051/0004-6361/202244916},
archivePrefix = {arXiv},
       eprint = {2211.10166},
 primaryClass = {astro-ph.EP},
       adsurl = {https://ui.adsabs.harvard.edu/abs/2022A&A...668A.178K}
}

@article{vehtari2024pareto,
  title={Pareto smoothed importance sampling},
  author={Vehtari, Aki and Simpson, Daniel and Gelman, Andrew and Yao, Yuling and Gabry, Jonah},
  journal={Journal of Machine Learning Research},
  volume={25},
  number={72},
  pages={1--58},
  year={2024}
}

@ARTICLE{2012AsBioSeager,
       author = {{Seager}, Sara and {Schrenk}, Matthew and {Bains}, William},
        title = "{An Astrophysical View of Earth-Based Metabolic Biosignature Gases}",
      journal = {Astrobiology},
         year = 2012,
        month = jan,
       volume = {12},
       number = {1},
        pages = {61-82},
          doi = {10.1089/ast.2010.0489},
       adsurl = {https://ui.adsabs.harvard.edu/abs/2012AsBio..12...61S}
}

@article{wilson2021mega,
  title={The Mega-MUSCLES spectral energy distribution of TRAPPIST-1},
  author={Wilson, David J and Froning, Cynthia S and Duvvuri, Girish M and France, Kevin and Youngblood, Allison and Schneider, P Christian and Berta-Thompson, Zachory and Brown, Alexander and Buccino, Andrea P and Hawley, Suzanne and others},
  journal={The Astrophysical Journal},
  volume={911},
  number={1},
  pages={18},
  year={2021},
  publisher={IOP Publishing}
}

@article{weiss2022architectures,
  title={Architectures of compact multi-planet systems: diversity and uniformity},
  author={Weiss, Lauren M and Millholland, Sarah C and Petigura, Erik A and Adams, Fred C and Batygin, Konstantin and Bloch, Anthony M and Mordasini, Christoph},
  journal={arXiv preprint arXiv:2203.10076},
  year={2022}
}

@article{walker2018exoplanet,
  title={Exoplanet biosignatures: future directions},
  author={Walker, Sara I and Bains, William and Cronin, Leroy and DasSarma, Shiladitya and Danielache, Sebastian and Domagal-Goldman, Shawn and Kacar, Betul and Kiang, Nancy Y and Lenardic, Adrian and Reinhard, Christopher T and others},
  journal={Astrobiology},
  volume={18},
  number={6},
  pages={779--824},
  year={2018},
  publisher={Mary Ann Liebert, Inc. 140 Huguenot Street, 3rd Floor New Rochelle, NY 10801 USA}
}

@article{moses2020atmospheric,
  title={Atmospheric chemistry on Uranus and Neptune},
  author={Moses, Julianne I and Cavali{\'e}, T and Fletcher, LN and Roman, MT},
  journal={Philosophical transactions of the royal society A},
  volume={378},
  number={2187},
  pages={20190477},
  year={2020},
  publisher={The Royal Society Publishing}
}

@article{calder2025abiotic,
  title={Abiotic ozone in the observable atmospheres of Venus and Venus-like exoplanets},
  author={Calder, Robb and Shorttle, Oliver and Jordan, Sean and Rimmer, Paul and Constantinou, Tereza},
  journal={Monthly Notices of the Royal Astronomical Society},
  volume={540},
  number={3},
  pages={2432--2450},
  year={2025},
  publisher={Oxford University Press}
}

@article{rimmer2021hydroxide,
  title={Hydroxide salts in the clouds of {Venus}: Their effect on the sulfur cycle and cloud droplet {pH}},
  author={Rimmer, Paul B and Jordan, Sean and Constantinou, Tereza and Woitke, Peter and Shorttle, Oliver and Hobbs, Richard and Paschodimas, Alessia},
  journal={The Planetary Science Journal},
  volume={2},
  number={4},
  pages={133},
  year={2021},
  publisher={IOP Publishing}
}

@misc{VehtariFAQ,
  author       = {Vehtari, Aki},
  title        = {CV-FAQ: Frequently Asked Questions about Computer Vision},
  year         = {2022},
  month        = {October},
  howpublished = {\url{https://users.aalto.fi/~ave/CV-FAQ.html}},
  note         = {Accessed on 2025-10-30}
}

@article{sivula2025uncertainty,
  title={Uncertainty in Bayesian leave-one-out cross-validation based model comparison},
  author={Sivula, Tuomas and Magnusson, M{\aa}ns and Matamoros, Asael Alonzo and Vehtari, Aki},
  journal={Bayesian Analysis},
  volume={1},
  number={1},
  pages={1--31},
  year={2025},
  publisher={International Society for Bayesian Analysis}
}

@article{tsai2017vulcan,
  title={VULCAN: an open-source, validated chemical kinetics Python code for exoplanetary atmospheres},
  author={Tsai, Shang-Min and Lyons, James R and Grosheintz, Luc and Rimmer, Paul B and Kitzmann, Daniel and Heng, Kevin},
  journal={The Astrophysical Journal Supplement Series},
  volume={228},
  number={2},
  pages={20},
  year={2017},
  publisher={IOP Publishing}
}

@article{lenton2018copse,
  title={COPSE reloaded: an improved model of biogeochemical cycling over Phanerozoic time},
  author={Lenton, Timothy M and Daines, Stuart J and Mills, Benjamin JW},
  journal={Earth-Science Reviews},
  volume={178},
  pages={1--28},
  year={2018},
  publisher={Elsevier}
}

@article{krissansen2022predictions,
  title={Predictions for observable atmospheres of trappist-1 planets from a fully coupled atmosphere--interior evolution model},
  author={Krissansen-Totton, Joshua and Fortney, Jonathan J},
  journal={The Astrophysical Journal},
  volume={933},
  number={1},
  pages={115},
  year={2022},
  publisher={IOP Publishing}
}

@article{schwieterman2018exoplanet,
  title={Exoplanet biosignatures: a review of remotely detectable signs of life},
  author={Schwieterman, Edward W and Kiang, Nancy Y and Parenteau, Mary N and Harman, Chester E and DasSarma, Shiladitya and Fisher, Theresa M and Arney, Giada N and Hartnett, Hilairy E and Reinhard, Christopher T and Olson, Stephanie L and others},
  journal={Astrobiology},
  volume={18},
  number={6},
  pages={663--708},
  year={2018},
  publisher={Mary Ann Liebert, Inc. 140 Huguenot Street, 3rd Floor New Rochelle, NY 10801 USA}
}

@article{Chin_2024,
doi = {10.3847/2041-8213/ad27d8},
url = {https://dx.doi.org/10.3847/2041-8213/ad27d8},
year = {2024},
month = {feb},
publisher = {The American Astronomical Society},
volume = {963},
number = {1},
pages = {L20},
author = {Laura Chin and Chuanfei Dong and Manasvi Lingam},
title = {Role of Planetary Radius on Atmospheric Escape of Rocky Exoplanets},
journal = {The Astrophysical Journal Letters}
}

@article{zhang2009new,
  title={A new and efficient estimation method for the generalized Pareto distribution},
  author={Zhang, Jin and Stephens, Michael A},
  journal={Technometrics},
  volume={51},
  number={3},
  pages={316--325},
  year={2009},
  publisher={Taylor \& Francis}
}

@article{sivula2020uncertainty,
  title={Uncertainty in Bayesian leave-one-out cross-validation based model comparison},
  author={Sivula, Tuomas and Magnusson, M{\aa}ns and Matamoros, Asael Alonzo and Vehtari, Aki},
  journal={arXiv preprint arXiv:2008.10296},
  year={2020}
}

@ARTICLE{2020ATremblay,
       author = {{Tremblay}, L. and {Line}, M.~R. and {Stevenson}, K. and {Kataria}, T. and {Zellem}, R.~T. and {Fortney}, J.~J. and {Morley}, C.},
        title = "{The Detectability and Constraints of Biosignature Gases in the Near- and Mid-infrared from Transit Transmission Spectroscopy}",
      journal = {\aj},
         year = 2020,
        month = mar,
       volume = {159},
       number = {3},
          eid = {117},
        pages = {117},
          doi = {10.3847/1538-3881/ab64dd},
archivePrefix = {arXiv},
       eprint = {1912.10939},
 primaryClass = {astro-ph.EP},
       adsurl = {https://ui.adsabs.harvard.edu/abs/2020AJ....159..117T}
}

@article{beichman2014observations,
  title={Observations of transiting exoplanets with the James Webb Space Telescope (JWST)},
  author={Beichman, Charles and Benneke, Bjoern and Knutson, Heather and Smith, Roger and Lagage, Pierre-Olivier and Dressing, Courtney and Latham, David and Lunine, Jonathan and Birkmann, Stephan and Ferruit, Pierre and others},
  journal={Publications of the Astronomical Society of the Pacific},
  volume={126},
  number={946},
  pages={1134},
  year={2014},
  publisher={IOP Publishing}
}

@article{glindemann1996free,
  title={Free phosphine from the anaerobic biosphere},
  author={Glindemann, Dietmar and Stottmeister, Ulrich and Bergmann, Armin},
  journal={Environmental Science and Pollution Research},
  volume={3},
  pages={17--19},
  year={1996},
  publisher={Springer}
}

@article{devai1988detection,
  title={Detection of phosphine: new aspects of the phosphorus cycle in the hydrosphere},
  author={D{\'e}vai, Istv{\'a}n and Felf{\"o}ldy, Lajos and Wittner, Ilona and Pl{\'o}sz, S{\'a}ndor},
  journal={Nature},
  volume={333},
  number={6171},
  pages={343--345},
  year={1988},
  publisher={Nature Publishing Group UK London}
}

@article{jenkins2000phosphine,
  title={Phosphine generation by mixed-and monoseptic-cultures of anaerobic bacteria},
  author={Jenkins, RO and Morris, T-A and Craig, P Jetal and Ritchie, AW and Ostah, N},
  journal={Science of the Total Environment},
  volume={250},
  number={1-3},
  pages={73--81},
  year={2000},
  publisher={Elsevier}
}

@article{bains2019new,
  title={New environmental model for thermodynamic ecology of biological phosphine production},
  author={Bains, William and Petkowski, Janusz J and Sousa-Silva, Clara and Seager, Sara},
  journal={Science of the total environment},
  volume={658},
  pages={521--536},
  year={2019},
  publisher={Elsevier}
}

@article{bregman1975observation,
  title={Observation of the nu-squared band of PH3 in the atmosphere of Saturn},
  author={Bregman, JD and Lester, DF and Rank, DM},
  journal={Astrophysical Journal, vol. 202, Nov. 15, 1975, pt. 2, p. L55, L56.},
  volume={202},
  pages={L55},
  year={1975}
}

@article{visscher2006atmospheric,
  title={Atmospheric chemistry in giant planets, brown dwarfs, and low-mass dwarf stars. II. Sulfur and phosphorus},
  author={Visscher, Channon and Lodders, Katharina and Fegley Jr, Bruce},
  journal={The Astrophysical Journal},
  volume={648},
  number={2},
  pages={1181},
  year={2006},
  publisher={IOP Publishing}
}

@ARTICLE{2023Padovani,
       author = {{Padovani}, Paolo and {Cirasuolo}, Michele},
        title = "{The Extremely Large Telescope}",
      journal = {Contemporary Physics},
         year = 2023,
        month = jan,
       volume = {64},
       number = {1},
        pages = {47-64},
          doi = {10.1080/00107514.2023.2266921},
archivePrefix = {arXiv},
       eprint = {2312.04299},
 primaryClass = {astro-ph.IM},
       adsurl = {https://ui.adsabs.harvard.edu/abs/2023ConPh..64...47P}
}

@article{prinn1975phosphine,
  title={Phosphine on Jupiter and implications for the Great Red Spot},
  author={Prinn, Ronald G and Lewis, John S},
  journal={Science},
  volume={190},
  number={4211},
  pages={274--276},
  year={1975},
  publisher={American Association for the Advancement of Science}
}

@ARTICLE{2021Greaves,
       author = {{Greaves}, Jane S. and {Richards}, Anita M.~S. and {Bains}, William and {Rimmer}, Paul B. and {Sagawa}, Hideo and {Clements}, David L. and {Seager}, Sara and {Petkowski}, Janusz J. and {Sousa-Silva}, Clara and {Ranjan}, Sukrit and {Drabek-Maunder}, Emily and {Fraser}, Helen J. and {Cartwright}, Annabel and {Mueller-Wodarg}, Ingo and {Zhan}, Zhuchang and {Friberg}, Per and {Coulson}, Iain and {Lee}, E'lisa and {Hoge}, Jim},
        title = "{Phosphine gas in the cloud decks of Venus}",
      journal = {Nature Astronomy},
         year = 2021,
        month = jan,
       volume = {5},
        pages = {655-664},
          doi = {10.1038/s41550-020-1174-4},
archivePrefix = {arXiv},
       eprint = {2009.06593},
 primaryClass = {astro-ph.EP},
       adsurl = {https://ui.adsabs.harvard.edu/abs/2021NatAs...5..655G}
}

@ARTICLE{2013ApJOwen,
       author = {{Owen}, James E. and {Wu}, Yanqin},
        title = "{Kepler Planets: A Tale of Evaporation}",
      journal = {\apj},
         year = 2013,
        month = oct,
       volume = {775},
       number = {2},
          eid = {105},
        pages = {105},
          doi = {10.1088/0004-637X/775/2/105},
archivePrefix = {arXiv},
       eprint = {1303.3899},
 primaryClass = {astro-ph.EP},
       adsurl = {https://ui.adsabs.harvard.edu/abs/2013ApJ...775..105O}
}

@article{bains2019trivalent,
  title={Trivalent phosphorus and phosphines as components of biochemistry in anoxic environments},
  author={Bains, William and Petkowski, Janusz Jurand and Sousa-Silva, Clara and Seager, Sara},
  journal={Astrobiology},
  volume={19},
  number={7},
  pages={885--902},
  year={2019},
  publisher={Mary Ann Liebert, Inc., publishers 140 Huguenot Street, 3rd Floor New~…}
}

@article{gassmann1993phosphane,
  title={Phosphane (PH3) in the biosphere},
  author={Gassmann, G{\"u}nter and Glindemann, Dietmar},
  journal={Angewandte Chemie International Edition in English},
  volume={32},
  number={5},
  pages={761--763},
  year={1993},
  publisher={Wiley Online Library}
}

@article{pridham1998determination,
  title={Determination of phosphine by packed column gas chromatography with alkali flame ionisation detection},
  author={Pridham, John{\'a}B and others},
  journal={Analytical Communications},
  volume={35},
  number={3},
  pages={109--111},
  year={1998},
  publisher={Royal Society of Chemistry}
}

@article{zhu2014penguins,
  title={Penguins significantly increased phosphine formation and phosphorus contribution in maritime Antarctic soils},
  author={Zhu, Renbin and Wang, Qing and Ding, Wei and Wang, Can and Hou, Lijun and Ma, Dawei},
  journal={Scientific reports},
  volume={4},
  number={1},
  pages={7055},
  year={2014},
  publisher={Nature Publishing Group UK London}
}

@article{sousa2019molecular,
  title={Molecular simulations for the spectroscopic detection of atmospheric gases},
  author={Sousa-Silva, Clara and Petkowski, Janusz J and Seager, Sara},
  journal={Physical Chemistry Chemical Physics},
  volume={21},
  number={35},
  pages={18970--18987},
  year={2019},
  publisher={Royal Society of Chemistry}
}

@article{mcmahon2022false,
  title={False biosignatures on Mars: anticipating ambiguity},
  author={McMahon, Sean and Cosmidis, Julie},
  journal={Journal of the Geological Society},
  volume={179},
  number={2},
  pages={jgs2021--050},
  year={2022},
  publisher={The Geological Society of London}
}

@INPROCEEDINGS{2015ESS350001S,
       author = {{Seager}, Sara and {Bains}, William and {Petkowski}, Janusz},
        title = "{Towards a List of Molecules as Potential Biosignature Gases for the Search for Life on Exoplanets}",
    booktitle = {AAS/Division for Extreme Solar Systems Abstracts},
         year = 2015,
       series = {AAS/Division for Extreme Solar Systems Abstracts},
       volume = {47},
        month = dec,
          eid = {500.01},
        pages = {500.01},
       adsurl = {https://ui.adsabs.harvard.edu/abs/2015ESS.....350001S}
}

@article{constantinou2023early,
  title={Early insights for atmospheric retrievals of exoplanets using JWST transit spectroscopy},
  author={Constantinou, Savvas and Madhusudhan, Nikku and Gandhi, Siddharth},
  journal={The Astrophysical Journal Letters},
  volume={943},
  number={2},
  pages={L10},
  year={2023},
  publisher={IOP Publishing}
}

@article{schwieterman2022evaluating,
  title={Evaluating the plausible range of N2O biosignatures on exo-earths: An integrated biogeochemical, photochemical, and spectral modeling approach},
  author={Schwieterman, Edward W and Olson, Stephanie L and Pidhorodetska, Daria and Reinhard, Christopher T and Ganti, Ainsley and Fauchez, Thomas J and Bastelberger, Sandra T and Crouse, Jaime S and Ridgwell, Andy and Lyons, Timothy W},
  journal={The Astrophysical Journal},
  volume={937},
  number={2},
  pages={109},
  year={2022},
  publisher={IOP Publishing}
}

@article{bains2021phosphine,
  title={Phosphine on Venus cannot be explained by conventional processes},
  author={Bains, William and Petkowski, Janusz J and Seager, Sara and Ranjan, Sukrit and Sousa-Silva, Clara and Rimmer, Paul B and Zhan, Zhuchang and Greaves, Jane S and Richards, Anita MS},
  journal={Astrobiology},
  volume={21},
  number={10},
  pages={1277--1304},
  year={2021},
  publisher={Mary Ann Liebert, Inc., publishers 140 Huguenot Street, 3rd Floor New~…}
}

@article{jones2015stable,
  title={Stable isotopes and iron oxide mineral products as markers of chemodenitrification.},
  author={Jones, L Camille and Peters, Brian and Lezama Pacheco, Juan S and Casciotti, Karen L and Fendorf, Scott},
  journal={Environmental science \& technology},
  volume={49},
  number={6},
  pages={3444--3452},
  year={2015},
  publisher={ACS Publications}
}

@article{samarkin2010abiotic,
  title={Abiotic nitrous oxide emission from the hypersaline Don Juan Pond in Antarctica},
  author={Samarkin, Vladimir A and Madigan, Michael T and Bowles, Marshall W and Casciotti, Karen L and Priscu, John C and McKay, Christopher P and Joye, Samantha B},
  journal={Nature Geoscience},
  volume={3},
  number={5},
  pages={341--344},
  year={2010},
  publisher={Nature Publishing Group UK London}
}

@article{seager2025prospects,
  title={Prospects for Detecting Signs of Life on Exoplanets in the JWST Era},
  author={Seager, Sara and Welbanks, Luis and Ellerbroek, Lucas and Bains, William and Petkowski, Janusz J},
  journal={arXiv preprint arXiv:2504.12946},
  year={2025}
}

@article{welbanks2025challenges,
  title={The Challenges of Detecting Gases in Exoplanet Atmospheres},
  author={Welbanks, Luis and Nixon, Matthew C and McGill, Peter and Tilke, Lana J and Wiser, Lindsey S and Rotman, Yoav and Mukherjee, Sagnick and Feinstein, Adina and Line, Michael R and Seager, Sara and others},
  journal={arXiv preprint arXiv:2504.21788},
  year={2025}
}

@article{taylor2025there,
  title={Are There Spectral Features in the MIRI/LRS Transmission Spectrum of K2-18b?},
  author={Taylor, Jake},
  journal={Research Notes of the AAS},
  volume={9},
  number={5},
  pages={118},
  year={2025},
  publisher={The American Astronomical Society}
}

@article{luque2025insufficient,
  title={Insufficient evidence for DMS and DMDS in the atmosphere of K2-18 b},
  author={Luque, R and Piaulet-Ghorayeb, C and Radica, M and Xue, Q and Zhang, M and Bean, JL and Samra, D and Steinrueck, ME},
  journal={Transit},
  volume={2250},
  number={2500},
  pages={2750},
  year={2025}
}

@article{madhusudhan2025new,
  title={New Constraints on DMS and DMDS in the Atmosphere of K2-18 b from JWST MIRI},
  author={Madhusudhan, Nikku and Constantinou, Savvas and Holmberg, M{\aa}ns and Sarkar, Subhajit and Piette, Anjali AA and Moses, Julianne I},
  journal={The Astrophysical Journal Letters},
  volume={983},
  number={2},
  pages={L40},
  year={2025},
  publisher={IOP Publishing}
}

@article{schmidt2025comprehensive,
  title={A Comprehensive Reanalysis of K2-18 b's JWST NIRISS+ NIRSpec Transmission Spectrum},
  author={Schmidt, Stephen P and MacDonald, Ryan J and Tsai, Shang-Min and Radica, Michael and Wang, Le-Chris and Ahrer, Eva-Maria and Bell, Taylor J and Fisher, Chloe and Thorngren, Daniel P and Wogan, Nicholas and others},
  journal={arXiv preprint arXiv:2501.18477},
  year={2025}
}

@article{truzzi1978extraordinary,
  title={On the extraordinary: an attempt at clarification},
  author={Truzzi, Marcello},
  journal={Zetetic Scholar},
  volume={1},
  pages={11},
  year={1978}
}

@book{sagan2011broca,
  title={Broca's brain: Reflections on the romance of science},
  author={Sagan, Carl},
  year={2011},
  publisher={Ballantine Books}
}

@article{ranjan2023importance,
  title={The Importance of the Upper Atmosphere to CO/O2 Runaway on Habitable Planets Orbiting Low-mass Stars},
  author={Ranjan, Sukrit and Schwieterman, Edward W and Leung, Michaela and Harman, Chester E and Hu, Renyu},
  journal={The Astrophysical Journal Letters},
  volume={958},
  number={1},
  pages={L15},
  year={2023},
  publisher={IOP Publishing}
}

@ARTICLE{2021MNRASAtri,
       author = {{Atri}, Dimitra and {Mogan}, Shane R. Carberry},
        title = "{Stellar flares versus luminosity: XUV-induced atmospheric escape and planetary habitability}",
      journal = {\mnras},
         year = 2021,
        month = jan,
       volume = {500},
       number = {1},
        pages = {L1-L5},
          doi = {10.1093/mnrasl/slaa166},
archivePrefix = {arXiv},
       eprint = {2009.04310},
 primaryClass = {astro-ph.EP},
       adsurl = {https://ui.adsabs.harvard.edu/abs/2021MNRAS.500L...1A}
}

@article{cleland2019moving,
  title={Moving beyond definitions in the search for extraterrestrial life},
  author={Cleland, Carol E},
  journal={Astrobiology},
  volume={19},
  number={6},
  pages={722--729},
  year={2019},
  publisher={Mary Ann Liebert, Inc., publishers 140 Huguenot Street, 3rd Floor New~…}
}

@article{sousa2020phosphine,
  title={Phosphine as a biosignature gas in exoplanet atmospheres},
  author={Sousa-Silva, Clara and Seager, Sara and Ranjan, Sukrit and Petkowski, Janusz Jurand and Zhan, Zhuchang and Hu, Renyu and Bains, William},
  journal={Astrobiology},
  volume={20},
  number={2},
  pages={235--268},
  year={2020},
  publisher={Mary Ann Liebert, Inc., publishers 140 Huguenot Street, 3rd Floor New~…}
}

@article{Oyama,
author = {Oyama, V. I. and Carle, G. C. and Woeller, F. and Pollack, J. B. and Reynolds, R. T. and Craig, R. A.},
title = {Pioneer Venus gas chromatography of the lower atmosphere of Venus},
journal = {Journal of Geophysical Research: Space Physics},
volume = {85},
number = {A13},
pages = {7891-7902},
doi = {https://doi.org/10.1029/JA085iA13p07891},
url = {https://agupubs.onlinelibrary.wiley.com/doi/abs/10.1029/JA085iA13p07891},
eprint = {https://agupubs.onlinelibrary.wiley.com/doi/pdf/10.1029/JA085iA13p07891},
year = {1980}
}

@article{skilling2004nested,
  title={Nested sampling},
  author={Skilling, John},
  journal={Bayesian inference and maximum entropy methods in science and engineering},
  volume={735},
  pages={395--405},
  year={2004}
}

@article{welbanks2023application,
  title={On the application of Bayesian leave-one-out cross-validation to exoplanet atmospheric analysis},
  author={Welbanks, Luis and McGill, Peter and Line, Michael and Madhusudhan, Nikku},
  journal={The Astronomical Journal},
  volume={165},
  number={3},
  pages={112},
  year={2023},
  publisher={IOP Publishing}
}

@article{vehtari2017practical,
  title={Practical Bayesian model evaluation using leave-one-out cross-validation and WAIC},
  author={Vehtari, Aki and Gelman, Andrew and Gabry, Jonah},
  journal={Statistics and computing},
  volume={27},
  pages={1413--1432},
  year={2017},
  publisher={Springer}
}

@article{Zhou2014,
author = {Zhou Lu  and Yih Chung Chang  and Qing-Zhu Yin  and C. Y. Ng  and William M. Jackson },
title = {Evidence for direct molecular oxygen production in CO<sub>2</sub> photodissociation},
journal = {Science},
volume = {346},
number = {6205},
pages = {61-64},
year = {2014},
doi = {10.1126/science.1257156},
URL = {https://www.science.org/doi/abs/10.1126/science.1257156},
eprint = {https://www.science.org/doi/pdf/10.1126/science.1257156}}

@article{Groller,
author = {Gröller, H. and Yelle, R. V. and Koskinen, T. T. and Montmessin, F. and Lacombe, G. and Schneider, N. M. and Deighan, J. and Stewart, A. I. F. and Jain, S. K. and Chaffin, M. S. and Crismani, M. M. J. and Stiepen, A. and Lefèvre, F. and McClintock, W. E. and Clarke, J. T. and Holsclaw, G. M. and Mahaffy, P. R. and Bougher, S. W. and Jakosky, B. M.},
title = {Probing the Martian atmosphere with MAVEN/IUVS stellar occultations},
journal = {Geophysical Research Letters},
volume = {42},
number = {21},
pages = {9064-9070},
doi = {https://doi.org/10.1002/2015GL065294},
year = {2015}
}

@ARTICLE{Schlawin2021,
       author = {{Schlawin}, Everett and {Leisenring}, Jarron and {McElwain}, Michael W. and {Misselt}, Karl and {Don}, Kenneth and {Greene}, Thomas P. and {Beatty}, Thomas and {Nikolov}, Nikolay and {Kelly}, Douglas and {Rieke}, Marcia},
        title = "{JWST Noise Floor. II. Systematic Error Sources in JWST NIRCam Time Series}",
      journal = {\aj},
         year = 2021,
        month = mar,
       volume = {161},
       number = {3},
          eid = {115},
        pages = {115},
          doi = {10.3847/1538-3881/abd8d4},
archivePrefix = {arXiv},
       eprint = {2010.03576},
 primaryClass = {astro-ph.IM},
       adsurl = {https://ui.adsabs.harvard.edu/abs/2021AJ....161..115S}
}

@article{seager2016,
author = {Seager, S. and Bains, W. and Petkowski, J.J.},
title = {Toward a List of Molecules as Potential Biosignature Gases for the Search for Life on Exoplanets and Applications to Terrestrial Biochemistry},
journal = {Astrobiology},
volume = {16},
number = {6},
pages = {465-485},
year = {2016},
doi = {10.1089/ast.2015.1404},
    note ={PMID: 27096351},
URL = { https://doi.org/10.1089/ast.2015.1404},
eprint = {  https://doi.org/10.1089/ast.2015.1404}
}

@ARTICLE{2023Spaargaren,
       author = {{Spaargaren}, Rob J. and {Wang}, Haiyang S. and {Mojzsis}, Stephen J. and {Ballmer}, Maxim D. and {Tackley}, Paul J.},
        title = "{Plausible Constraints on the Range of Bulk Terrestrial Exoplanet Compositions in the Solar Neighborhood}",
      journal = {\apj},
         year = 2023,
        month = may,
       volume = {948},
       number = {1},
          eid = {53},
        pages = {53},
          doi = {10.3847/1538-4357/acac7d},
archivePrefix = {arXiv},
       eprint = {2211.01800},
 primaryClass = {astro-ph.EP},
       adsurl = {https://ui.adsabs.harvard.edu/abs/2023ApJ...948...53S}
}

@article{France_2016,
doi = {10.3847/0004-637X/820/2/89},
url = {https://dx.doi.org/10.3847/0004-637X/820/2/89},
year = {2016},
month = {mar},
publisher = {The American Astronomical Society},
volume = {820},
number = {2},
pages = {89},
author = {Kevin France and R. O. Parke Loyd and Allison Youngblood and Alexander Brown and P. Christian Schneider and Suzanne L. Hawley and Cynthia S. Froning and Jeffrey L. Linsky and Aki Roberge and Andrea P. Buccino and James R. A. Davenport and Juan M. Fontenla and Lisa Kaltenegger and Adam F. Kowalski and Pablo J. D. Mauas and Yamila Miguel and Seth Redfield and Sarah Rugheimer and Feng Tian and Mariela C. Vieytes and Lucianne M. Walkowicz and Kolby L. Weisenburger},
title = {THE MUSCLES TREASURY SURVEY. I. MOTIVATION AND OVERVIEW*},
journal = {The Astrophysical Journal}
}

@ARTICLE{2012Owen,
       author = {{Owen}, James E. and {Jackson}, Alan P.},
        title = "{Planetary evaporation by UV \& X-ray radiation: basic hydrodynamics}",
      journal = {\mnras},
         year = 2012,
        month = oct,
       volume = {425},
       number = {4},
        pages = {2931-2947},
          doi = {10.1111/j.1365-2966.2012.21481.x},
archivePrefix = {arXiv},
       eprint = {1206.2367},
 primaryClass = {astro-ph.EP},
       adsurl = {https://ui.adsabs.harvard.edu/abs/2012MNRAS.425.2931O}
}

@ARTICLE{2001Tarter,
       author = {{Tarter}, Jill},
        title = "{The Search for Extraterrestrial Intelligence (SETI)}",
      journal = {\araa},
         year = 2001,
        month = jan,
       volume = {39},
        pages = {511-548},
          doi = {10.1146/annurev.astro.39.1.511},
       adsurl = {https://ui.adsabs.harvard.edu/abs/2001ARA&A..39..511T}
}

@article{zhu2015sources,
  title={Sources and impacts of atmospheric NH 3: Current understanding and frontiers for modeling, measurements, and remote sensing in North America},
  author={Zhu, Liye and Henze, Daven K and Bash, Jesse O and Cady-Pereira, Karen E and Shephard, Mark W and Luo, Ming and Capps, Shannon L},
  journal={Current Pollution Reports},
  volume={1},
  pages={95--116},
  year={2015},
  publisher={Springer}
}

@article{seager2013biomass,
  title={A biomass-based model to estimate the plausibility of exoplanet biosignature gases},
  author={Seager, S and Bains, W and Hu, R},
  journal={The Astrophysical Journal},
  volume={775},
  number={2},
  pages={104},
  year={2013},
  publisher={IOP Publishing}
}

@article{beatty2024sulfur,
  title={Sulfur dioxide and other molecular species in the atmosphere of the sub-Neptune GJ 3470 b},
  author={Beatty, Thomas G and Welbanks, Luis and Schlawin, Everett and Bell, Taylor J and Line, Michael R and Murphy, Matthew and Edelman, Isaac and Greene, Thomas P and Fortney, Jonathan J and Henry, Gregory W and others},
  journal={The Astrophysical Journal Letters},
  volume={970},
  number={1},
  pages={L10},
  year={2024},
  publisher={IOP Publishing}
}

@article{davenport2025toi,
  title={TOI-421 b: A Hot Sub-Neptune with a Haze-free, Low Mean Molecular Weight Atmosphere},
  author={Davenport, Brian and Kempton, Eliza M-R and Nixon, Matthew C and Ih, Jegug and Deming, Drake and Fu, Guangwei and May, EM and Bean, Jacob L and Gao, Peter and Rogers, Leslie and others},
  journal={The Astrophysical Journal Letters},
  volume={984},
  number={2},
  pages={L44},
  year={2025},
  publisher={IOP Publishing}
}

@article{benneke2024jwst,
  title={JWST Reveals CH $ \_4 $, CO $ \_2 $, and H $ \_2 $ O in a Metal-rich Miscible Atmosphere on a Two-Earth-Radius Exoplanet},
  author={Benneke, Bj{\"o}rn and Roy, Pierre-Alexis and Coulombe, Louis-Philippe and Radica, Michael and Piaulet, Caroline and Ahrer, Eva-Maria and Pierrehumbert, Raymond and Krissansen-Totton, Joshua and Schlichting, Hilke E and Hu, Renyu and others},
  journal={arXiv preprint arXiv:2403.03325},
  year={2024}
}

@article{bell2023methane,
  title={Methane throughout the atmosphere of the warm exoplanet WASP-80b},
  author={Bell, Taylor J and Welbanks, Luis and Schlawin, Everett and Line, Michael R and Fortney, Jonathan J and Greene, Thomas P and Ohno, Kazumasa and Parmentier, Vivien and Rauscher, Emily and Beatty, Thomas G and others},
  journal={Nature},
  volume={623},
  number={7988},
  pages={709--712},
  year={2023},
  publisher={Nature Publishing Group UK London}
}

@inbook{Kartal,
author = {Kartal, Boran and Keltjens, Jan T and Jetten, Mike SM},
publisher = {John Wiley \& Sons, Ltd},
isbn = {9780470015902},
title = {The Metabolism of Anammox},
booktitle = {Encyclopedia of Life Sciences},
chapter = {},
pages = {},
doi = {https://doi.org/10.1002/9780470015902.a0021315}
}

@ARTICLE{Huang2022,
       author = {{Huang}, Jingcheng and {Seager}, Sara and {Petkowski}, Janusz J. and {Ranjan}, Sukrit and {Zhan}, Zhuchang},
        title = "{Assessment of Ammonia as a Biosignature Gas in Exoplanet Atmospheres}",
      journal = {Astrobiology},
         year = 2022,
        month = feb,
       volume = {22},
       number = {2},
        pages = {171-191},
          doi = {10.1089/ast.2020.2358},
archivePrefix = {arXiv},
       eprint = {2107.12424},
 primaryClass = {astro-ph.EP},
       adsurl = {https://ui.adsabs.harvard.edu/abs/2022AsBio..22..171H}
}

@article{airapetian2016prebiotic,
  title={Prebiotic chemistry and atmospheric warming of early Earth by an active young Sun},
  author={Airapetian, VS and Glocer, A and Gronoff, G and Hebrard, Eric and Danchi, W},
  journal={Nature Geoscience},
  volume={9},
  number={6},
  pages={452--455},
  year={2016},
  publisher={Nature Publishing Group}
}

@article{Jackson_2020,
doi = {10.1088/1748-9326/ab9ed2},
url = {https://dx.doi.org/10.1088/1748-9326/ab9ed2},
year = {2020},
month = {jul},
publisher = {IOP Publishing},
volume = {15},
number = {7},
pages = {071002},
author = {R B Jackson and M Saunois and P Bousquet and J G Canadell and B Poulter and A R Stavert and P Bergamaschi and Y Niwa and A Segers and A Tsuruta},
title = {Increasing anthropogenic methane emissions arise equally from agricultural and fossil fuel sources},
journal = {Environmental Research Letters}
}

@article{BUICK,
author = {Buick, R.},
title = {Did the Proterozoic ‘Canfield Ocean’ cause a laughing gas greenhouse?},
journal = {Geobiology},
volume = {5},
number = {2},
pages = {97-100},
doi = {https://doi.org/10.1111/j.1472-4669.2007.00110.x},
url = {https://onlinelibrary.wiley.com/doi/abs/10.1111/j.1472-4669.2007.00110.x},
eprint = {https://onlinelibrary.wiley.com/doi/pdf/10.1111/j.1472-4669.2007.00110.x},
year = {2007}
}

@article{Marais2002,
author = {Des Marais, David J. and Harwit, Martin O. and Jucks, Kenneth W. and Kasting, James F. and Lin, Douglas N.C. and Lunine, Jonathan I. and Schneider, Jean and Seager, Sara and Traub, Wesley A. and Woolf, Neville J.},
title = {Remote Sensing of Planetary Properties and Biosignatures on Extrasolar Terrestrial Planets},
journal = {Astrobiology},
volume = {2},
number = {2},
pages = {153-181},
year = {2002},
doi = {10.1089/15311070260192246},
    note ={PMID: 12469366},

URL = { https://doi.org/10.1089/15311070260192246},
eprint = {  https://doi.org/10.1089/15311070260192246}
}

@article{Leger2011,
author = {Leger, A. and Fontecave, M and Labeyrie, Antoine and Samuel, Bunmi and Demangeon, Olivier and Valencia, Diana},
year = {2011},
month = {05},
pages = {335-41},
title = {Is the Presence of Oxygen on an Exoplanet a Reliable Biosignature?},
volume = {11},
journal = {Astrobiology},
doi = {10.1089/ast.2010.0516}
}

@article{berndt1996reduction,
  title={Reduction of CO2 during serpentinization of olivine at 300 C and 500 bar},
  author={Berndt, Michael E and Allen, Douglas E and Seyfried Jr, William E},
  journal={Geology},
  volume={24},
  number={4},
  pages={351--354},
  year={1996},
  publisher={Geological Society of America}
}

@article{zahnle2020creation,
  title={Creation and evolution of impact-generated reduced atmospheres of early Earth},
  author={Zahnle, Kevin J and Lupu, Roxana and Catling, David C and Wogan, Nick},
  journal={The Planetary Science Journal},
  volume={1},
  number={1},
  pages={11},
  year={2020},
  publisher={IOP Publishing}
}

@article{mccollom2016abiotic,
  title={Abiotic methane formation during experimental serpentinization of olivine},
  author={McCollom, Thomas M},
  journal={Proceedings of the National Academy of Sciences},
  volume={113},
  number={49},
  pages={13965--13970},
  year={2016},
  publisher={National Acad Sciences}
}

@article{grozeva2017experimental,
  title={Experimental study of carbonate formation in oceanic peridotite},
  author={Grozeva, Niya G and Klein, Frieder and Seewald, Jeffrey S and Sylva, Sean P},
  journal={Geochimica et Cosmochimica Acta},
  volume={199},
  pages={264--286},
  year={2017},
  publisher={Elsevier}
}

@ARTICLE{1959NaturCocconi,
       author = {{Cocconi}, Giuseppe and {Morrison}, Philip},
        title = "{Searching for Interstellar Communications}",
      journal = {\nat},
         year = 1959,
        month = sep,
       volume = {184},
       number = {4690},
        pages = {844-846},
          doi = {10.1038/184844a0},
       adsurl = {https://ui.adsabs.harvard.edu/abs/1959Natur.184..844C}
}

@article{mccollom2016generation,
  title={Generation of hydrogen and methane during experimental low-temperature reaction of ultramafic rocks with water},
  author={McCollom, Thomas M and Donaldson, Christopher},
  journal={Astrobiology},
  volume={16},
  number={6},
  pages={389--406},
  year={2016},
  publisher={Mary Ann Liebert, Inc. 140 Huguenot Street, 3rd Floor New Rochelle, NY 10801 USA}
}

@ARTICLE{2024AJChoza,
       author = {{Choza}, Carmen and {Bautista}, Daniel and {Croft}, Steve and {Siemion}, Andrew P.~V. and {Brzycki}, Bryan and {Bhattaram}, Krishnakumar and {Czech}, Daniel and {de Pater}, Imke and {Gajjar}, Vishal and {Isaacson}, Howard and {Lacker}, Kevin and {Lacki}, Brian and {Lebofsky}, Matthew and {MacMahon}, David H.~E. and {Price}, Danny and {Schoultz}, Sarah and {Sheikh}, Sofia and {Varghese}, Savin Shynu and {Morgan}, Lawrence and {Drew}, Jamie and {Worden}, S. Pete},
        title = "{The Breakthrough Listen Search for Intelligent Life: Technosignature Search of 97 Nearby Galaxies}",
      journal = {\aj},
         year = 2024,
        month = jan,
       volume = {167},
       number = {1},
          eid = {10},
        pages = {10},
          doi = {10.3847/1538-3881/acf576},
archivePrefix = {arXiv},
       eprint = {2312.03943},
 primaryClass = {astro-ph.IM},
       adsurl = {https://ui.adsabs.harvard.edu/abs/2024AJ....167...10C}
}

@article{guzman2013abiotic,
  title={Abiotic production of methane in terrestrial planets},
  author={Guzm{\'a}n-Marmolejo, Andr{\'e}s and Segura, Ant{\'\i}gona and Escobar-Briones, Elva},
  journal={Astrobiology},
  volume={13},
  number={6},
  pages={550--559},
  year={2013},
  publisher={Mary Ann Liebert, Inc. 140 Huguenot Street, 3rd Floor New Rochelle, NY 10801 USA}
}

@article{etiope2013abiotic,
  title={Abiotic methane on Earth},
  author={Etiope, Giuseppe and Sherwood Lollar, Barbara},
  journal={Reviews of Geophysics},
  volume={51},
  number={2},
  pages={276--299},
  year={2013},
  publisher={Wiley Online Library}
}

@article{Thompson2022,
author = {Maggie A. Thompson  and Joshua Krissansen-Totton  and Nicholas Wogan  and Myriam Telus  and Jonathan J. Fortney },
title = {The case and context for atmospheric methane as an exoplanet biosignature},
journal = {Proceedings of the National Academy of Sciences},
volume = {119},
number = {14},
pages = {e2117933119},
year = {2022},
doi = {10.1073/pnas.2117933119}}

@ARTICLE{Kozakis2022,
       author = {{Kozakis}, Thea and {Mendon{\c{c}}a}, Jo{\~a}o M. and {Buchhave}, Lars A.},
        title = "{Is ozone a reliable proxy for molecular oxygen?. I. The O$_{2}$-O$_{3}$ relationship for Earth-like atmospheres}",
      journal = {\aap},
         year = 2022,
        month = sep,
       volume = {665},
          eid = {A156},
        pages = {A156},
          doi = {10.1051/0004-6361/202244164},
archivePrefix = {arXiv},
       eprint = {2208.09415},
 primaryClass = {astro-ph.EP},
       adsurl = {https://ui.adsabs.harvard.edu/abs/2022A&A...665A.156K}
}

@ARTICLE{1993Sagan,
       author = {{Sagan}, Carl and {Thompson}, W. Reid and {Carlson}, Robert and {Gurnett}, Donald and {Hord}, Charles},
        title = "{A search for life on Earth from the Galileo spacecraft}",
      journal = {\nat},
         year = 1993,
        month = oct,
       volume = {365},
       number = {6448},
        pages = {715-721},
          doi = {10.1038/365715a0},
       adsurl = {https://ui.adsabs.harvard.edu/abs/1993Natur.365..715S}
}

@article{liggins2022growth,
  title={Growth and evolution of secondary volcanic atmospheres: I. Identifying the geological character of hot rocky planets},
  author={Liggins, Philippa and Jordan, Sean and Rimmer, Paul B and Shorttle, Oliver},
  journal={Journal of Geophysical Research: Planets},
  volume={127},
  number={7},
  pages={e2021JE007123},
  year={2022},
  publisher={Wiley Online Library}
}

@article{mancinelli1988evolution,
  title={The evolution of nitrogen cycling},
  author={Mancinelli, Rocco L and McKay, Christopher P},
  journal={Origins of Life and Evolution of the Biosphere},
  volume={18},
  pages={311--325},
  year={1988},
  publisher={Springer}
}

@article{sander2015compilation,
  title={Compilation of Henry's law constants (version 4.0) for water as solvent},
  author={Sander, Rolf},
  journal={Atmospheric Chemistry and Physics},
  volume={15},
  number={8},
  pages={4399--4981},
  year={2015},
  publisher={Copernicus Publications G{\"o}ttingen, Germany}
}

@ARTICLE{1982JGRKasting,
       author = {{Kasting}, J.~F.},
        title = "{Stability of ammonia in the primitive terrestrial atmosphere}",
      journal = {\jgr},
         year = 1982,
        month = apr,
       volume = {87},
       number = {C4},
        pages = {3091-3098},
          doi = {10.1029/JC087iC04p03091},
       adsurl = {https://ui.adsabs.harvard.edu/abs/1982JGR....87.3091K}
}

@ARTICLE{2012ApJHu,
       author = {{Hu}, Renyu and {Seager}, Sara and {Bains}, William},
        title = "{Photochemistry in Terrestrial Exoplanet Atmospheres. I. Photochemistry Model and Benchmark Cases}",
      journal = {\apj},
         year = 2012,
        month = dec,
       volume = {761},
       number = {2},
          eid = {166},
        pages = {166},
          doi = {10.1088/0004-637X/761/2/166},
archivePrefix = {arXiv},
       eprint = {1210.6885},
 primaryClass = {astro-ph.EP},
       adsurl = {https://ui.adsabs.harvard.edu/abs/2012ApJ...761..166H}
}

@ARTICLE{2017MNRAS470187A,
       author = {{Ardaseva}, Aleksandra and {Rimmer}, Paul B. and {Waldmann}, Ingo and {Rocchetto}, Marco and {Yurchenko}, Sergey N. and {Helling}, Christiane and {Tennyson}, Jonathan},
        title = "{Lightning chemistry on Earth-like exoplanets}",
      journal = {\mnras},
         year = 2017,
        month = sep,
       volume = {470},
       number = {1},
        pages = {187-196},
          doi = {10.1093/mnras/stx1012},
archivePrefix = {arXiv},
       eprint = {1704.07917},
 primaryClass = {astro-ph.EP},
       adsurl = {https://ui.adsabs.harvard.edu/abs/2017MNRAS.470..187A}
}

@article{seager2013biosignature,
  title={Biosignature gases in {H2}-dominated atmospheres on rocky exoplanets},
  author={Seager, S and Bains, W and Hu, R},
  journal={The Astrophysical Journal},
  volume={777},
  number={2},
  pages={95},
  year={2013},
  publisher={IOP Publishing}
}

@article{archer1996atlas,
  title={An atlas of the distribution of calcium carbonate in sediments of the deep sea},
  author={Archer, David E},
  journal={Global Biogeochemical Cycles},
  volume={10},
  number={1},
  pages={159--174},
  year={1996},
  publisher={Wiley Online Library}
}

@article{bednarvsek2012global,
  title={The global distribution of pteropods and their contribution to carbonate and carbon biomass in the modern ocean},
  author={Bednar{\v{s}}ek, Nina and Mo{\v{z}}ina, Jasna and Vogt, Meike and O'Brien, Colleen and Tarling, Geraint A},
  journal={Earth System Science Data},
  volume={4},
  number={1},
  pages={167--186},
  year={2012},
  publisher={Copernicus GmbH}
}

@article{Harada_2024,
doi = {10.3847/1538-4365/ad3e81},
url = {https://dx.doi.org/10.3847/1538-4365/ad3e81},
year = {2024},
month = {may},
publisher = {The American Astronomical Society},
volume = {272},
number = {2},
pages = {30},
author = {Caleb K. Harada and Courtney D. Dressing and Stephen R. Kane and Bahareh Adami Ardestani},
title = {Setting the Stage for the Search for Life with the Habitable Worlds Observatory: Properties of 164 Promising Planet-survey Targets},
journal = {The Astrophysical Journal Supplement Series},
}

@article{quanz2022large,
  title={Large Interferometer For Exoplanets (LIFE)-I. Improved exoplanet detection yield estimates for a large mid-infrared space-interferometer mission},
  author={Quanz, Sascha Patrick and Ottiger, Marcel and Fontanet, E and Kammerer, Jens and Menti, Franziska and Dannert, Felix and Gheorghe, A and Absil, Olivier and Airapetian, Vladimir S and Alei, Eleonora and others},
  journal={Astronomy \& Astrophysics},
  volume={664},
  pages={A21},
  year={2022},
  publisher={EDP Sciences}
}

@article{caldeira2003anthropogenic,
  title={Anthropogenic carbon and ocean pH},
  author={Caldeira, Ken and Wickett, Michael E},
  journal={Nature},
  volume={425},
  number={6956},
  pages={365--365},
  year={2003},
  publisher={Nature Publishing Group UK London}
}

@article{Fennel,
author = {Fennel, Katja and Mattern, Jann and Doney, Scott and Bopp, Laurent and Moore, Andrew and Yu, Liuqian and Wang, Bin},
year = {2022},
month = {09},
pages = {},
title = {Ocean biogeochemical modelling},
journal = {Nature Reviews Methods Primers},
doi = {10.1038/s43586-022-00154-2}
}

@article{claire2006biogeochemical,
  title={Biogeochemical modelling of the rise in atmospheric oxygen},
  author={Claire, Mark W and Catling, David C and Zahnle, Kevin J},
  journal={Geobiology},
  volume={4},
  number={4},
  pages={239--269},
  year={2006},
  publisher={Wiley Online Library}
}

@article{gebauer2017evolution,
  title={Evolution of earth-like extrasolar planetary atmospheres: assessing the atmospheres and biospheres of early earth analog planets with a coupled atmosphere biogeochemical model},
  author={Gebauer, S and Grenfell, JL and Stock, Joachim Wolfgang and Lehmann, Ralph and Godolt, M and von Paris, Philip and Rauer, H},
  journal={Astrobiology},
  volume={17},
  number={1},
  pages={27--54},
  year={2017},
  publisher={Mary Ann Liebert, Inc. 140 Huguenot Street, 3rd Floor New Rochelle, NY 10801 USA}
}

@article{xue2024jwst,
  title={JWST Thermal Emission of the Terrestrial Exoplanet GJ 1132b},
  author={Xue, Qiao and Bean, Jacob L and Zhang, Michael and Mahajan, Alexandra S and Ih, Jegug and Eastman, Jason D and Lunine, Jonathan I and Mansfield, Megan Weiner and Coy, Brandon P and Kempton, Eliza M-R and others},
  journal={arXiv preprint arXiv:2408.13340},
  year={2024}
}

@article{zahnle2017cosmic,
  title={The cosmic shoreline: The evidence that escape determines which planets have atmospheres, and what this may mean for Proxima Centauri B},
  author={Zahnle, Kevin J and Catling, David C},
  journal={The Astrophysical Journal},
  volume={843},
  number={2},
  pages={122},
  year={2017},
  publisher={IOP Publishing}
}

@article{gillen2023call,
  title={The call for a new definition of biosignature},
  author={Gillen, Catherine and Jeancolas, Cyrille and McMahon, Sean and Vickers, Peter},
  journal={Astrobiology},
  volume={23},
  number={11},
  pages={1228--1237},
  year={2023},
  publisher={Mary Ann Liebert, Inc., publishers 140 Huguenot Street, 3rd Floor New~…}
}

@article{stanton2018nitrous,
  title={Nitrous oxide from chemodenitrification: A possible missing link in the Proterozoic greenhouse and the evolution of aerobic respiration},
  author={Stanton, Chloe L and Reinhard, Christopher T and Kasting, James F and Ostrom, Nathaniel E and Haslun, Joshua A and Lyons, Timothy W and Glass, Jennifer B},
  journal={Geobiology},
  volume={16},
  number={6},
  pages={597--609},
  year={2018},
  publisher={Wiley Online Library}
}

@article{vickers2023confidence,
  title={Confidence of life detection: The problem of unconceived alternatives},
  author={Vickers, Peter and Cowie, Christopher and Dick, Steven J and Gillen, Catherine and Jeancolas, Cyrille and Rothschild, Lynn J and McMahon, Sean},
  journal={Astrobiology},
  volume={23},
  number={11},
  pages={1202--1212},
  year={2023},
  publisher={Mary Ann Liebert, Inc., publishers 140 Huguenot Street, 3rd Floor New~…}
}

@ARTICLE{1993Kasting,
       author = {{Kasting}, James F. and {Whitmire}, Daniel P. and {Reynolds}, Ray T.},
        title = "{Habitable Zones around Main Sequence Stars}",
      journal = {\icarus},
         year = 1993,
        month = jan,
       volume = {101},
       number = {1},
        pages = {108-128},
          doi = {10.1006/icar.1993.1010},
       adsurl = {https://ui.adsabs.harvard.edu/abs/1993Icar..101..108K}
}

@article{green2021call,
  title={Call for a framework for reporting evidence for life beyond Earth},
  author={Green, James and Hoehler, Tori and Neveu, Marc and Domagal-Goldman, Shawn and Scalice, Daniella and Voytek, Mary},
  journal={Nature},
  volume={598},
  number={7882},
  pages={575--579},
  year={2021},
  publisher={Nature Publishing Group UK London}
}

@article{roberson2011greenhouse,
  title={Greenhouse warming by nitrous oxide and methane in the Proterozoic Eon},
  author={Roberson, April L and Roadt, J and Halevy, I and Kasting, JF},
  journal={Geobiology},
  volume={9},
  number={4},
  pages={313--320},
  year={2011},
  publisher={Wiley Online Library}
}

@ARTICLE{2004Icar,
       author = {{Kress}, Monika E. and {McKay}, Christopher P.},
        title = "{Formation of methane in comet impacts: implications for Earth, Mars, and Titan}",
      journal = {\icarus},
         year = 2004,
        month = apr,
       volume = {168},
       number = {2},
        pages = {475-483},
          doi = {10.1016/j.icarus.2003.10.013},
       adsurl = {https://ui.adsabs.harvard.edu/abs/2004Icar..168..475K}
}

@article{schaefer2010chemistry,
  title={Chemistry of atmospheres formed during accretion of the Earth and other terrestrial planets},
  author={Schaefer, Laura and Fegley Jr, Bruce},
  journal={Icarus},
  volume={208},
  number={1},
  pages={438--448},
  year={2010},
  publisher={Elsevier}
}

@article{thompson2021composition,
  title={Composition of terrestrial exoplanet atmospheres from meteorite outgassing experiments},
  author={Thompson, Maggie A and Telus, Myriam and Schaefer, Laura and Fortney, Jonathan J and Joshi, Toyanath and Lederman, David},
  journal={Nature Astronomy},
  volume={5},
  number={6},
  pages={575--585},
  year={2021},
  publisher={Nature Publishing Group UK London}
}

@ARTICLE{1967Hitchcock,
       author = {{Hitchcock}, Dian R. and {Lovelock}, James E.},
        title = "{Life detection by atmospheric analysis}",
      journal = {\icarus},
         year = 1967,
        month = jan,
       volume = {7},
       number = {1-3},
        pages = {149-159},
          doi = {10.1016/0019-1035(67)90059-0},
       adsurl = {https://ui.adsabs.harvard.edu/abs/1967Icar....7..149H}
}

@ARTICLE{1965Lovelock,
       author = {{Lovelock}, J.~E.},
        title = "{A Physical Basis for Life Detection Experiments}",
      journal = {\nat},
         year = 1965,
        month = aug,
       volume = {207},
       number = {4997},
        pages = {568-570},
          doi = {10.1038/207568a0},
       adsurl = {https://ui.adsabs.harvard.edu/abs/1965Natur.207..568L}
}

@ARTICLE{1975Lovelock,
       author = {{Lovelock}, J.~E.},
        title = "{Thermodynamics and the Recognition of Alien Biospheres}",
      journal = {Proceedings of the Royal Society of London Series B},
         year = 1975,
        month = may,
       volume = {189},
       number = {1095},
        pages = {167-180},
          doi = {10.1098/rspb.1975.0051},
       adsurl = {https://ui.adsabs.harvard.edu/abs/1975RSPSB.189..167L}
}

@ARTICLE{Lovelock1974,
       author = {{Lovelock}, James E. and {Margulis}, Lynn},
        title = "{Atmospheric homeostasis by and for the biosphere: the gaia hypothesis}",
      journal = {Tellus},
         year = 1974,
        month = feb,
       volume = {26},
       number = {1-2},
        pages = {2-10},
          doi = {10.1111/j.2153-3490.1974.tb01946.x10.3402/tellusa.v26i1-2.9731},
       adsurl = {https://ui.adsabs.harvard.edu/abs/1974Tell...26....2L}
}

@article{schlichting2015atmospheric,
  title={Atmospheric mass loss during planet formation: the importance of planetesimal impacts},
  author={Schlichting, Hilke E and Sari, Re’em and Yalinewich, Almog},
  journal={Icarus},
  volume={247},
  pages={81--94},
  year={2015},
  publisher={Elsevier}
}

@ARTICLE{2022arXiv221014293M,
       author = {{Meadows}, Victoria and {Graham}, Heather and {Abrahamsson}, Victor and {Adam}, Zach and {Amador-French}, Elena and {Arney}, Giada and {Barge}, Laurie and {Barlow}, Erica and {Berea}, Anamaria and {Bose}, Maitrayee and {Bower}, Dina and {Chan}, Marjorie and {Cleaves}, Jim and {Corpolongo}, Andrea and {Currie}, Miles and {Domagal-Goldman}, Shawn and {Dong}, Chuanfei and {Eigenbrode}, Jennifer and {Enright}, Allison and {Fauchez}, Thomas J. and {Fisk}, Martin and {Fricke}, Matthew and {Fujii}, Yuka and {Gangidine}, Andrew and {Gezer}, Eftal and {Glavin}, Daniel and {Grenfell}, Lee and {Harman}, Sonny and {Hatzenpichler}, Roland and {Hausrath}, Libby and {Henderson}, Bryana and {Johnson}, Sarah Stewart and {Jones}, Andrea and {Hamilton}, Trinity and {Hickman-Lewis}, Keyron and {Jahnke}, Linda and {Kacar}, Betul and {Kopparapu}, Ravi and {Kempes}, Christopher and {Kish}, Adrienne and {Krissansen-Totton}, Joshua and {Leavitt}, Wil and {Komatsu}, Yu and {Lichtenberg}, Tim and {Lindsay}, Melody and {Maggiori}, Catherine and {Des Marais}, David and {Mathis}, Cole and {Morono}, Yuki and {Neveu}, Marc and {Ni}, Grace and {Nixon}, Conor and {Olson}, Stephanie and {Parenteau}, Niki and {Perl}, Scott and {Quinn}, Richard and {Raj}, Chinmayee and {Rodriguez}, Laura and {Rutter}, Lindsay and {Sandora}, McCullen and {Schmidt}, Britney and {Schwieterman}, Eddie and {Segura}, Antigona and {Sekerci}, Fatih and {Seyler}, Lauren and {Smith}, Harrison and {Soares}, Georgia and {Som}, Sanjoy and {Suzuki}, Shino and {Teece}, Bonnie and {Weber}, Jessica and {Wolfe-Simon}, Felisa and {Wong}, Michael and {Yano}, Hajime and {Young}, Liza},
        title = "{Community Report from the Biosignatures Standards of Evidence Workshop}",
      journal = {arXiv e-prints},
         year = 2022,
        month = oct,
          eid = {arXiv:2210.14293},
        pages = {arXiv:2210.14293},
          doi = {10.48550/arXiv.2210.14293},
archivePrefix = {arXiv},
       eprint = {2210.14293},
 primaryClass = {astro-ph.IM},
       adsurl = {https://ui.adsabs.harvard.edu/abs/2022arXiv221014293M}
}

@ARTICLE{Krissansen2022,
       author = {{Krissansen-Totton}, Joshua and {Thompson}, Maggie and {Galloway}, Max L. and {Fortney}, Jonathan J.},
        title = "{Understanding planetary context to enable life detection on exoplanets and test the Copernican principle}",
      journal = {Nature Astronomy},
         year = 2022,
        month = feb,
       volume = {6},
        pages = {189-198},
          doi = {10.1038/s41550-021-01579-7},
archivePrefix = {arXiv},
       eprint = {2202.10333},
 primaryClass = {astro-ph.EP},
       adsurl = {https://ui.adsabs.harvard.edu/abs/2022NatAs...6..189K}
}

@article{des2008nasa,
  title={The NASA astrobiology roadmap},
  author={Des Marais, David J and Nuth III, Joseph A and Allamandola, Louis J and Boss, Alan P and Farmer, Jack D and Hoehler, Tori M and Jakosky, Bruce M and Meadows, Victoria S and Pohorille, Andrew and Runnegar, Bruce and others},
  journal={Astrobiology},
  volume={8},
  number={4},
  pages={715--730},
  year={2008},
  publisher={Mary Ann Liebert, Inc. 140 Huguenot Street, 3rd Floor New Rochelle, NY 10801~…}
}



\appendix

\section{Examples of Abundance-Space Comparative Biosignatures}\label{appendix:biosig}
\label{sec:bio}  
Comparative biosignatures offer a framework to address the challenge of distinguishing biotic from abiotic signals in data; abiotic patterns would only emerge if we consider all relevant abiotic processes. (This otherwise serves as a diagnostic test, see Section \ref{sec:models})

\begin{figure*}
	\includegraphics[width=\textwidth]{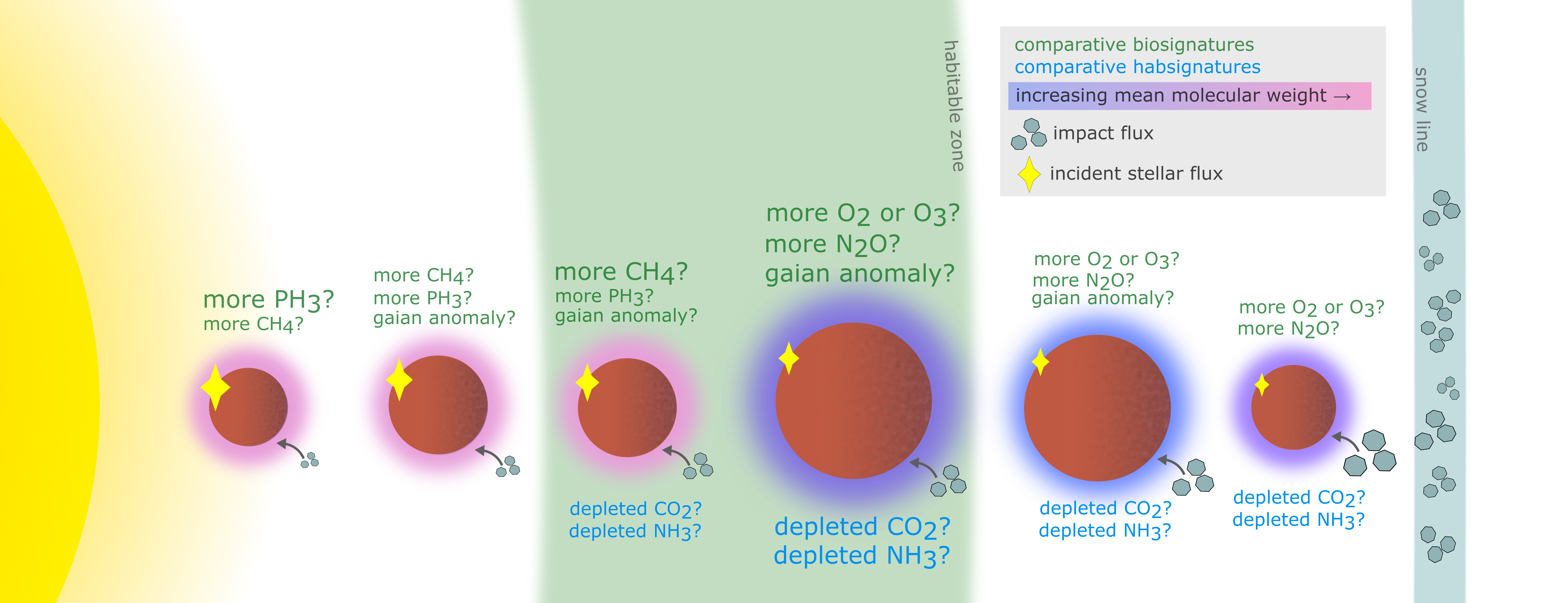}
    \caption{Illustration of comparative biosignatures and habsignatures in an extrasolar system. Potential comparative biosignatures (green text) and habsignatures (blue text) as deviations from the abiotic baseline are annotated, with larger fonts indicating stronger candidates due to larger priors $p(\text{life | context})$; non-HZ (outside the green annulus sector) planets are also considered, but have smaller priors. Systemic abiotic processes setting biosignature baseline expectations include incident stellar flux (yellow stars), impactor flux from the snow line (blue pebbles; blue annulus sector), and atmospheric mean molecular weight (pink to blue spectrum, only applicable to \ce{CH4} and \ce{PH3}). The snow line includes any molecular ice line (e.g., \ce{CO2}, \ce{H2O}). Orbital distances, colours, and trend of mean molecular weight are illustrative.
}
    \label{fig:patterns}
\end{figure*}

Here, we review examples of comparative biosignatures that would theoretically exhibit a $ \Delta \mathcal{E}_L > 4$, as illustrated in Figure \ref{fig:patterns}. In the context of exoplanets, the framework considers fitting the model atmosphere to the full transmission/emission spectra, where anomalies represent failures of the model to fit the data, rather than specific misfits to certain molecules. We focus here instead focus on simplified, single-molecule cases within abundance-space. We assume all relevant dominant abiotic processes have been accounted for when calibrating for the abiotic baseline (i.e., a reliable Step 2 in Fig. \ref{fig:build}), effectively treating the planetary properties as ``all else equal''. For each case study, we first introduce the biosignature that is expected to vary predictably or deterministically with systemic parameters under purely abiotic conditions. We then examine the extremes of these parameters where biological processes would most sharply juxtapose with the established abiotic trends, offering the strongest potential indicators of life.

\subsection{More \ce{O2} and/or \ce{O3} on an outer and/or less massive planet}
\label{sec:oxygen}

\emph{The nature of \ce{O2} and \ce{O3} biosignatures: }Among all proposed signatures for life detection, molecular oxygen (\ce{O2}) is notable as a strong biosignature \citep[e.g.,][and references therein]{Meadows2018}. It has been historically regarded as a reliable marker of biological activity due to its production from oxygenic photosynthesis, and corresponding lack of abiotic source: volcanic gases are much too reduced to release significant \ce{O2} \citep{holland2002volcanic}. Though anoxygenic photosynthesis predates oxygenic \citep{blankenship2010early}, oxygenic photosynthesis is the dominant metabolism on modern Earth, powered by incident stellar flux; the dominant source of energy at our planet's surface. As starlight is expected to be a ubiquitous source of energy on terrestrial habitable zone planets, similar metabolic pathways could plausibly operate on extraterrestrial worlds. 

Ozone (\ce{O3}) can also act as a biosignature for photosynthetic life, as it can be a proxy for molecular oxygen in when probing mid-infrared wavelengths where \ce{O2} is not spectrally active \citep[e.g.,][]{Marais2002, Segura2003, Leger2011}.  Abundances of \ce{O3} correlate with \ce{O2}, especially for planets with lower incident UV \citep{Kozakis2022}. 

\emph{Ambiguities in \ce{O2} and \ce{O3} interpretation: }Despite its biological association, interpreting oxygen or ozone as definitive biosignatures is challenging: Surface processes can consume oxygen, masking its biological production, and abiotic processes can cause it to accumulate.

On Earth, the delayed rise of atmospheric \ce{O2}, despite the presence of oxygenic photosynthetic organisms, serves as an example of a false negative. Geological sinks counteracted the biological production of \ce{O2}, delayed its accumulation to detectable levels \citep{sleep2005dioxygen, lyons2014rise, planavsky2014low, reinhard2017false} for at-least half a billion years \citep{planavsky2014evidence}.

One mechanism for abiotic \ce{O2} production involves photolysis of water vapour and subsequent hydrogen escape after ocean vaporisation. The remaining oxygen radicals can then re-combine to form \ce{O2}. This scenario is expected to be more prominent for planets with high incident stellar fluxes, e.g., orbiting young super-luminous M dwarf stars \citep{LandB2015}.

Planetary atmospheres with low inventories of non-condensible gases can lead to high stratospheric water vapour abundances due to an elevated tropospheric cold trap, exposing the vapour to a higher UV flux and promoting photolysis \citep{Wordsworth2014}. This mechanism is not restricted to specific stellar types and depends primarily on the planet's atmospheric composition.

The photolysis of carbon dioxide was also thought to be a source of atmospheric \ce{O2} on M-dwarf planets \citep{gao2015stability}, depending on the effectiveness of stellar UV radiation in photolysing \ce{CO2}, and the presence of mechanisms that prevent the rapid recombination of carbon monoxide and oxygen \citep{domagal2014abiotic, tian2014high, gao2015stability, harman2015abiotic}. However, recent work suggests that \ce{O2} remains a trace gas under more accurate atmospheric modelling, as prior models underestimated the altitude of the \ce{CO2} photolysis peak. This reduces the likelihood of \ce{O2} as an abiotic by-product on \ce{CO2}-rich M-dwarf planets, though \ce{O3} still poses false-positive risks \citep{ranjan2023importance}.

\emph{The abiotic baseline for \ce{O2} and \ce{O3}: }The aforementioned photochemical processes are driven by the incident stellar spectrum, which, all else being equal, creates a system-scale expectation for the abundance of \ce{O2} and \ce{O3} as incident flux drops off as the inverse-square of orbital separation. The higher flux incident on planets closer to the host star shortens the photochemical lifetimes of O-bearing molecules, promoting their photolytic destruction and the subsequent formation of oxygen radicals (O) that readily combine to form \ce{O2} \citep{Zhou2014}. 

Abiotic ozone has an even stronger dependence with the distance from the host star, as its production is based on a stellar flux of sufficient intensity for photolysis of \ce{O2} \citep{Kozakis2022}. Therefore, if ozone is exclusively observed in a terrestrial HZ farther away from the host star, it implies a substantial source flux of \ce{O2}.

The abiotic build-up of \ce{O2} is amplified by higher rates of hydrogen escape closer to the host star. The more XUV radiation heats the upper atmosphere, increasing thermally-driven atmospheric escape: more energetic hydrogen molecules can reach escape velocity (Jeans escape), and the heating causes the atmosphere to dynamically expand and escape into space \citep[hydrodynamic escape;][]{1982Hunten, 2008Johnson}. A net loss of hydrogen (and potentially carbon) from escape limits the reformation of O-bearing molecules and drives atmospheric oxidation. Non-thermal escape processes, encompassing mechanisms driven by photochemical reactions or charged particle interactions, are often also {dependent} on stellar radiation for photoionisation \citep[i.e., photochemical escape,][]{shematovich2018escape}, solar wind for kinetic energy \citep[sputtering exchange, e.g.,][]{lundin2007planetary, Gronoff} or charge-exchange \citep[charge escape,][]{shematovich2018escape}. Consequently, planets closer to their host stars emerge as favourable candidates for abiotic oxygen build-up. 

For H-loss, planetary mass and radius play a similarly critical role. Hydrodynamic escape negatively correlates with gravity \citep{Kubyshkina}. The evolution of low-mass planets ($\lesssim$10 Earth masses) is particularly dominated by atmospheric hydrodynamic loss due to a combination of high thermal energy and low gravity \citep{Kubyshkina}. This atmospheric loss intensifies with increasing planetary radius and thus with increasing planetary core temperature and atmospheric mass fraction. Non-thermal atmospheric escape rates also have a monotonic function of the planetary radius. Magnetohydrodynamic modelling has revealed the trade-off between the cross-sectional area of a planet (which increases with size, boosting escape rates) and its associated escape velocity (which also increases with size but diminishes escape rates), with {escape rates} peaking at 0.7 Earth radii \citep{Chin_2024}. Planets that are more massive and/or are closer to their host star are thus anticipated to accumulate more abiotic \ce{O2}. 

If significant amounts of \ce{O2} are detected on an outer and/or massive planet within a system, especially one containing water and carbon dioxide, while an inner/less massive planet exhibits little to no \ce{O2}, it becomes challenging to attribute this solely to abiotic mechanisms, hinting at a biotic origin. This situation is exemplified by our solar system, where Earth's abundant atmospheric \ce{O2} (Fig. \ref{fig:solar_syst}) is an outlier compared to the baseline set-up by Venus and Mars, both of which have $\ll$1\% atmospheric \ce{O2} \citep{Oyama, Groller}.

The expected abiotic \ce{O2} build-up on a planet could be predicted by atmospheric models incorporating photochemistry and atmospheric escape \citep[as in][]{ranjan2023importance}. However, unknown parameters affecting planetary evolution, such as original O, C and H budget of the atmosphere and the past behaviour of stochastic stellar events classify \ce{O2}/\ce{O3} as a Class II biosignature. The abiotic baseline would provide the calibration against past events like stochastic stellar flares (influenced by semi-major axis) or the initial volatile budget (linked to disk chemistry, orbital architecture, and planetary mass distributions that influence volatile delivery).

\subsection{More \ce{CH4} (with \ce{CO2}) on an inner planet or planet with high molecular weight atmosphere} \label{sec:ch4}

\emph{The nature of \ce{CH4} as a biosignature: }Methane (\ce{CH4}) has long been regarded as a potential biosignature gas, due to its large terrestrial surface fluxes \citep{Jackson_2020}.  Most of the Earth's atmospheric methane is biogenic in origin: produced directly from life, or released from metamorphic reactions of organic matter \citep{jackson2020increasing}.  While anthropogenic sources are the majority of methane production on modern-day Earth \citep{Jackson_2020}, the Archaean Earth had a biosphere even more abundant in methane, >100 times more than modern-day Earth, sourced from methanogens \citep{KASTING2005119, catling2020archean, catling2001biogenic}.  See \citet{Thompson2022} for a thorough review of methane as a biosignature gas. 

Methanogenesis may be widespread across other habitable worlds due to the likely ubiquity of the \ce{CO2} and \ce{H2} redox couple in the atmosphere of terrestrial planets. Methane is also one of the few biosignatures that may be readily detectable with JWST for Earth-like biogenic fluxes, as evidenced by its first discovery in an exoplanet atmosphere \citep{madhusudhan2023carbon}. {\ce{CH4} has since also been detected in the atmospheres of WASP-80 b \citep{bell2023methane}, TOI-270 d \citep{benneke2024jwst} and GJ 3470 b \citep{beatty2024sulfur}. If the planetary context can be used to rule out abiotic sources, methane could be the first indication of life beyond Earth -- for these examples or others forthcoming. }

However, methane is sometimes dismissed as irredeemably ambiguous due to its ubiquity in planetary environments and potential for non-biological production \citep{kasting2005methane, Marais2002}. It is thus imperative to first understand the atmospheric, geochemical, and evolutionary context of the planet. 

\emph{Ambiguities in \ce{CH4} detection: }Abiotic methane can be sourced from volcanism, metamorphism, and exogenically through impacts. Through the hydration and metamorphic transformation of ferromagnesian minerals, serpentinisation produces \ce{H2} \citep{berndt1996reduction}. Subsequent abiotic Fischer–Tropsch reactions use up the produced \ce{H2} to reduce inorganic carbon from \ce{CO2} to methane and other abiotic organic compounds \citep{guzman2013abiotic, etiope2013abiotic}. However, serpentinisation laboratory experiments did not produce detectable \ce{CH4} \citep{grozeva2017experimental, mccollom2016generation}, suggesting that \ce{CH4} may have arisen in earlier experiments from organic contamination \citep{mccollom2016abiotic}. 

In oxidising atmospheres, methane faces a short photochemical lifetime, oxidised by photolytically produced OH radicals. In environments where \ce{CO2} dominates over methane, oxidants drive methane destruction, converting it to \ce{CO2} while producing H that can escape to space. In reduced environments, where \ce{CH4} dominates over \ce{CO2}, methane polymerises into aerosols, which fall to the ground and remove the atmospheric \ce{CH4} \citep{Thompson2022}. While thermal decomposition can then release \ce{CH4} back into the atmosphere, H released during methane photolysis is irretrievably lost to space, such that the rising C:H ratio of condensate material cannot reproduce methane. Regions abundant in hydrogen could replenish the H to facilitate the recombination of hydrogen and carbon to form methane, extending its photochemical lifetime. Inner system regions with a greater incident stellar flux are subject to greater photochemical destruction fluxes of \ce{CH4}, and are less likely to possess significant quantities of \ce{H2} due to greater hydrogen escape. An abiotic gradient is thus set up, with decreasing \ce{CH4} with decreasing orbital radius. 

The biosignature pairings for \ce{CH4} could resolve some of the ambiguities; the detection of \ce{CH4} with \ce{CO2} could indicate the biogenic atmospheric disequilibrium \citep{Krissansen2018}, a broader biosignature than a single gas. Abiotic mechanisms typically have difficulty forming atmospheres abundant in \ce{CH4} and \ce{CO2} but deficient in \ce{CO} due to the strong redox imbalance between \ce{CH4} and \ce{CO2}. Without life continuously replenishing atmospheric \ce{CH4}, this redox imbalance could not persist because \ce{CH4} would be photolytically destroyed. Methane is thus more likely to be of biogenic origin when detected in high abundance with \ce{CO2} and little \ce{CO} \citep{Krissansen2018, Thompson2022}. 

Asteroid and cometary impacts could also lead to the generation of abiotic methane in a planetary atmosphere. Asteroids deliver significant reducing power in the form of iron, which, upon impact, reacts with the water vapour in the atmosphere to produce \ce{H2} and iron oxide (\ce{FeO}) \citep{zahnle2020creation}. Subsequently, \ce{H2} and \ce{FeO} can react with atmospheric \ce{CO2} or \ce{CO} to form \ce{CH4}. The amount of methane generated depends on the carbon availability before impact, the quantity of iron delivered by the impactor, and the presence of catalysts facilitating methane production \citep{Itcovitz_2022} -- all of which are challenging to constrain for exoplanets. Such post-impact atmospheres, while only transient, could provide a false positive, potentially distinguished if accompanied by \ce{H2}-dominated atmospheres \citep{zahnle2020creation} (as could be inferred from the planet's density). 

Comets vaporise upon entry or impact, potentially forming methane during the subsequent cooling phase. The effectiveness of condensing dust from cometary impactors as catalysts for methane production remains uncertain, with models suggesting high \ce{CH4} yields \citep{2004Icar, kasting2005methane}, and experimental outgassing of chondritic materials producing little methane \citep{schaefer2010chemistry, thompson2021composition}.  Despite this uncertainty, the potential of comets to generate \ce{CH4} warrants consideration as a potential source of methane if detected. 

\emph{The abiotic baseline for \ce{CH4}: }Methane derived from impacts is expected to correlate with planets experiencing greater bombardment fluxes. For planetary systems with `peas in a pod' architectures, outer-disk planets would be subject to a greater influx of impactors. For the Trappist-1 system, models reveal that innermost planets experience a higher rate of cometary impacts and retain more volatiles, leading to the subsequent build-up of new secondary atmospheres \citep{Kral2018}. We expect this pattern to be prominent in other similarly-structured planetary systems, where secondary or transient post-impact signatures, like \ce{CH4}, will be correlated with the outer planets. Abiotic methane is thus more likely to accumulate at detectable levels on outer planets: shielded from photochemically produced OH radicals, smaller loss rates of H, and greater meteoritic flux as an exogenic source of reducing power. If methane is detected exclusively or in a much higher abundance in an inner planet, an alternative origin therefore becomes a more plausible explanation. 

For more solar system-like orbital geometries, the impact rate experienced by terrestrial planets would be positively correlated with planetary mass and radius, or be a function of proximity to the system's exo-Jupiter \citep{2009Horner,2010Horner, 2012IHorner}. Such systems might be harder to fit a baseline to, and might be better defined by the stochasticity of the impact events, rather than the frequency or mass influx. 

Methane is especially unstable in high mean molecular weight atmospheres \citep{Thompson2022}, possessing an intrinsically short photochemical lifetime in such settings, regardless of the presence of \ce{CO2}. A trend is thus expected across atmospheric molecular weight, where planets with {lower} molecular weight atmospheres are expected to have more \ce{CH4} be built up abiotically. To maintain detectable levels of methane in a high mean molecular weight atmosphere, then a high continuous replenishment is needed, which is unlikely to be sustained by abiotic processes alone. The most compelling {case} for methane as a comparative biosignature is thus presented: \ce{CH4} with \ce{CO2} (in the absence of \ce{CO}) on an inner planet with a high mean-molecular-weight atmosphere. {However, CO is significantly more difficult to detect with JWST spectroscopy than either \ce{CO2} or \ce{CH4}, due to its weaker spectral features \citep[e.g.][]{davenport2025toi}. Therefore, the absence of a CO detection should not be misinterpreted as evidence that CO is truly absent.}

Self-consistent atmospheric models can describe photochemistry and thermochemistry-derived estimates for abiotic \ce{CH4} abundances. However, atmospheric methane levels also depend heavily on surface-atmosphere processes that cannot be observationally constrained, such as weathering rates and metamorphism, which constitute observationally `hidden' parameters related to tectonic regime. Methane would therefore be most accurately classified as a Class II biosignature. Impact-derived methane is likely to be either probabilistic, depending on the system's architecture, or, for the case of large impacts, stochastic. 

\subsection{More \ce{N2O} (with \ce{O2}) on an outer planet}

\emph{The nature of \ce{N2O} as a biosignature: }Nitrous oxide (\ce{N2O}) is another compelling exoplanet biosignature gas, as a product of microbial nitrogen metabolism \citep[see review of \ce{N2O} as a biosignature][]{schwieterman2022evaluating}. Earth's atmospheric \ce{N2O} is primarily of biological origin.  The presence of \ce{N2O} is readily detectable in the near-IR and mid-IR in the Earth’s spectral energy distribution \citep{gordon2022hitran2020}. 

Denitrification --- a relatively ubiquitous metabolism on Earth --- transforms \ce{NO3-} to \ce{N2} gas, with \ce{N2O} as an intermediate product \citep{offre2013archaea}.  Biological denitrification occurs during the decay of plant matter, thus the detection of \ce{N2O} alongside photosynthesis signatures \ce{H2O} and \ce{CO2} suggests the presence of decaying (and live) plant life.  

Biogenic \ce{N2O} fluxes from denitrification may have been more pronounced earlier in Earth's history. This is because modern-day earth has copper catalysing the last step of denitrification from \ce{N2O} to \ce{N2}, which was in limited availability in the Proterozoic \citep{BUICK} rendering denitrification less efficient.  

Despite its primarily biological origin, abiotic sources of \ce{N2O} can lead to false positive detections. Lightning can amplify \ce{N2O} through disequilibrium nitrogen fixation. However, lightning-produced \ce{N2O} can be distinguished from a biogenic source if other lightning by-products can be detected, such as \ce{NO2} \citep{schwieterman2022evaluating}. 

\emph{The abiotic baseline for \ce{N2O}: }Stellar activity and stellar proton events {can also produce \ce{N2O}}, which can create a system-scale expectation for {its abundance} as incident flux drops off with increasing orbital separation. The secondary production of energetic electrons can generate NO in planetary atmospheres \citep{crutzen1975solar}, which, in the presence of sufficient H-bearing species, can react to form \ce{N2O} \citep{airapetian2016prebiotic}. Though solar particle events are stochastic, the effect on each planet is a function of semi-major axis: planets closer to the host star, subject to higher fluxes of energetic protons and electrons, are therefore more prone to larger abiotic fluxes of \ce{N2O}. A gradient of abiotic \ce{N2O} is thus formed, with increasing abundances in the atmospheres of planets closer to their host star. Therefore, if \ce{N2O} is detected in greater abundance on a planet farther from the host star (and all else being equal), it is unlikely to have formed exclusively from stellar activity. 

Similarly to biological denitrification, chemodenitrification involves the abiotic reduction of oxidised nitrogen species such as \ce{NO3-} by using chemical reductants like ferrous iron (\ce{Fe2+}), producing \ce{N2O} gas abiotically \citep{samarkin2010abiotic, jones2015stable, stanton2018nitrous}. This process needs reductants and oxidants present simultaneously in ocean layers, which is unlikely with an abiotic \ce{O2} atmosphere, which requires a completely oxidised ocean for long-term stability. The simultaneous presence of \ce{N2O} and \ce{O2} would thus be a stronger biosignature than \ce{N2O} alone. Though lightning can create small amounts of \ce{NOx} that are chemodenitrified without biotic \ce{O2}, detectable levels likely need biological production \citep{schwieterman2022evaluating}. 

Nitrous oxide is susceptible to photochemical degradation, but oxidised atmospheres with \ce{O2}/\ce{O3} or \ce{CO2} can shield it from incident stellar UV, protecting against destructive photolysis \citep{stanton2018nitrous, roberson2011greenhouse, schwieterman2022evaluating}. As such, a large replenishing flux would be required to sustain nitrous oxide to observable levels, particularly in the upper atmosphere. This challenges the build-up of both biotic and abiotic \ce{N2O} (especially for inner planets), with the potential of a false negative when biogenic \ce{N2O} cannot reach detectable levels. So another added advantage of looking for \ce{N2O} on planets farther out is that the increased photochemical lifetime reduces the odds of false negatives. 

The atmospheric profile of abiotic \ce{N2O} could be predicted with self-consistent photochemical and thermochemical atmospheric models, incorporating observed host star activity and detections of related molecules like \ce{N2}, \ce{CH4}, \ce{H2O}, and \ce{H2} as constraints for the atmospheric chemistry. Lightning could be more challenging to incorporate given the difficulty of remotely constraining lightning behaviour on a planet, or indirectly constraining parameters to tune lightning models such as cloud compositions and wind speeds to help modulate expectations. Here, other lightning signatures like \ce{NO2} could be used to constrain lightning power in atmospheric models. However, stellar particle events that influence \ce{N2O} abundance are inherently stochastic, and a single snapshot observation of the star-planet system cannot fully capture their historical activity. Instead of a direct, deterministic relationship between observed parameters and \ce{N2O} levels, the semi-major axis can serve as a proxy for stellar exposure, as the impact of past stellar events on each planet is consistently modulated by orbital distance. This treatment of past stellar stochastic behaviour as a `hidden' variable classifies \ce{N2O} as a Class II biosignature. This is especially the case when \ce{N2O} is detected with \ce{O2} to account for the otherwise unfettered chemodenitrification.

\subsection{More \ce{PH3} on planet with higher mean molecular weight}

\emph{The nature of \ce{PH3} as a biosignature: }While phosphine (\ce{PH3}) is a trace gas in Earth's atmosphere, biospheres on other planets, particularly anoxic ones, could potentially accumulate detectable levels as anoxic life forms might produce \ce{PH3} in significantly higher quantities \citep{bains2019trivalent}.

Anoxic life forms have two potential pathways for \ce{PH3} production. In the first, less favoured pathway \citep[as argued by][]{sousa2020phosphine}, acidic products of anoxic fermentation of organic matter could react with trace metal phosphides (potentially present in the environment) to generate \ce{PH3}. The widespread detection of \ce{PH3} in animal faeces in various environments argues against this route \citep{gassmann1993phosphane, pridham1998determination, zhu2014penguins}, as consistent metal phosphide contamination is improbable. The second pathway involves direct conversion of environmental phosphorus into \ce{PH3} by anaerobic bacteria during their metabolic processes. Supporting evidence comes from studies where \ce{PH3} was detected in bacterial cultures devoid of metal phosphides \citep[e.g.,][]{devai1988detection, glindemann1996free, jenkins2000phosphine, bains2019new}, negating the indirect production route. Despite the debate over precise mechanisms, robust evidence for biological \ce{PH3} production (citations above) solidifies its candidacy as a biosignature.

\emph{The abiotic baseline for \ce{PH3}: }The abiotic formation pathways for phosphine are fundamentally different for gas giants and terrestrial planets, meaning a direct comparison of their atmospheric abundances is not meaningful. Instead, in our framework, the value of observing gas giants is not for direct comparison, but to use them as probes of the system’s primordial elemental abundances; an otherwise inaccessible, latent parameter.

The abundance of \ce{PH3} in the deep, hot atmospheres of gas giants is governed by thermochemical equilibrium, which can be robustly modelled \citep{visscher2006atmospheric}. Observations of \ce{PH3} on a system's gas giants, such as Jupiter \citep{prinn1975phosphine} and Saturn \citep{bregman1975observation}, can therefore be used to infer the primordial phosphorus budget from the natal disk. Second, the architecture of the entire planetary system, including the locations of other planets and potential asteroid or cometary belts, is crucial for modelling the later exogenic delivery of phosphorus via impacts over geological time.

This inferred phosphorus budget then serves as an informed prior for modelling the system's terrestrial planets. It allows for a much more robust calculation of the expected abiotic \ce{PH3} abundance on a rocky planet, considering plausible abiotic sources like volcanism or surface chemistry, which are ultimately limited by the total phosphorus available. A terrestrial planet with a detected \ce{PH3} abundance that significantly exceeds this system-calibrated abiotic baseline would represent a powerful anomaly. This approach strengthens the inference of biogenicity by replacing a broad, unconstrained prior on the planetary phosphorus budget with a system-specific, observationally-informed one.

The uncertainty surrounding potential abiotic pathways on terrestrial planets, such as the putative \ce{PH3} on Venus \citep{2021Greaves, bains2021phosphine}, highlights the need for such constraints. Because its abiotic production on terrestrial worlds depends on the observationally-inaccessible phosphorus inventory (a system-wide latent variable) which must be inferred from other planets, phosphine is best described as a Class II biosignature within our framework.

\section{A Practical Guide to Interpreting Bayesian LOO-CV}\label{appendix:limitations}
{The expected log pointwise predictive density (elpd) and its approximation via Pareto Smoothed Importance Sampling (PSIS–LOO) provide a practical measure of out-of-sample predictive performance. Interpreting differences in elpd between models ($\Delta \mathcal{E}_L$) requires care and an awareness of the method’s assumptions. This appendix summarises practical guidance for interpreting these diagnostics, following \citet{VehtariFAQ}.}

\subsection{Guidelines for Model Comparison and Selection}
{The elpd$_\mathrm{LOO}$ metric should be viewed not as a binary test but as a continuous measure of evidence. It quantifies the relative support for competing models, typically abiotic versus biotic, without enforcing rigid thresholds. Numerical guidelines, adapted from \citet{VehtariFAQ}, offer useful reference points:}

\begin{itemize}
    \item \textbf{Magnitude of Difference:} If $|\Delta\mathcal{E}_L| < 4$, the evidence favouring one model over another is weak; predictive performances are statistically indistinguishable.
    \item \textbf{Significance Estimate:} For $|\Delta\mathcal{E}_L| > 4$, the difference may be meaningful. Comparing this value to its standard error $\text{SE}_\Delta$, a ratio $|\Delta\mathcal{E}_L| / \text{SE}_\Delta > 2$ indicates a substantial improvement in predictive capability.
\end{itemize}

{Two caveats qualify this interpretation. First, the SE assumes independent data points; the hierarchical structure of our framework explicitly models planetary dependencies, partially satisfying this condition. Second, the SE relies on the Central Limit Theorem, valid only for large $N_p$ (typically $>100$). In most exoplanet systems, $N_p$ is small, making $|\Delta\mathcal{E}_L| < 4$ the more reliable heuristic. Consequently, elpd metrics are best used as indicators of evidence strength, not as strict detection criteria.}

\subsection{The Pareto $k$ Diagnostic as a Model Checking Tool}
{The Pareto $k$ diagnostic is central to assessing PSIS–LOO reliability. It quantifies how influential each data point is on the posterior, serving primarily as a tool for model checking. Large $k$ values identify data points for which the PSIS approximation is unreliable and the model likely misspecified.}

{The interpretation of the Pareto $k$ value provides a detailed diagnostic:}
\begin{itemize}
    \item \textbf{$k < 0.5$:} {Good. The approximation is reliable.}
    \item \textbf{$0.5 \le k < 0.7$:} {Acceptable. The approximation is reliable, but the observation has a notable influence, and the uncertainty of elpd$_{\text{LOO}}$ can be slightly underestimated.}
    \item \textbf{$0.7 \le k < 1$:} {Bad. The approximation is unreliable. The planet is a statistical outlier that is highly influential. This is a strong warning that the model struggles to predict this data point.}
    \item \textbf{$k \ge 1$:} {Very bad. The approximation has failed. The model is likely severely misspecified for this data point.}
\end{itemize}
{In our framework, a planet with a high Pareto $k$ ($k \ge 0.7$) is an ``influential world". This is not merely a computational issue; it is a scientific signal that the model's assumptions (i.e., its underlying physics and chemistry) are failing for that specific planet. Such an observation demands further investigation into what makes it so anomalous.}

\subsection{Connecting Diagnostics to Scientific Scenarios}
{This statistical guidance directly informs the interpretation of the three scientific outcomes outlined in Figure 7 of the main text.}
\begin{enumerate}
    \item \textbf{Models with Similar Predictions:} {When $|\Delta\mathcal{E}_L| < 4$, we cannot confidently distinguish between the abiotic and biotic models. This provides the statistical justification for our agnostic stance in Case 1 (all planets follow the baseline) and contributes to the ambiguity in Case 3.}
    \item \textbf{Model Misspecification and the "Unknown Unknown":} {An anomalous planet that is poorly described by \emph{both} the abiotic model $M$ and the biotic model $M_L$ provides the clearest diagnostic for our "unknown unknown" scenario. This situation will be flagged by the anomalous planet yielding a \textbf{high Pareto $k$ value for both models}. The high $k$ values act as a definitive warning that \emph{neither} model is adequate, reinforcing the conclusion that we are observing a phenomenon not captured by our current understanding.}
    \item \textbf{Small Number of Data Points ($N_p$):} {As stated above, with a small number of planets, all statistical comparisons are challenging. The variance of the elpd$_{\text{LOO}}$ estimate is large, making it difficult to resolve anything but very large differences between models. This means our framework is most statistically powerful when applied to planet-rich systems.}
\end{enumerate}
{Ultimately, LOO-CV and its diagnostics should not be viewed as a life detection algorithm. They are a set of tools to quantify evidence, perform model checking, and identify the most scientifically interesting and anomalous worlds, and guide the next steps of scientific inquiry.}

\section{On the Computation of the elpd}\label{appendix:compute}

As the number of planets increases, computing all $\mathcal{E}(D^p | \mathbf{D}^{-p})$ terms becomes computationally demanding, since each requires a full Bayesian refit. Pareto Smoothed Importance Sampling (PSIS) \citep{vehtari2017practical} provides an efficient alternative by re-weighting posterior samples rather than refitting the model for each planet. Details are given by \citet{vehtari2017practical, vehtari2024pareto} and its application to exoplanet retrievals by \citet{welbanks2023application}.

PSIS approximates LOO–CV by fitting the model once to the complete dataset and using importance sampling to estimate the effect of omitting each planet’s data. Full refits are required only when the approximation fails. Instead of using unstable raw importance weights, PSIS smooths them by fitting a Pareto distribution to the upper tail, which both stabilises the estimate and provides a reliability diagnostic via the Pareto shape parameter $k$.

The PSIS procedure yields the $k$ diagnostic through the following steps:
\begin{enumerate}
    \item \textbf{Compute Importance Weights:} For each planet $p$, and each posterior sample $s$, calculate
    \begin{equation}
        w_s^p \propto \frac{1}{p(D^p | \theta^s, \mathcal{M})}.
    \end{equation}
    Large weights arise when a sample $\theta^s$ makes $D^p$ highly improbable under the full posterior, signalling high influence.
    \item \textbf{Model the Tail Behaviour:} The heaviest weights (typically the top 20\%) are fitted with a Generalised Pareto Distribution (GPD) \citep{zhang2009new}, which models the tail of the weight distribution.
    \item \textbf{Extract the Pareto Shape Parameter:} The estimated shape parameter $\hat{k}$ from the GPD fit quantifies the heaviness of the tail and therefore the influence of that data point.
\end{enumerate}

{\citet{vehtari2017practical} determined empirically that PSIS estimates are reliable for $k \lessapprox 0.7$ whilst cases where $k \gtrapprox 0.7$ indicate the approximation is unreliable, necessitating a full model refit.} This adaptive mechanism allows elpd$_\mathrm{LOO}$ to be computed efficiently across multi-planet datasets while maintaining robust diagnostics of model reliability.

\section{Examples of empirical Abundance-Space habsignatures}\label{appendix:habsig}
{Here, we examine specific habsignature candidates through the lens of Abundance-Space Inference. We assume that high-fidelity abundances have been retrieved (Method 1) and focus on the systemic abundance trends expected under purely abiotic conditions. Deviations from these trends constitute our primary targets for comparative biosignature detection.}

\subsection{Depleted \ce{CO2} on an outer terrestrial planet}

\emph{The nature of \ce{CO2} as a habsignature: }The depletion of \ce{CO2}, or a generally low carbon abundance in the atmosphere of a temperate rocky planet compared to others within the same system, has been proposed by \citet{Triaud2024} as an indicator of the presence of either biomass, extensive liquid water, or active plate tectonics, or a combination of these factors. This is showcased by our solar system, where Earth is uniquely depleted in atmospheric \ce{CO2}, compared to its neighbouring terrestrial planets (Fig. \ref{fig:solar_syst}), showing a clear anomaly suggestive of liquid water.

Life contributes to the biogeochemical carbon cycle on Earth by consuming \ce{CO2} in oxygenic photosynthesis, where carbon is sequestered in soils and hydrocarbon deposits and by creating shells of calcite and aragonite (\ce{CaCO3}) that are stored in sediments \citep{archer1996atlas, bednarvsek2012global}. 

\emph{The abiotic baseline for \ce{CO2}: }Abiotically, \ce{CO2} may be depleted due to several factors, including a nightside cold trap \citep{turbet2018modeling}, a reduced interior, and photodissociation. Among these depletion mechanisms, photodissociation is notably a predictable systemic process, expected to monotonically diminish quadratically with increasing orbital separation. Assuming comparable carbon inventories, a baseline emerges with rising \ce{CO2} abundances at greater distances from the host star, as \ce{CO2} undergoes less destruction. Consequently, an outer planet exhibiting depleted \ce{CO2}, like Earth's depleted \ce{CO2} compared to Venus' (Fig. \ref{fig:solar_syst}), indicates the presence of an alternative depletion process.

\ce{CO2} is outgassed more from oxidised interior outgassing \citep{liggins2022growth}, so is sensitive to planetary redox state. Planetary bulk compositions can be increasingly reduced with orbital distance from their host star \citep{wordsworth2018redox}. Opposing the above-mentioned trend, in secondary planetary atmospheres, higher concentrations of \ce{CO2} could be expected on more oxidised planets, that would be closer to the host star. Though this trend is expected, it remains challenging to remotely constrain planetary interior oxygen fugacity, due to the stochasticity of planetary history; volcanic or not.Nonetheless, there is still a case to be made for a notable anomaly in reduced \ce{CO2} on an inner planet.

Distinguishing between \ce{CO2} depletion as a ``habsignature" and ``biosignature" presents a challenge. To discern between habsignatures and biosignatures, searches for \ce{CO2} depletion should be complemented by the detection of other comparative biosignatures, unique to life. The identification of multiple comparative biosignatures would provide more robust constraints on the processes shaping atmospheric composition, considering all known abiotic processes that could plausibly explain observed features.

\subsection{Depleted \ce{NH3} in outer planet}

\emph{The nature of \ce{NH3} as a habsignature: }Similarly to \ce{CO2}, depletion of ammonia (\ce{NH3}) in the atmosphere of a temperate rocky planet compared to others within the same system, could serve as a habsignature, a biosignature, or a habiosignature. \ce{NH3} can both be produced by life as a product of metabolisms, and consumed due to its high bio-usability, as plants and various microorganisms readily absorb \ce{NH3}. Previous studies have focused on detecting the presence of \ce{NH3} as a biosignature, where life acts as a net source of \ce{NH3}, saturating surface sinks \citep[e.g.,][]{seager2013biomass}. For a comprehensive review of ammonia as a biosignature gas in exoplanet atmospheres, see \citet{Huang2022}. In this study, we instead emphasise the depletion of \ce{NH3} as a comparative habsignature and biosignature.

Ammonia on Earth is primarily produced from anthropogenic sources; industrial by-products, agriculture-related biomass decomposition and motor vehicles \citep{zhu2015sources}. Abiotic sources include lightning \citep{2017MNRAS470187A, mancinelli1988evolution} and volcanic outgassing \citep{schaefer2010chemistry, liggins2022growth} that, although negligible compared to biogenic production fluxes, could dominate on other planets. Large impacts on water vapour atmospheres can also create a reducing environment favourable for the formation of \ce{NH3} \citep{Itcovitz_2022}.

There are limited depletion pathways to remove \ce{NH3} from the atmosphere. Ammonia can be destroyed directly by photolysis \citep{1982JGRKasting, seager2013biosignature}, or depending on the atmospheric background composition, \ce{NH3} can react with photochemically produced radicals such as O, H, and OH \citep{2012ApJHu}. Ammonia is also very soluble in water \citep{sander2015compilation} and is therefore susceptible to depletion from dissolution in oceans.

Biogenically, \ce{NH3} can be produced by ammonification (converting organic nitrogen to ammonia within organic matter), dissimilatory nitrate reduction to ammonium, and biological nitrogen fixation (bacteria and archaea convert atmospheric \ce{N2} to \ce{NH3}). On Earth, nitrogen fixation dominates production rates \citep{rascio2008biological}, although its energetically costly in nature and the requirement for strictly anoxic conditions \citep{burgess1996mechanism} present challenges to its development on other planets.

In addition to acting as a source, life is also a significant sink for \ce{NH3} on Earth. \ce{NH3} serves as an ideal nitrogen source for life, being easily integrated into various amino acids and organic molecules without the need to break the strong triple bond \ce{N2}. Microbial life can also metabolise \ce{NH3} as an energy source through its anaerobic oxidation to \ce{N2} \citep{Kartal}, further acting as a sink for ammonia. 

Earth metabolisms currently serve as a net source of \ce{NH3} \citep[][and references therein]{Huang2022}, which complicate the use of \ce{NH3} depletion as a  ha\emph{bio}signature if biotic sources are offsetting oceanic \ce{NH3} depletion. Large biotic \ce{NH3} sources could lead to false negatives in habiosignature assessments, as \ce{NH3} would not appear depleted despite the presence of an ocean and active life. In such cases, combining \ce{NH3} data with other indicators, like \ce{CO2} depletion, becomes essential for strengthening evidence of habitability.

\emph{The abiotic baseline for \ce{NH3}: }The comparative approach offers a means to distinguish between \ce{NH3} depletion from photochemistry, or depletion from the presence of an ocean or life, from the relative abundance of \ce{NH3} as a function of orbital separation. \ce{NH3} would experience higher rates of photochemical depletion in planets closer to the host star due to increased UV energy (directly photolysing ammonia, or producing more reactive radicals) whereas outer planets may allow \ce{NH3} to accumulate more easily in their atmospheres. However, caution is needed: on planets sufficiently far from the star, \ce{NH3} would freeze out and thus remain undetectable. Here, atmospheric models are essential to align \ce{NH3} observations with predictions, to determine whether the non-detection of \ce{NH3} is consistent with expectations. Here, key anchor point for calibrating the \ce{NH3} baseline is planets at orbital distances where \ce{NH3} remains gaseous, yet accumulates to detectable abundances.

It may perhaps be challenging to imagine multiple exoplanets developing sufficiently high \ce{NH3} levels to notice a depletion, considering Earth's low abiotic \ce{NH3} production.  Instabilities in a planetary system can raise impact rates across all planets \citep[similar to the  solar system's early instability;][]{CLEMENT2019778}, distributing sufficient reducing power to form \ce{NH3} \citep{Itcovitz_2022}. When observing young systems, their atmospheres thus could be in such a global transient post-impact environment. More impacts likely occur on outer system planets in `peas in a pod' architectures (Section \ref{sec:ch4}), leading to higher \ce{NH3} accumulation. As such, a depletion of \ce{NH3} in such outer planets, relative to the baseline mapped out by other planets, may suggest removal by an ocean.


\bsp	
\label{lastpage}
\end{document}